\documentclass[acmtog,authorversion,nonacm]{acmart}

\setcopyright{acmcopyright}
\acmJournal{TOG}
\copyrightyear{2022}
\acmYear{2022}
\acmDOI{10.1145/1122445.1122456}

\definecolor{darkpastelgreen}{rgb}{0.01, 0.75, 0.24}
\definecolor{darkpastelyellow}{rgb}{0.70, 0.70, 0.01}

\newcommand{\secref}[1]{Sec.~\ref{#1}}

\newcommand{\figref}[1]{Fig.~\ref{#1}}

\newcommand{\eqnref}[1]{Eq.~(\ref{#1})}

\usepackage{dsfont}
\newcommand{\R}{\mathds{R}}
\usepackage{multirow,tabularx}
\usepackage{wrapfig}
\usepackage{enumitem}

\usepackage{caption}
\usepackage{subcaption}

\begin{document}

\title{Computational Pattern Making from 3D Garment Models}

\author{Nico Pietroni}
 \affiliation{
   \institution{University of Technology Sydney}
   \city{Sydney}
   \country{Australia}
 }
\email{nico.pietroni@uts.edu.au}

\author{Corentin Dumery}
\affiliation{
  \institution{ETH Zurich}
  \city{Zurich}
  \country{Switzerland}
}
\email{corentin.dumery@gmail.com}

\author{Raphael Guenot-Falque}
\affiliation{
  \institution{University of Technology Sydney}
  \city{Sydney}
  \country{Australia}}
\email{raphael.guenot-falque@uts.edu.au}

\author{Mark Liu}
\affiliation{
  \institution{University of Technology Sydney}
  \city{Sydney}
  \country{Australia}}
\email{Mark.Liu@uts.edu.au}

\author{Teresa Vidal-Calleja}
\affiliation{
  \institution{University of Technology Sydney}
  \city{Sydney}
  \country{Australia}}
\email{teresa.vidalcalleja@uts.edu.au}

\author{Olga Sorkine-Hornung}
\affiliation{
  \institution{ETH Zurich}
  \city{Zurich}
  \country{Switzerland}
}
\email{sorkine@inf.ethz.ch}

\renewcommand{\shortauthors}{Pietroni et al.}

\begin{abstract}

We propose a method for computing a sewing pattern of a given 3D garment model. Our algorithm segments an input 3D garment shape into patches and computes their 2D parameterization, resulting in pattern pieces that can be cut out of fabric and sewn together to manufacture the garment. Unlike the general state-of-the-art approaches for surface cutting and flattening, our method explicitly targets garment fabrication. It accounts for the unique properties and constraints of tailoring, such as seam symmetry, the usage of darts, fabric grain alignment, and a flattening distortion measure that models woven fabric deformation, respecting its anisotropic behavior. We bootstrap a recent patch layout approach developed for quadrilateral remeshing and adapt it to the purpose of computational pattern making, ensuring that the deformation of each pattern piece stays within prescribed bounds of cloth stress. While our algorithm can automatically produce the sewing patterns, it is fast enough to admit user input to creatively iterate on the pattern design. Our method can take several target poses of the 3D garment into account and integrate them into the sewing pattern design. We demonstrate results on both skintight and loose garments, showcasing the versatile application possibilities of our approach.

\end{abstract}


\begin{teaserfigure}
\centering
  \includegraphics[width=0.95\textwidth]{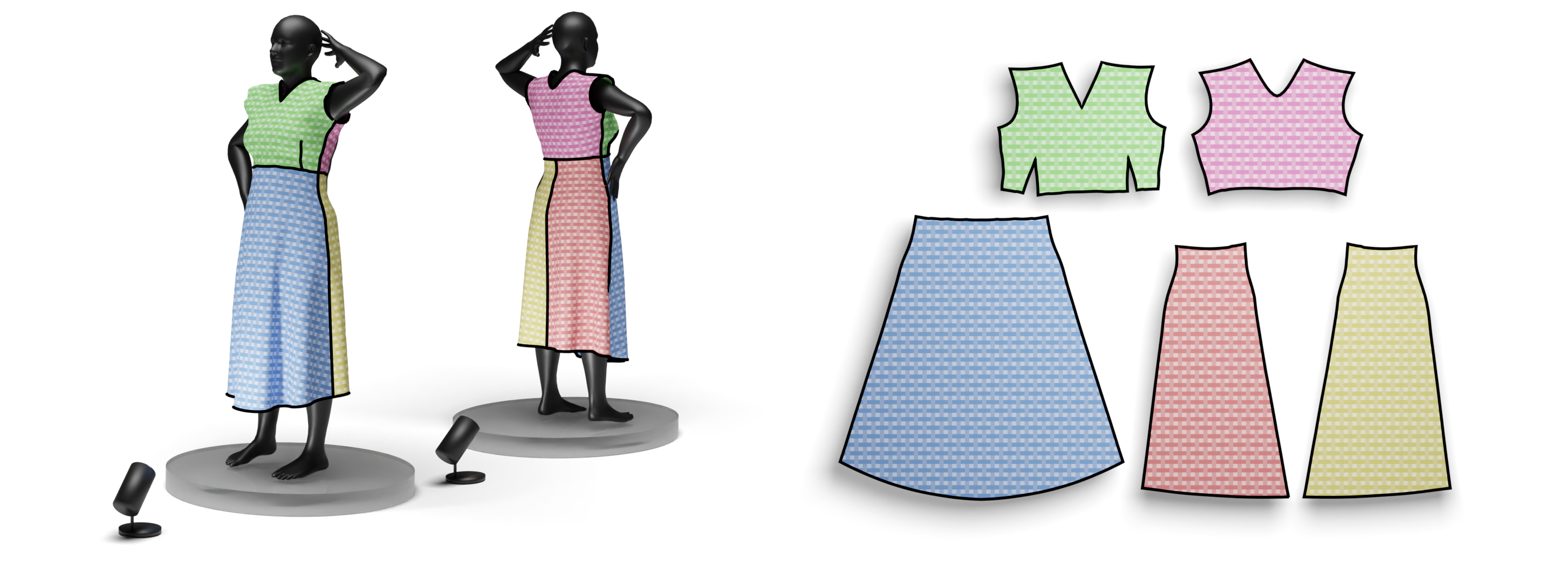}
    \caption{Our technique receives a 3D mesh of a given garment as input, automatically segments the shape into patches (left and middle) and computes the 2D parametrerization required to produce a pattern (right) that can be cut out and sewn to fabricate the actual garment.}
  \label{fig:teaser}
\end{teaserfigure}

\maketitle

\section{Introduction}
\label{sec:introduction}

In this work, we propose a method for automatically creating a sewing pattern for a given 3D model of a garment. 

In the fashion industry, the garment creation process starts from the 2D domain: The pattern maker creates the 2D sewing pattern using traditional, often tacit knowledge \cite{Chen1998}, established templates and a few standard measurements, such as waist circumference, shoulder width, etc. The cut fabric pieces are then sewn together to form the garment. Designers may work with mannequins to experiment with the desired shape and draping of the fabric in the physical 3D space, but the ultimate determination of the garment shape comes from the 2D pattern. Also, in digital garment design tools such as Clo3D~\shortcite{CLO}, the 2D pattern is required to simulate and drape the garment on a virtual mannequin or avatar, and thus the design process is centered around the sewing pattern. Pattern making is currently a manual, time-consuming task requiring much experience and skill. One of the consequences is that custom made clothing that perfectly fits the intended wearer is a luxury available to the privileged few. The majority of clothing is mass-produced using standard sizes based on averaged measurements that do not fit well for most people \cite{SizeGermany}.

There are many powerful techniques to model shapes directly in 3D, with various approaches specifically developed for 3D garment modeling, e.g.\ with sketch based user interfaces \cite{Katja:2021,Rose:DevelopableSurfaces:2007,decaudin2006,Turquin:SketchInterface:2007,Robson:ContextAwareGarments:2011}, input gestures in the physical~\cite{Wibowo:Dressup:2012} or virtual reality~\cite{TiltBrush}, via 3D scanning~\cite{pons2017clothcap} or data driven modeling~\cite{Wang:GarmentShapeSpace:2018}. Advances in human body modeling~\cite{bogo2014faust,SMPL:2015,osman2020star-body,imGHUM:2021} facilitate custom digital tailoring of garments that fit the personalized body avatar without having to rely on standard sizes. However, the 2D sewing pattern of the garment is still necessary for accurate cloth simulation and the manufacturing of the physical garment.

To create the 2D pattern, the 3D garment surface is divided into patches, also called panels, and each panel is flattened onto the 2D domain. Many  methods exist for surface segmentation and low-distortion parameterization, as we discuss in \secref{sec:related}. State-of-the-art approaches are able to compute optimal cuts that balance the flattening distortion and cut length~\cite{Sharp:2018:VSC,Li:2018:OptCuts,Poranne:Autocuts:2017}. Recent work~\cite{Katja:2021} applies such a technique to create the sewing pattern, but the resulting panel shapes are visually far from common practice in fashion and challenging to sew. The problem is that most segmentation techniques are general and do not account for the specific setting of cloth pattern making and the fabrication constraints unique to it. In particular, parameterizing a 3D shape to be made of woven textile requires a special deformation model aware of the thread properties and fabric grain alignment, as well as shape and length constraints on the seams and garment aesthetics. For example, to sew together two pieces of cloth, in practice, the two matching seams must be of the same length, as straight or at least smooth as possible, and ideally they should be reflection-symmetric in 2D, so that the tailor can put the panels one on top of the other and sew them together along one planar curve. 

In this work, we propose a shape segmentation and patch flattening method specifically intended for pattern making (see \figref{fig:teaser}). Our core contributions are a 3D patch layout creation approach informed by a geometric measure of woven fabric distortion and tailoring fabrication requirements, as well as a patch parameterization method that minimizes the textile deformation measure and ensures that the pattern pieces are sewable. Our method incorporates mechanisms of classical pattern making, such as the usage of darts, preference towards global symmetry and vertical grain alignment by default. Our algorithm can work fully automatically, but it is efficient enough to admit interactive input, allowing creative design iteration in real time. The designer can sketch on the 3D garment model to hint at desired seams and textile grain alignment and vary the flattening parameters according to the physical properties of the desired fabric. As an additional option, if multiple poses of the garment corresponding to various body poses are available, our algorithm can adapt the pattern by integrating the information from all the poses. We demonstrate results on a number of tight-fitting and loose garments, showcasing the usability and versatility of our method. To foster future research on digital fashion, we will publicly release our software implementation.

\section{Related work}
\label{sec:related}

Digital garment design and fabrication pose several fundamental, interconnected research problems in geometry processing and physics based simulation. We give a brief review of the methods for surface modeling, patch decomposition and flattening in the context of computational pattern making.  

\vspace{-3pt}
\paragraph{Garment shape design.}
The digitalization of the fashion industry in general and garment design in particular bears multiple economical, ecological and societal advantages and poses fascinating research challenges, sparking significant interest \cite{nayak:2017:automation}. 
Clo3D \shortcite{CLO} and Optitex~\shortcite{Optitex} are examples of common CAD tools used in the industry. Such interactive CAD editors allow the designer to create sewing patterns in 2D and simulate their physical appearance and draping on an avatar in 3D. Adjusting the 3D garment shape requires changing the 2D pattern via a trial-and-error process. In the research community, Umetani et al.~\shortcite{Umetani:2011} propose bidirectional interactive garment editing, leveraging fast cloth simulation to enable users to work in 2D and 3D simultaneously and observe the effects of changes in both modes.  

\begin{figure}[t]
    \begin{tabular}{@{}c@{}}
        \includegraphics[width=0.9\linewidth]{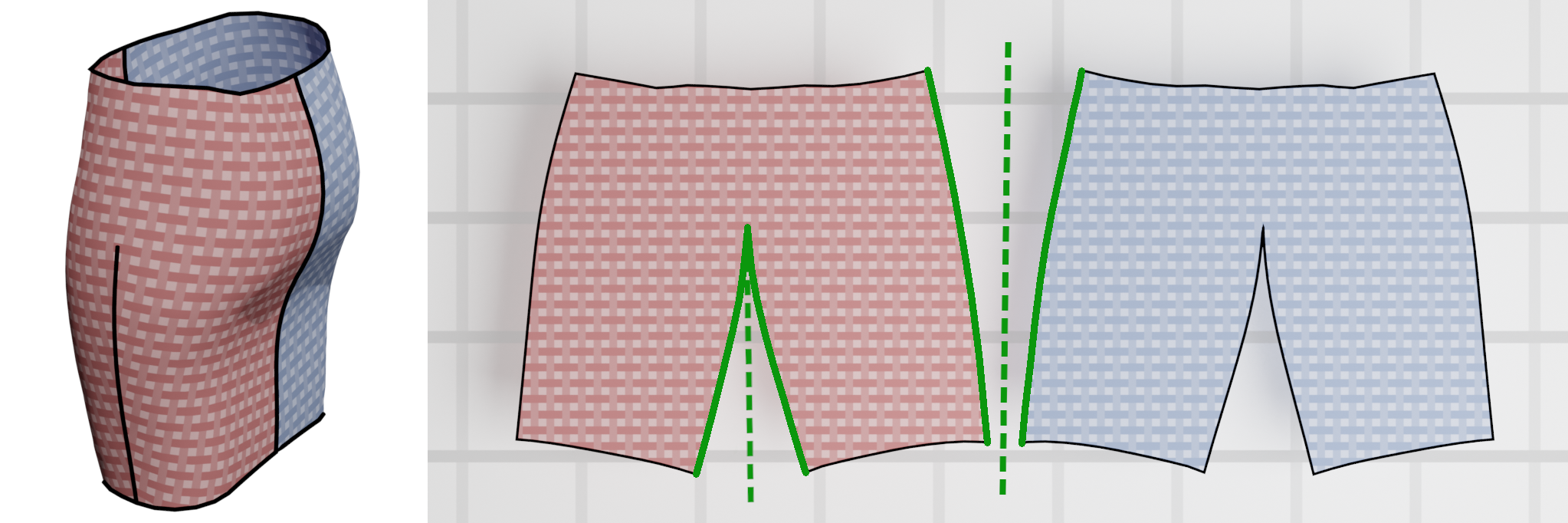}\\
        \includegraphics[width=0.9\linewidth]{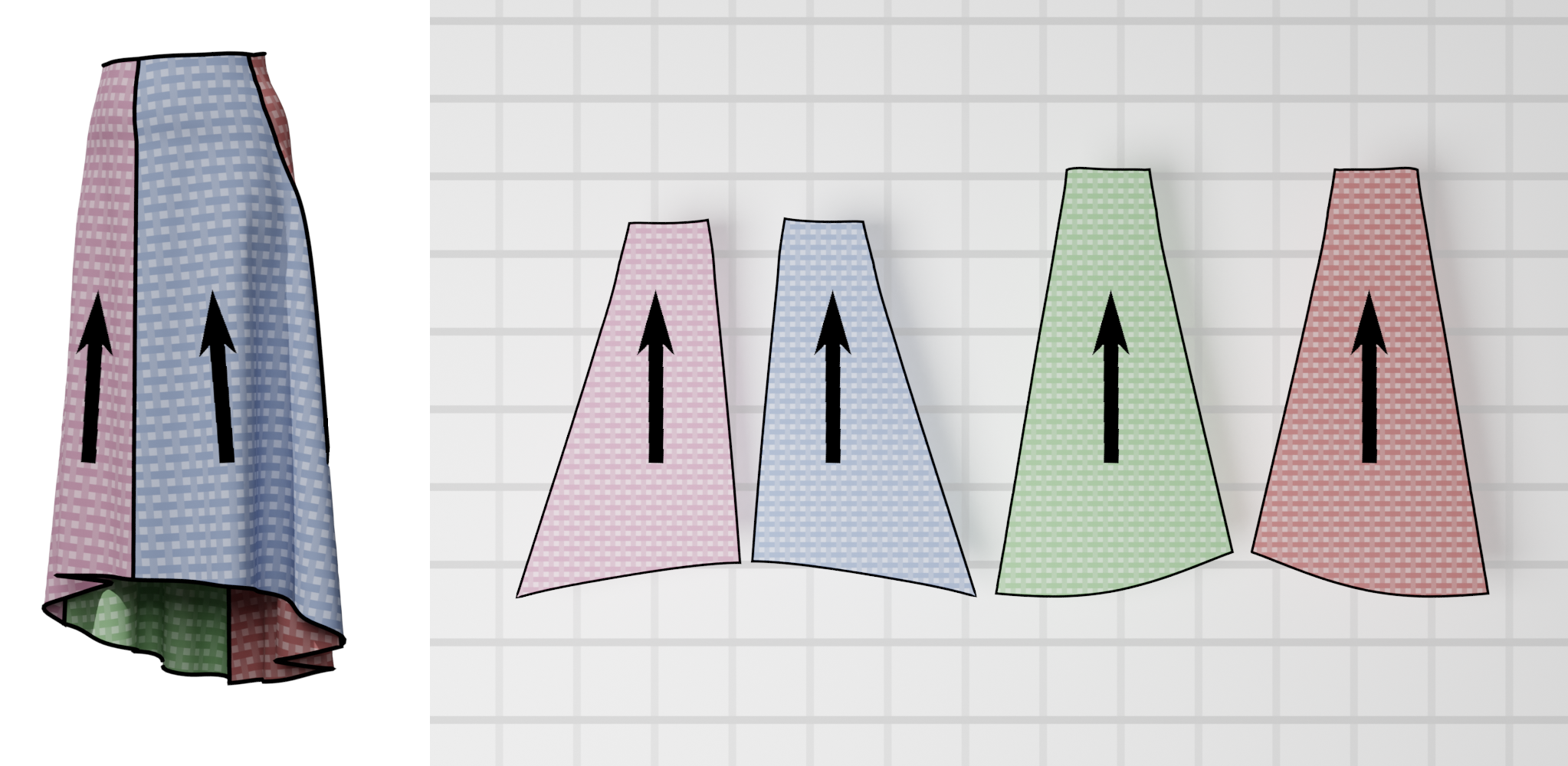}
    \end{tabular}
    \caption{Top: matching seams on two patches to be stitched together, as well as the sides of a dart must be of equal length and ideally reflection-symmetric. Bottom: the fabric grain of pattern pieces aims to align with the vertical direction on the worn garment.  
    }
    \label{fig:requirements}
\end{figure}

\begin{figure*}[t]
    \begin{tabular}{@{}c@{}c@{}c@{}c@{}}
        \includegraphics[height=0.2\linewidth]{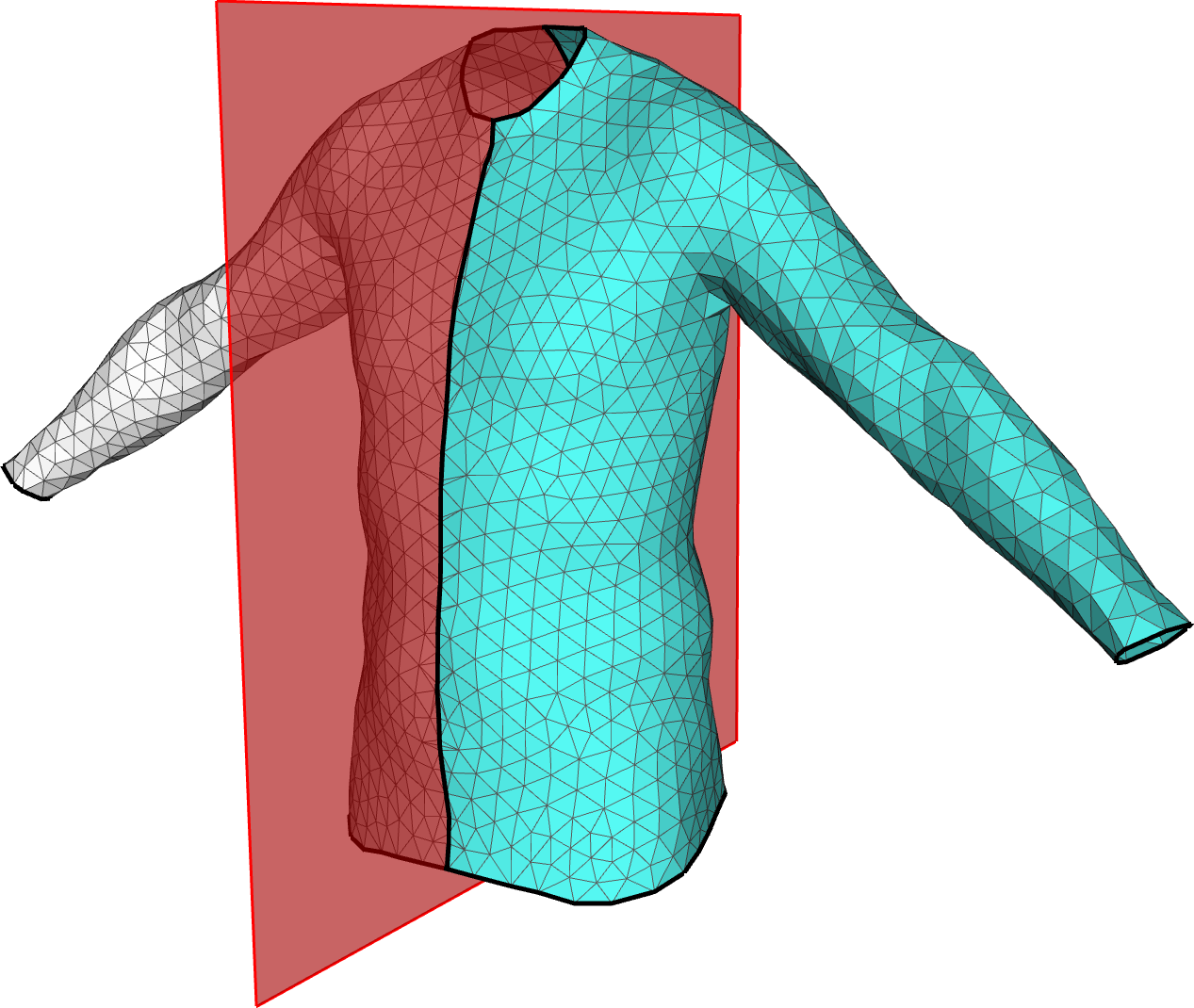}&
        \includegraphics[height=0.18\linewidth]{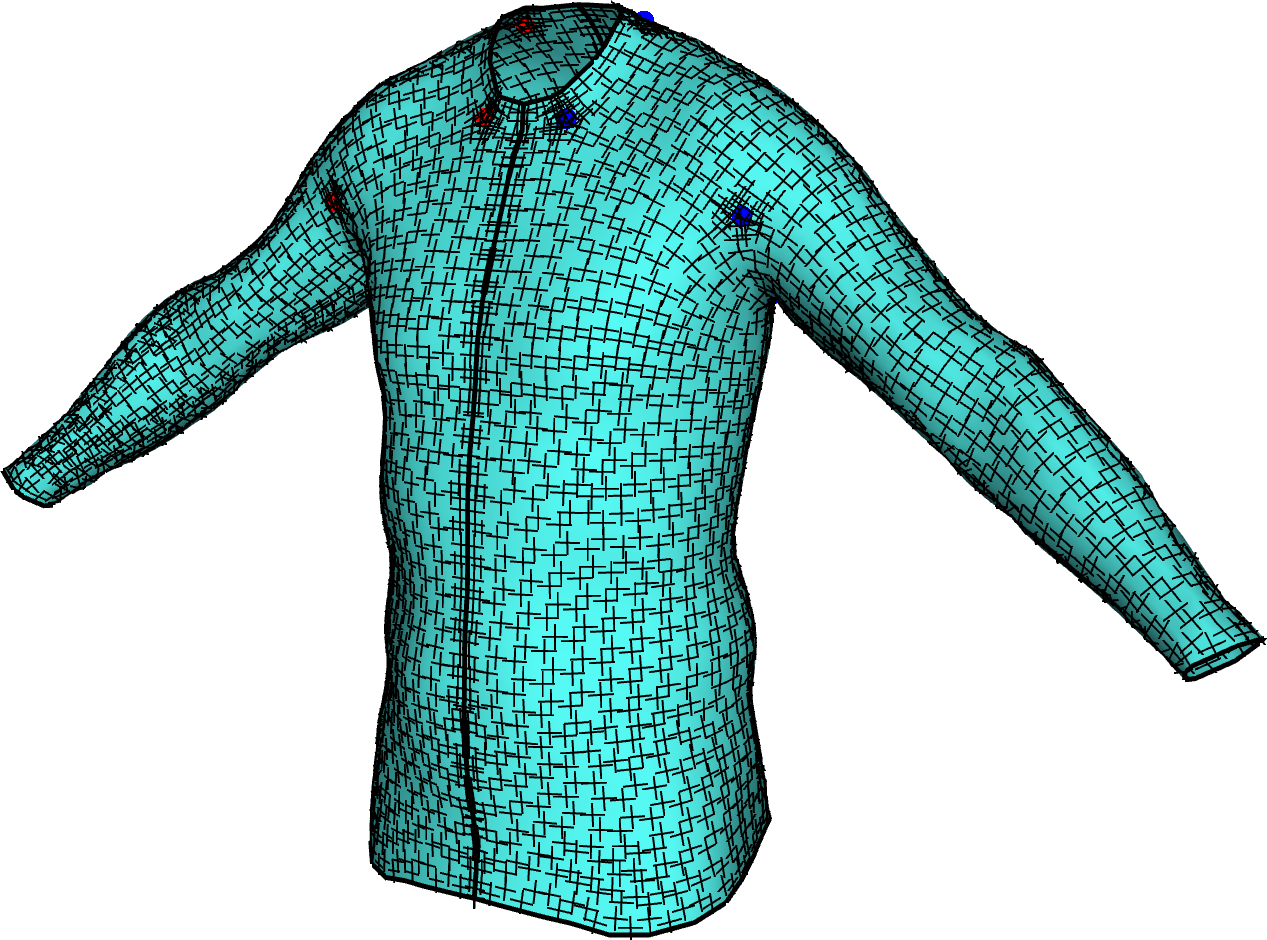}&
        \includegraphics[height=0.18\linewidth]{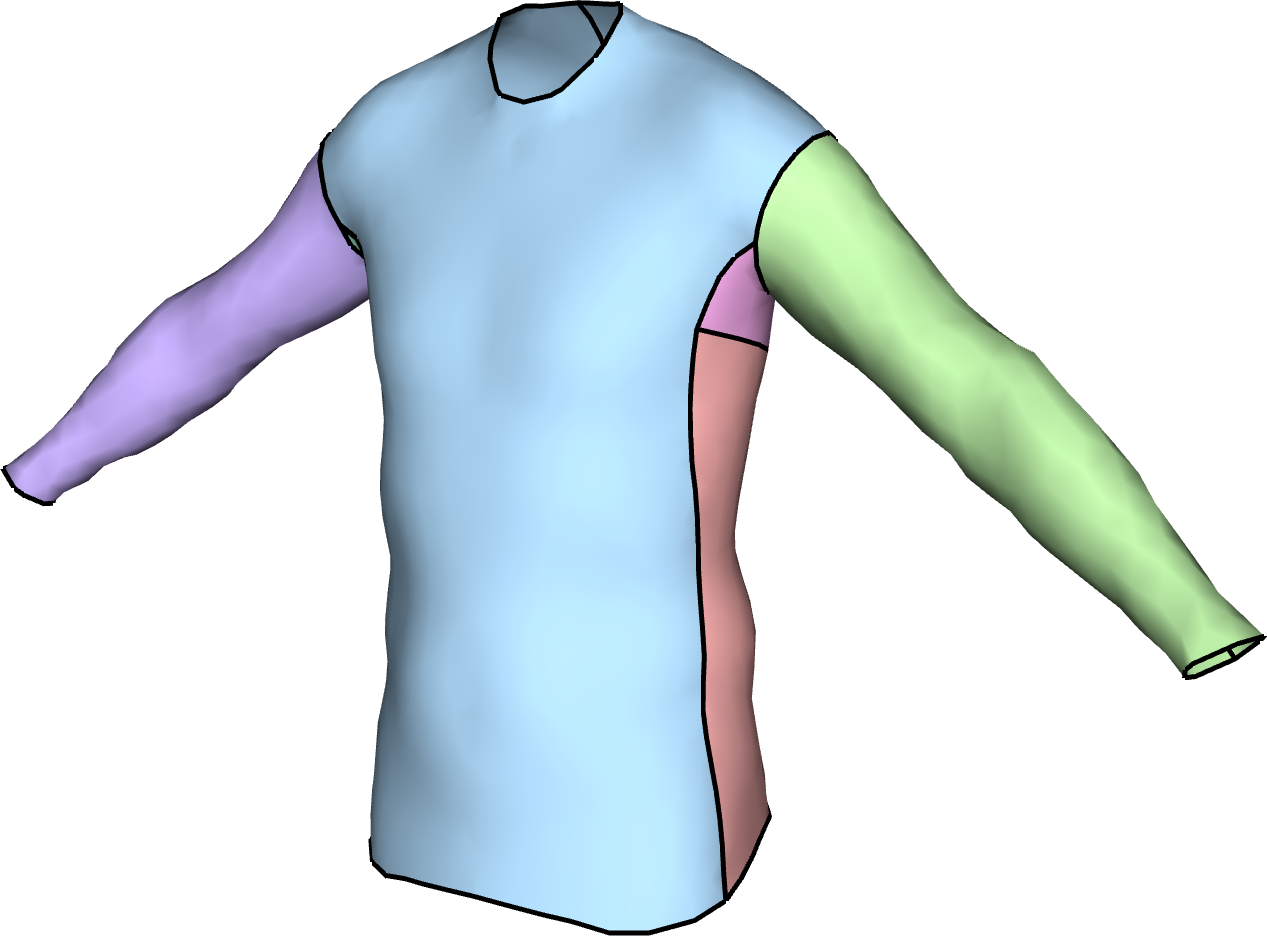}&
        \includegraphics[height=0.22\linewidth]{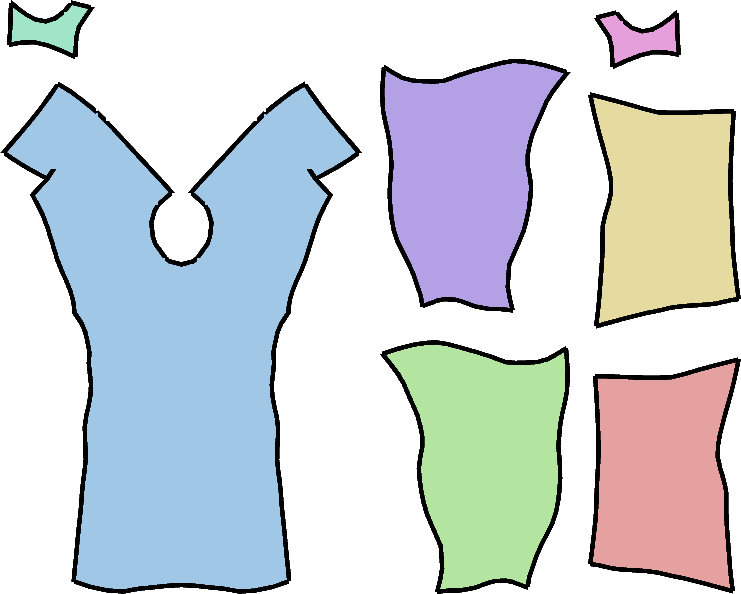}\\
        \small (a) symmetrization & 
        \small (b) cross-field &
        \small (c) path tracing \& patch decomposition & 
        \small (d) patch parameterization
    \end{tabular}
    \caption{An overview of the different steps of our pipeline. The shape is first symmetrized (a), a smooth cross-field is computed on the mesh (b), paths are then traced on the shape by following the cross-field (c), and the resulting patches are then flattened onto the $UV$ plane to create a 2D sewing pattern (d).}
    \label{fig:pipeline}
\end{figure*}

Various computational approaches create 3D garment shapes from contours, seam-, fold- and boundary curves, sketched by the user on an avatar, see e.g.~\cite{Turquin:SketchInterface:2007,decaudin2006,Rose:DevelopableSurfaces:2007,Robson:ContextAwareGarments:2011,Katja:2021}.
Other approaches leverage data and neural networks to construct 3D garments from 3D scans~\cite{Chen2015,pons2017clothcap,Bang2021} or by using parametric models and learning~\cite{Wang:GarmentShapeSpace:2018,Vidarre2020}; a recent data set of 3D garments with corresponding 2D patterns can be used for supervision~\cite{GarmentDataset:2021}. To simulate and further process the garment models, most methods require 2D patterns, which are either provided as input or computed using surface segmentation and parameterization, as discussed below. Dedicated approaches to knitwear design and fabrication are explored in  \cite{McCann2016,Narayanan*2019,Yuksel2012,Narayanan2018,Wu2019}, where a sewing pattern is not needed, since knitting machines can continuously knit various topologies. By contrast, in this paper we focus on pattern based garment construction from woven fabric (or knitted fabric such as jersey, treated similarly), which is an overwhelmingly widespread practice in the garment industry.

\vspace{-3pt}
\paragraph{Sewing pattern design and optimization.}
Several methods \emph{modify given} garment patterns to fit a particular body shape \cite{cordier2003made,Wang2005,Brouet2012,meng2012flexible,Bartle2016,Wang2018patterns,Bang2021,LIU2018113}, optimize stress, pressure and seam traction \cite{Montes2020} or generate user-defined target folds \cite{Li:2018:FoldSketch}. They do not modify the given topological patch layout. In contrast, our method \emph{generates} a  sewing pattern directly from the 3D garment model while respecting fabric stress bounds and fabrication constraints. 

The evolutionary algorithm by Kwok et al.\ \shortcite{Kwok2015} generates a sewing pattern from 3D body geometry, but it does not account for fit or fabric stress and manufacturing requirements.

\vspace{-3pt}
\paragraph{Patch layout decomposition.}
General surface segmentation methods for the purpose of low-distortion flattening aim at creating a small number of patches (or short cuts) to reduce Gaussian curvature. A number of works compute the patches and cuts jointly with the parameterization in order to optimize or bound both~\cite{BoundedDistortParam:2002,Poranne:Autocuts:2017,Li:2018:OptCuts}. Variational surface cutting~\cite{Sharp:2018:VSC} optimizes the tradeoff between a given distortion measure and cut length. Another class of works computes approximations of a given shape with (nearly) developable surfaces~\cite{D-Charts:2005,Stein:2018,Ion:ApproximatingDOGs:2020,BinningerVerhoeven:GaussThinning:2021}. However, the seams generated by all such methods are not appropriate for garment fabrication because they are far too complex and difficult to sew and do not necessarily fulfill even the basic requirement of equal length for matching seams, nor symmetry.
Huang  et al.\ \shortcite{HUANG2012680} use predefined cuts based on an anthropomorphic subdivision, which do not generalize to other garments or body shapes.

Methods for patch layout generation for the purpose of globally smooth parameterization and semiregular remeshing focus on the quality of the layout and its combinatorial structure \cite{Campen2017survey}. State-of-the-art techniques compute a smooth tangent vector field on the surface~\cite{Vaxman2017} aligned to features and principal directions as the basis for tracing the patch decomposition~\cite{RazaR15,Pietroni2016,QuadMixer,LivesuPPSC20,Pietroni2021}. The topology of these layouts is restricted by the remeshing application, e.g.\ four corners per patch for quadrangulation, which is too stringent for pattern making. We adapt the method by Pietroni et al.~\shortcite{Pietroni2021} to suit the different setting of garment design and fabrication, where the topology requirements are different and textile based distortion measures must be bounded.

\vspace{-3pt}
\paragraph{Surface parameterization and textile distortion measures.}

General flattening methods take a surface of disk topology and find its mapping to 2D that minimizes a distortion measure \cite{hormann:MPT:2007}. The used objective most typically measures conformal distortion (e.g.\ \cite{Levy2002,sheffer:inria-00105689}) or deviation from isometry (e.g.\ \cite{Liu2008,Rabinovich:SLIM:2017}). Guaranteeing locally and globally injective parameterization is possible~\cite{Sawhney2017,Jiang2017} but can be computationally costly. Most importantly, the general distortion measures are isotropic and do not account for the particular behavior of woven fabric, which is almost inextensible along the yarn directions and more stretchable in the diagonal direction.  In other words, developable surfaces are global minimizers of all these distortion measures, but non-vanishing locally minimal energy states do not necessarily model plausible fabric configurations.

McCartney et al.~\shortcite{McCartney2000,MCCARTNEY2005} and Wang et al.~\shortcite{Wang:WovenMesh:2005} look specifically at models for woven fabric and flatten the mesh by discriminatively optimizing the yarn stretch and shear. Wang et al.~\shortcite{Wang:WovenMesh:2005} create a discrete orthogonal grid to represent the textile and wrap it around a target 3D mesh, thereby establishing the 2D-to-3D mapping. McCartney et al.\ opt for a continuous approach, where the intextensible yarn directions are represented by the UV-axes in the flat domain. Their energy optimization is nonlinear and its optimization could be  impractical for realtime interactive design iteration framework. Our textile deformation model is inspired by these works, but we use a different optimization strategy to achieve fast performance, and we incorporate seam symmetry~\cite{Wolff:Symmetry:VMV2019} and grain alignment constraints for fabrication feasibility.

\section{Overview}
\label{sec:overview}

Our pipeline takes as input a triangle mesh representing the target garment in 3D and {automatically} produces a 2D sewing pattern. The designer can  influence the final layout by sketching some curves on the surface, indicating desired seams. In addition to the rest shape in 3D, our approach can also consider additional target poses that the garment should assume when worn.

Our method is intended for woven fabrics that consist of warp and weft yarns. The warp direction is the main (longitudinal) direction of the fabric, also termed \emph{grain}, and the weft direction is orthogonal to it. Depending on the material, the textile threads can be elastic or nearly inextensible, determining the stretching resistance of the fabric along the warp and weft directions. The fabric can usually stretch more along the diagonal direction, resulting in shear of the woven structure. The main parameters of our framework are therefore the maximal admissible stretch and shear, determining the maximum deformation that the material undergoes from the flat configuration to the draped 3D shape. Note that excessive shear is undesirable, even if physically possible, because it leads to wrinkles and impression of bad fit. Ideally, the deformation of the pattern pieces is close to isometry. We assume  the fabric is sufficiently thin and its resistance to bending too small influence the pattern making.

\subsection{Requirements}
\label{sec:requirements}

To compute a usable, manufacturable sewing pattern for a garment, our algorithm should fulfill several (interdependent) requirements.

\emph{Patch shape.} The pattern should preferably consist of few large patches with straight or at least smooth boundary segments. Practical patterns pieces have few corners (between 6 and 8), and their angles should be approximately orthogonal. 
If a patch contains darts, their opening angles should also be sufficiently large, and the dart length usually should not exceed a fraction of the patch length, as it is then easier in practice to cut through and sew two separate pieces together.

\emph{Bounded fabric strain.} According to the mentioned fabric parameters, the textile cannot exceed the prescribed deformation threshold when draped into the target 3D shape. The lower this bound, the more pattern pieces are required for doubly curved target shapes.

\emph{Seam sewing feasibility.} Matching seams to be sewn together must have equal length. In practice, straight seams are easiest to sew, and it is desirable that seams can be sewn flat, meaning that the matching seams must be reflection-symmetric, so that the two pieces of fabric can be placed on top of each other to sew (\figref{fig:requirements}, top). 

\emph{Layout symmetry.} Aesthetics is strongly associated with symmetry, and therefore symmetric sewing patterns are desirable, in particular when the targeted garment design is symmetric.

\emph{Grain alignment.} A pattern piece should be oriented in 2D such that the fabric grain is roughly aligned with the vertical (gravity) direction when the garment is worn, or with some other axis, such as along the arm (see \figref{fig:requirements}, bottom). Grain alignment ensures predictable, symmetric behavior when the garment is draped and subject to gravity, and pressure of the body, as well as when shrinkage due to washing occurs. Sometimes the designer may wish to prescribe a special grain direction, as in bias cut, where the grain is at 45 degrees to the vertical direction, e.g.\ in some skirt designs.

\emph{Efficiency.} Constructing an efficient pipeline allows the designer to creatively iterate and adjust the patch layout by providing simple input by interactively sketching on the 3D surface.

\subsection{Pipeline structure}

\figref{fig:pipeline} summarizes our processing pipeline. 

\emph{Input.} We assume the input garment mesh to be a manifold and with well shaped triangles (otherwise we uniformly remesh with standard tools).
If additional poses of the 3D garment are available, we assume their meshes share the connectivity and are in full correspondence with the rest pose. 
If the input garment is symmetric, we can enforce symmetry throughout the pipeline by splitting the mesh along the symmetry plane, running the method on one side, and then reflecting the result (\figref{fig:pipeline}a).

\emph{Cross-field construction.} We compute a smooth 4-rotational-symmetric (4-RoSy) tangent vector field on the surface (\figref{fig:pipeline}b) aligned with principal curvature directions and boundaries. If the input includes multiple poses, we integrate the contribution of the curvature of all the poses in the derived field (see~\secref{phaseCrossfield}). 

\emph{Layout construction.} With the goal of producing pattern pieces that fulfill the requirements above, we trace a set of \emph{paths} across the mesh to partition the surface into the different panels (see \secref{sec:layout} and \figref{fig:pipeline}c). The paths are oriented to follow the underlying cross-field and include the curves sketched by the designer. The patch formation is controlled by the textile distortion measure, and the designer can specify the maximum number of corners allowed in a patch, to control patch complexity. At the end of this step, the seam on the symmetry plane can be optionally removed, if possible.

\emph{Patch flattening.} 
Finally, each patch is flattened onto the 2D space and packed with the others into a single textile sheet (see \figref{fig:pipeline}d). This step is achieved by a novel parameterization method that mimics textile physics using geometric measures.
We impose constraints on the seams to ensure that they can be physically sewn. The parameterization step is also deployed during patch layout construction to individually uphold the distortion threshold of each patch.

\section{Method}

The following sections detail each phase of our processing pipeline.

\label{sec:methods}
\subsection{Cross-field construction}
\label{phaseCrossfield}

\begin{figure}[b]
    \begin{tabular}{cc}
        \includegraphics[height=0.5\linewidth]{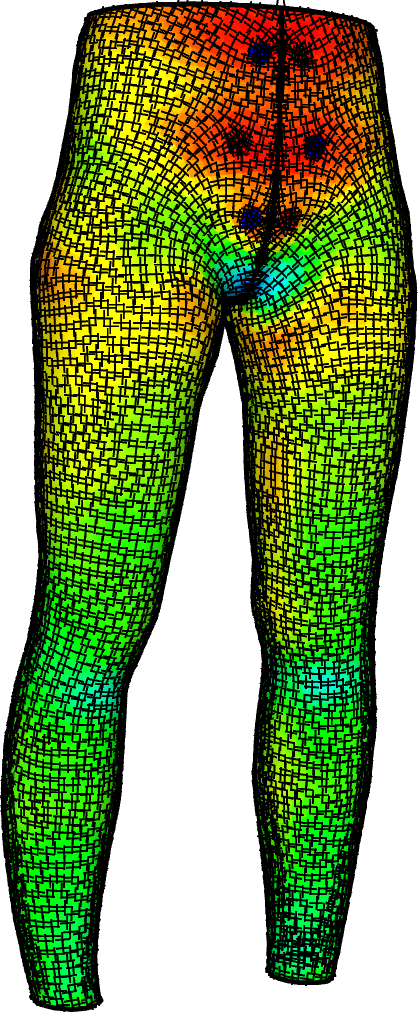}&
        \includegraphics[height=0.5\linewidth]{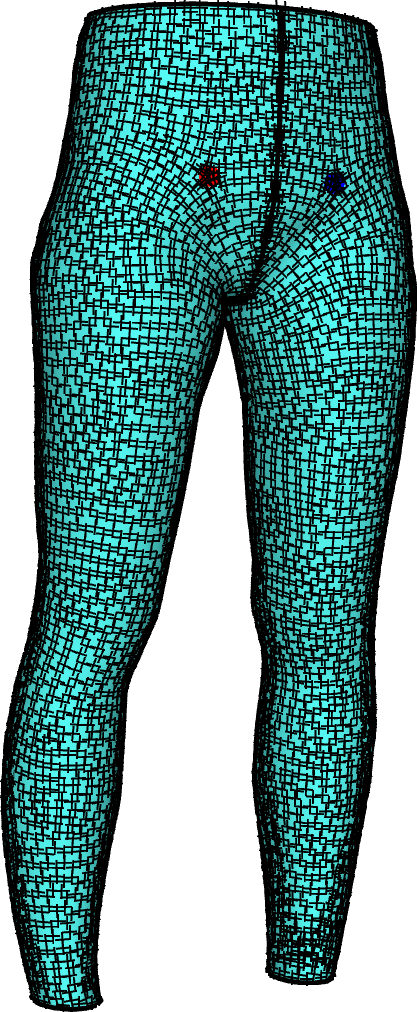}
    \end{tabular}
    \caption{Cross-field computation: The multi-scale main curvature directions used as soft constraint to retrieve a smooth field.}
    \label{fig:field}
\end{figure}

Given a proper input, we initially construct a cross-field (a 4-RoSy tangent-vector field \cite{Vaxman2017}) on the input surface. The cross-field is crucial in our framework as it drives the entire tracing process; the cuts follow the directions expressed by the cross-field. Intuitively, aligning the seams to the main curvature directions is an excellent strategy to maximize the developability of the resulting patches. A significant collection of quadrangulation methods have demonstrated the strong correlation that exists between developability and curvature alignment \cite{Bommes2009,Bommes2013,Pietroni2021}. 

We initialize the field by extracting curvature directions at a low scale using the method proposed in \cite{Pan2010}. Similarly to \cite{Bommes2009} we use the anisotropy of the curvature directions
to detect the regions where these curvatures are essential (see \figref{fig:field}, left). 

Then we run the globally smooth method proposed by \citet{Diamanti2014}, adding soft constraints to align the field to the principal curvature directions, similarly to \cite{PanozzoLPZ12}. The anisotropy weighs each soft constraint value we previously computed to adhere to the principal curvature directions where those are relevant, and smooth on the rest. In addition to these soft constraints, we impose the field to align simultaneously with the user-defined constraints and the boundary edges of the 3D mesh. This way, the path will hit the boundary orthogonally, and we implicitly avoid creating artifacts or long stripes (\figref{fig:sleeves}). 
The resulting field is shown in \figref{fig:field}, right. 

When multiple frames are provided, we first compute the multiscale curvature and the anisotropy for each frame. Then, we average the field directions for each face in the rest shape (we transport the field on the best shape and perform $\pi/2$ invariant interpolation of the cross-field, weighted by the anisotropy values). Lastly, we smooth all the fields together in the rest position and average the anisotropy among all frames.

 \subsection{Patch layout creation}
 \label{sec:layout}
This pipeline step obtains a patch layout as the byproduct of an iterative process that traces field-aligned paths over the target surface. The network of paths designs a patch layout composed of rectangular or non-rectangular patches. Distinct paths can only intersect orthogonally on the surface. Two intersecting paths can cross or stop when they meet, forming a T-junction. We aim at inserting the optimal number of paths in the right locations to obtain patches that satisfy the requirements we introduced in \secref{sec:overview}.

 \paragraph{Path tracing}
 
 \begin{figure}[t]
    \begin{tabular}{cc}
        \includegraphics[height=0.25\linewidth]{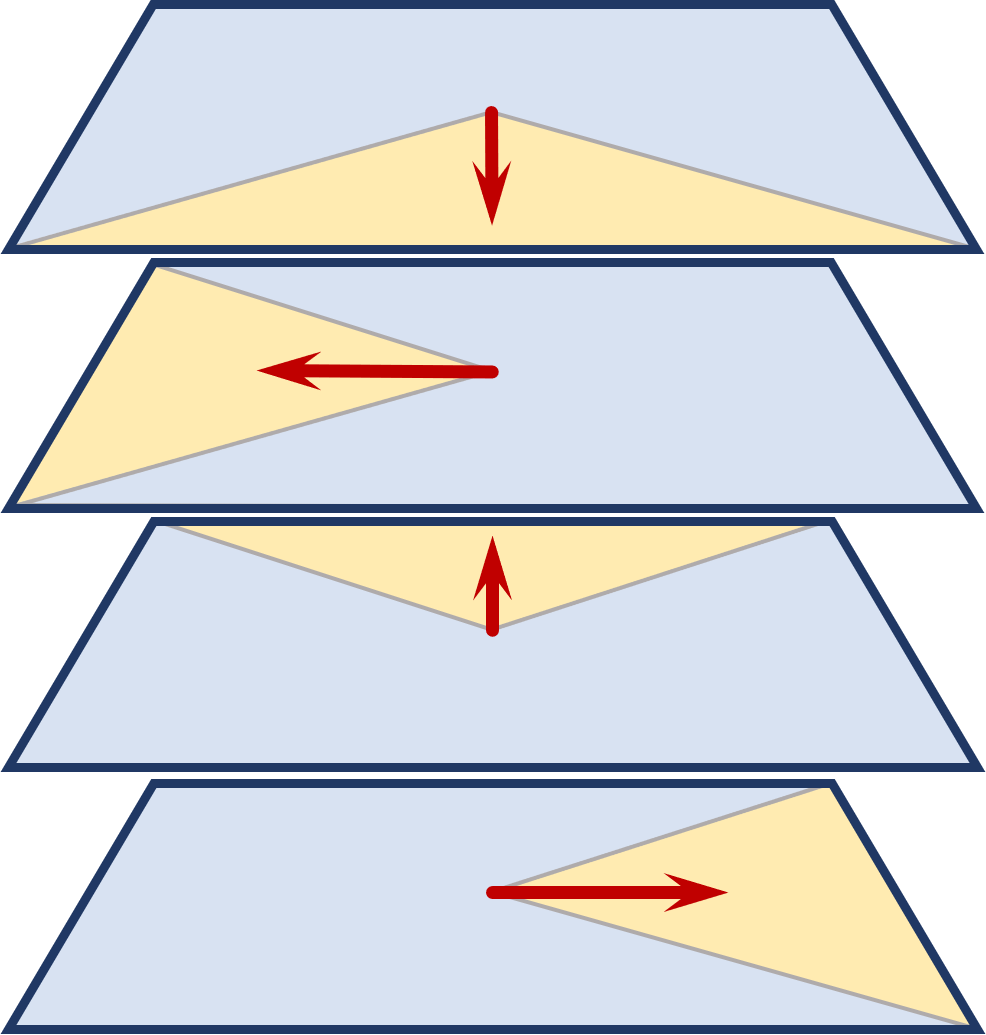}&
        \includegraphics[height=0.3\linewidth]{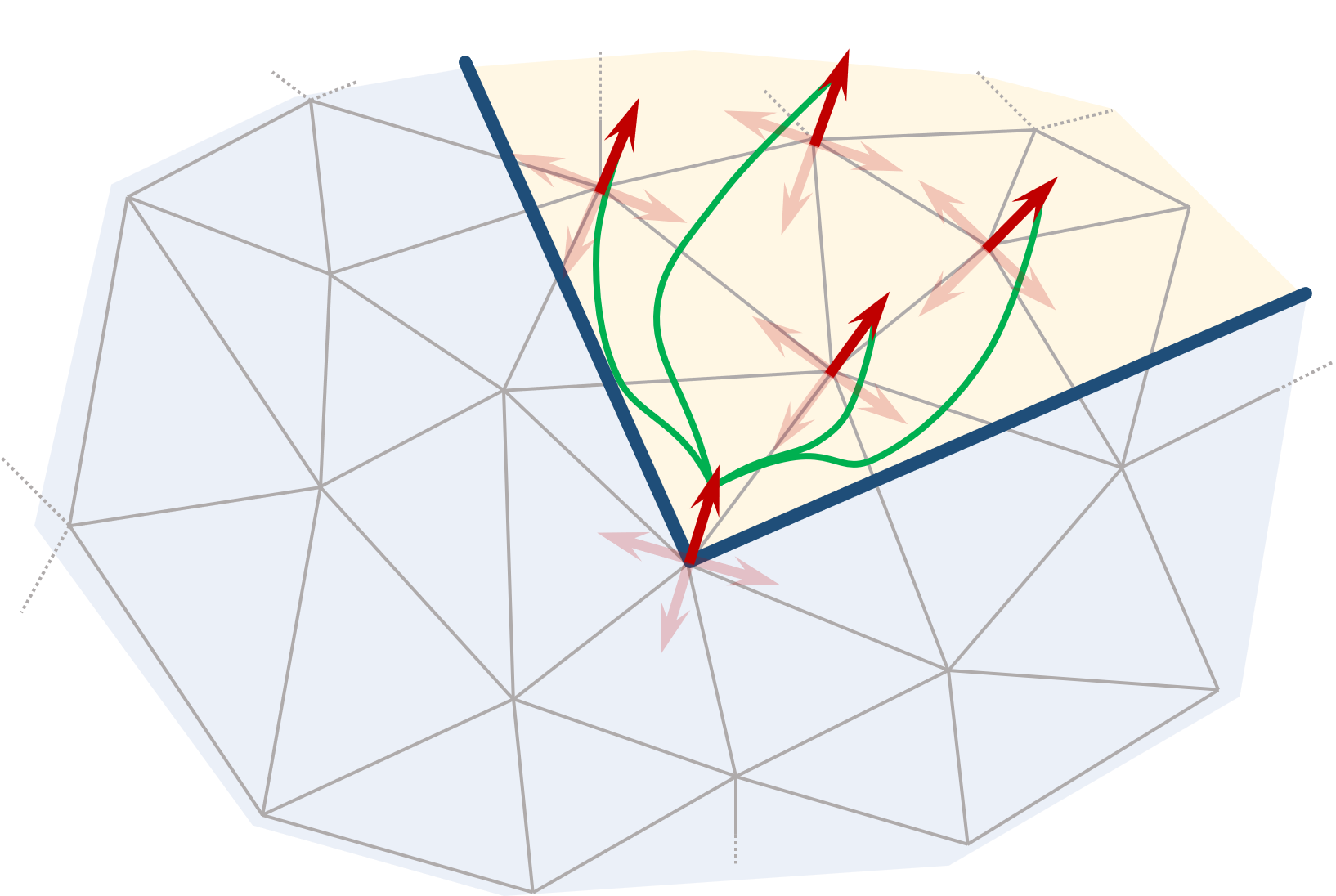}
    \end{tabular}
    \caption{The construction of the tracing graph. Each mesh vertex spawns 4 nodes in the graph, corresponding to the 4 components of the cross-field. Edges in the graph connect nodes of adjacent vertices that represent matching cross-field directions.}
    \label{fig:graph}
\end{figure}

To trace field-oriented paths, we use the graph-based approach proposed by \citet{QuadMixer} and \citet{Pietroni2021}. First, we interpolate the cross-field on vertices (considering their invariance to 90-degree rotations). We create a graph having four nodes for each mesh vertex, one for each direction of its cross-field. Then, we connect adjacent nodes with matching cross-field direction.
Finally, we associate to each connection of the graph a weight that depends on how much it drifts from the direction defined at the adjacent nodes (see \figref{fig:graph}). For details on the graph creation we refer to \cite{Pietroni2021,Pietroni2016}.

\paragraph{Path classification.}
\begin{figure}[t]
   \includegraphics[width=0.9\linewidth]{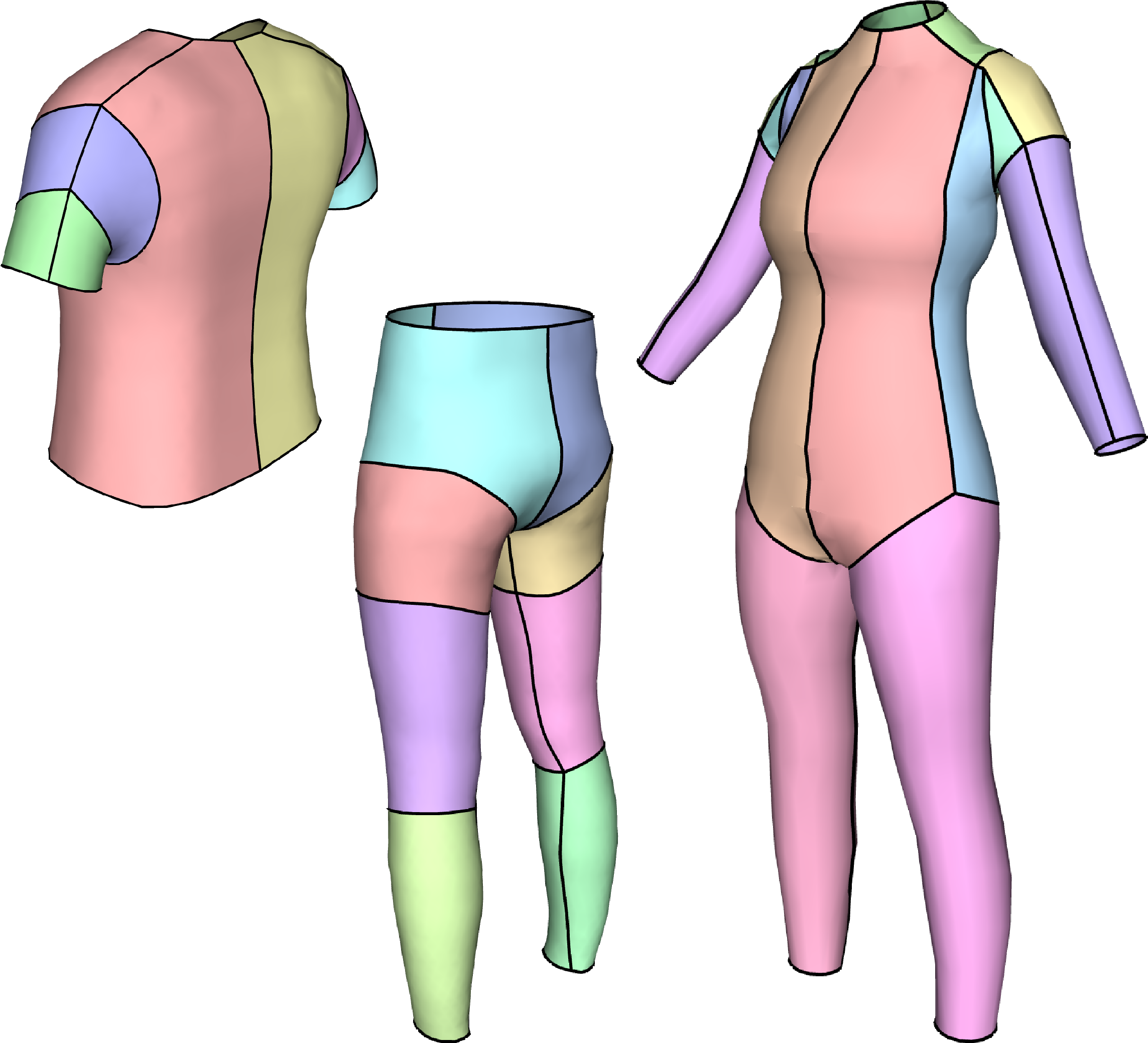}
    \caption{Loops and border-to-border paths are both useful in most of the cases to obtain a proper patch decomposition.}
    \label{fig:loop_vs_boundary}
\end{figure}

\begin{figure}[t]
   \includegraphics[width=0.7\linewidth]{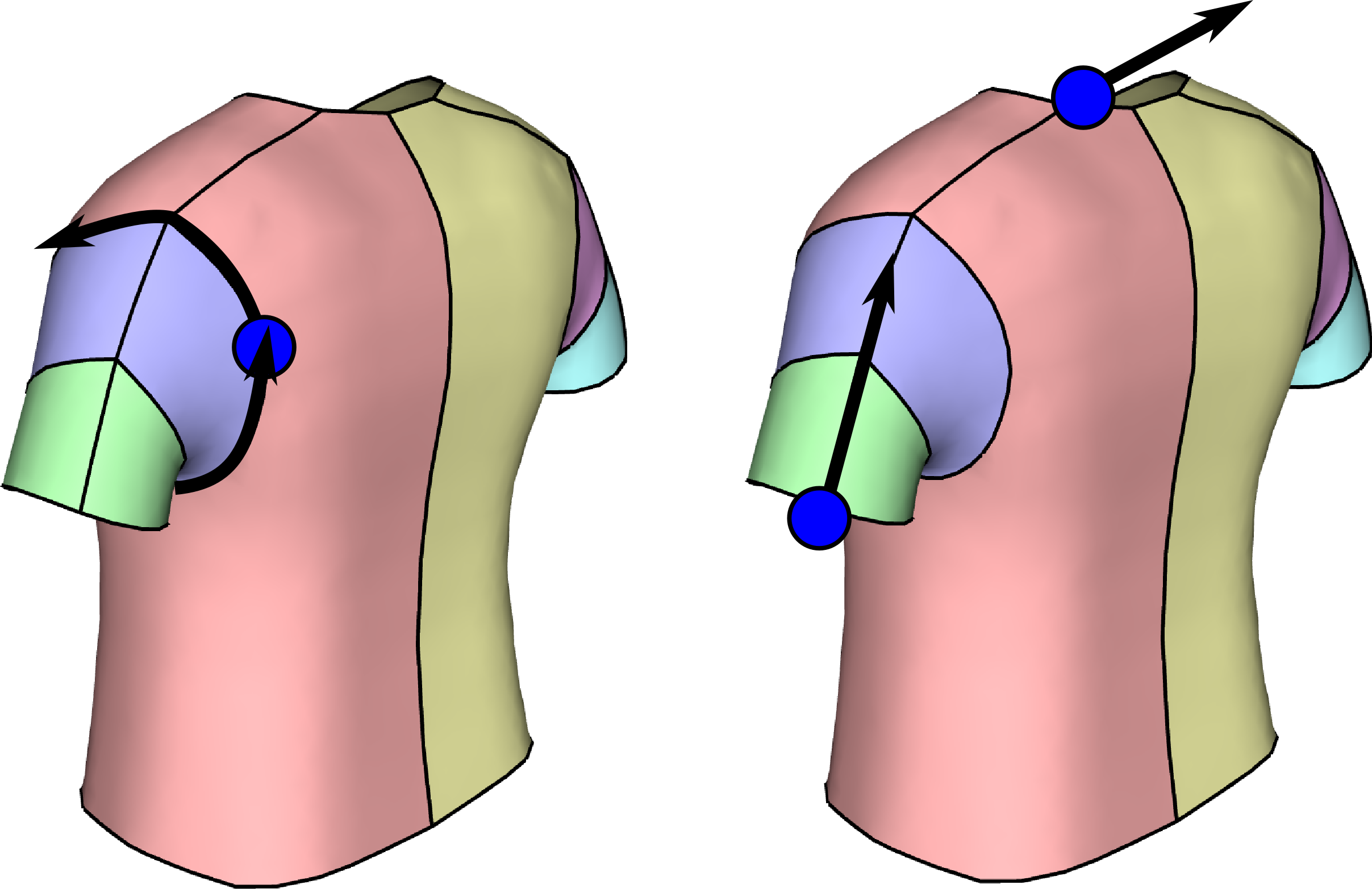}
    \caption{Loop tracing (left) and border-to-border tracing (right).}
    \label{fig:tracing}
\end{figure}

We trace two classes of paths: paths connecting border-to-border and paths connecting to themselves, i.e., loops. We found that both categories of these paths are fundamental to assembling a proper patch layout (see \figref{fig:loop_vs_boundary}). We formulate the problem of path tracing as a shortest-path search between a given \emph{source node} and a set of potential \emph{destination nodes} in the graph. In particular, when tracing a \emph{loop} from a specific vertex, we select an internal node as the source and we aim at coming back to the same node (see \figref{fig:tracing}, left).

\begin{figure*}[t]
\begin{tabular}{@{}c|c@{}}
    \begin{tabular}{cccc}
        \includegraphics[width=0.11\linewidth]{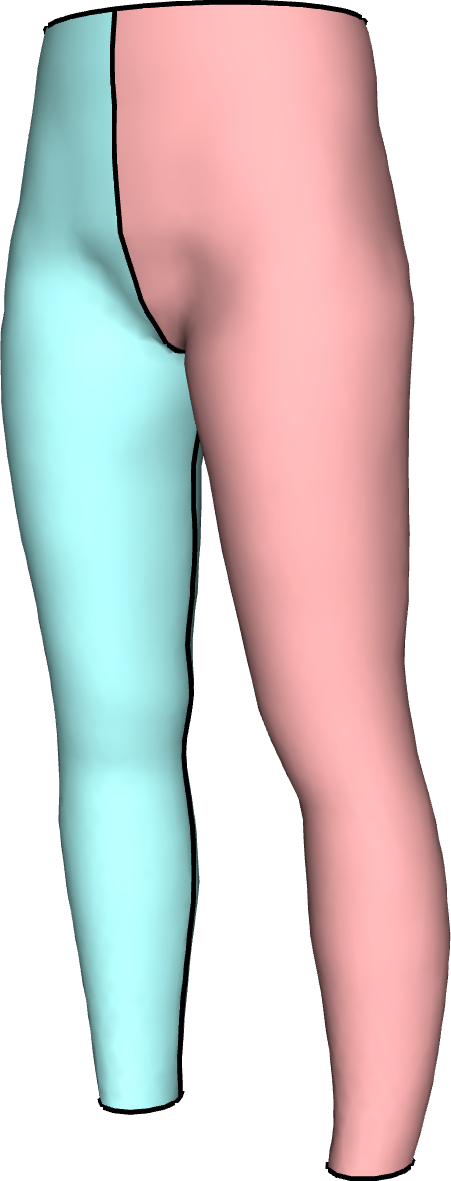}&
        \includegraphics[width=0.11\linewidth]{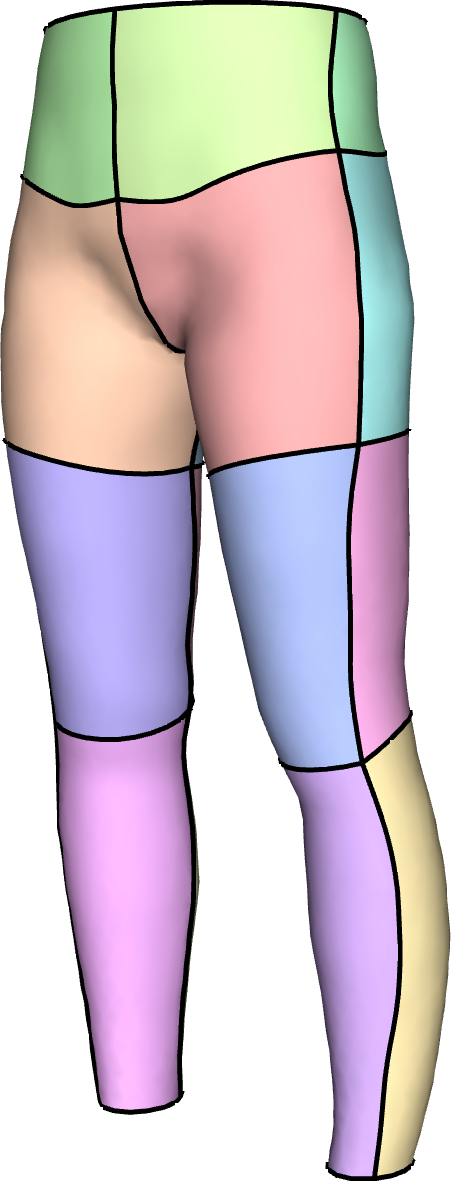}&
        \includegraphics[width=0.11\linewidth]{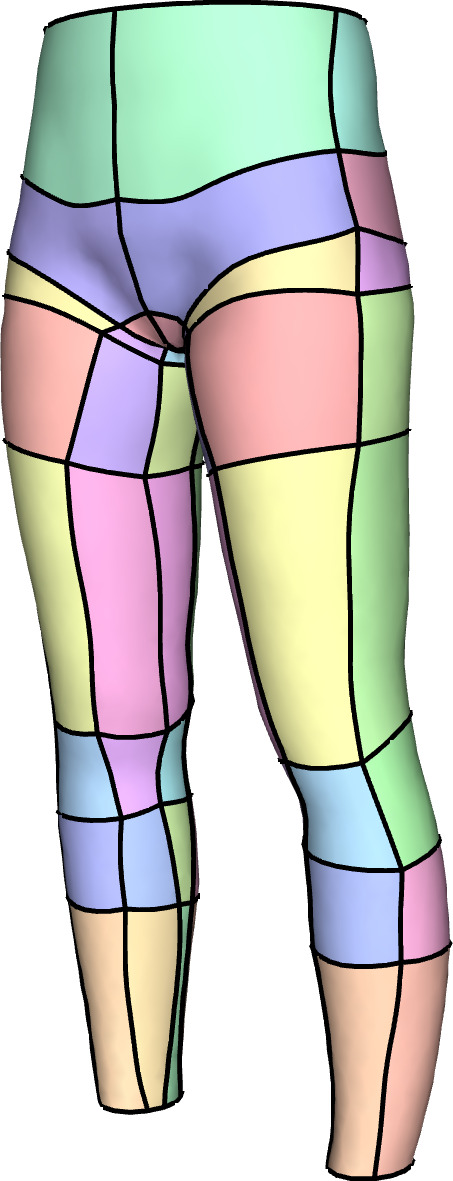}&
        \includegraphics[width=0.11\linewidth]{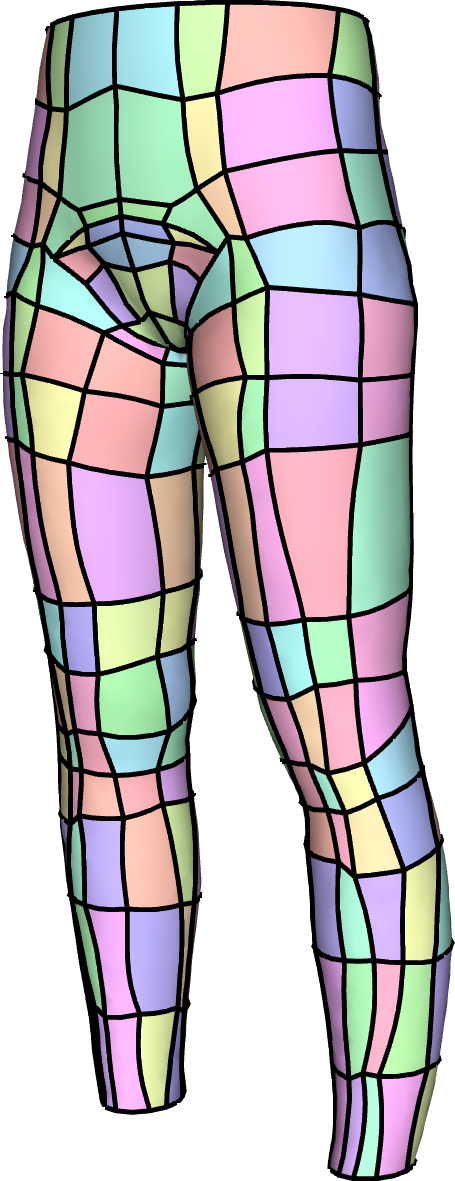}
    \end{tabular}&
    \begin{tabular}{cc}
        \includegraphics[width=0.11\linewidth]{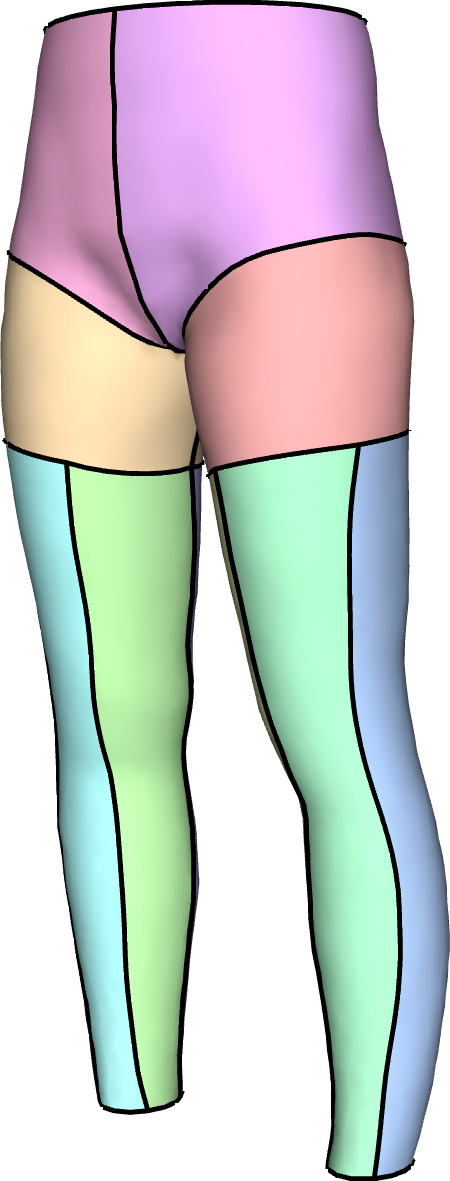}&
        \includegraphics[width=0.11\linewidth]{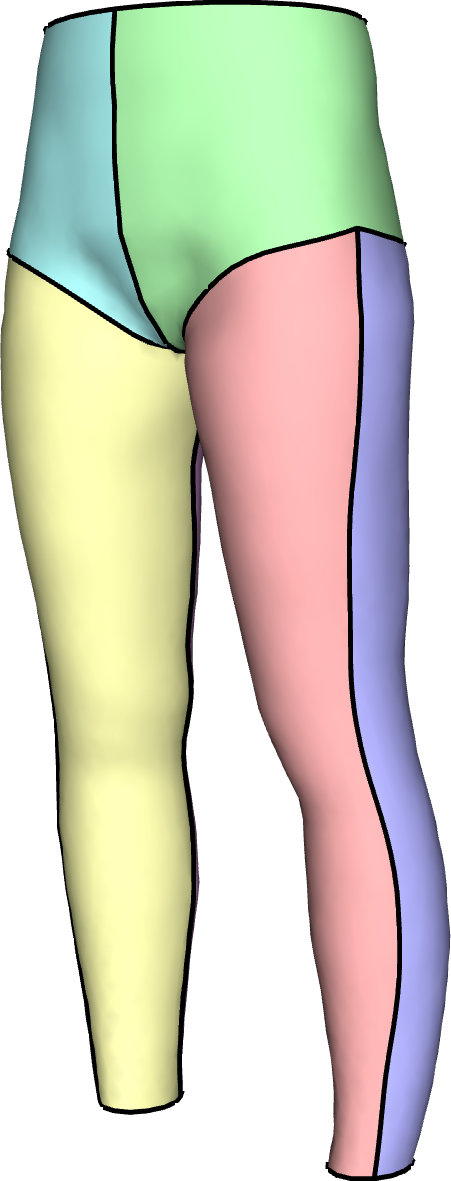}
    \end{tabular}\\
     insertion &  removal
\end{tabular}
    \caption{
    Loops and border-to-border paths are inserted iteratively until the goals for each patch are satisfied. In the removal step, patches are fused by removing paths if the fused patch still satisfies our goals.}
    \label{fig:sampling}
\end{figure*}

\begin{figure}[h]
   \includegraphics[width=0.5\linewidth]{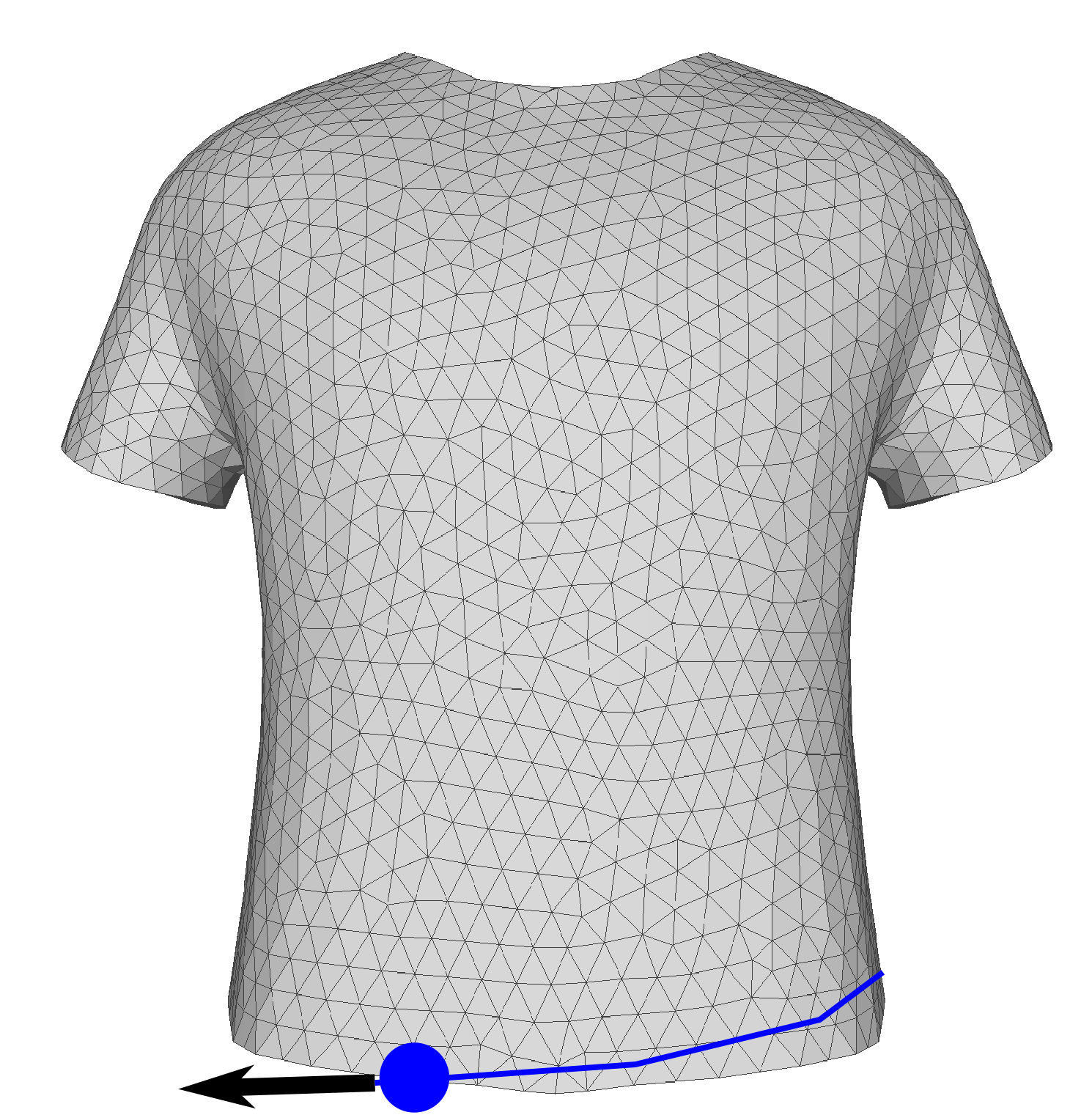}
    \caption{
    The insertion of border-to-border paths that would generate long thin strips, such as the one shown in blue, is avoided by selecting vertices that \emph{enter} and \emph{exit} the mesh w.r.t.\ the cross-field.}
    \label{fig:sleeves}
\end{figure}

Similarly, for border-to-border paths, we select as source a node lying on a boundary vertex whose direction \emph{enters} the mesh, and as a destination all the border nodes whose directions \emph{exit} the mesh (see \figref{fig:tracing}, right). This trick avoids any path to form strange, long strips when touching the border (see \figref{fig:sleeves}). In a preprocessing step, we mark each boundary node with a label to determine if it is an exit or an entrance node. Since we align the field to boundaries, we  always have these two kinds of nodes for each boundary vertex.  

Paths are sequences of adjacent edges that belong to the triangle mesh. So, they usually have irregular shapes. We smooth the paths and reproject them over the surface to make them more regular and we update the rest of the mesh as a consequence. 

 \paragraph{Path insertion.}

We initially sample a set of candidate paths by tracing border-to-border paths and loops. For the sake of efficiency, we subsample uniformly the source nodes on the border and in the interior of the mesh. Similarly to \cite{LivesuPPSC20}, we insert a path using a greedy strategy that favors the furthest path from the previously inserted one.
As in \cite{Pietroni2021,Pietroni2016}, the distance is computed for each node using the M4 stratification of the graph \cite{Campen2012}, and averaged for each path. Notice that two paths crossing orthogonally can be very far. In contrast, parallel paths tend to be close. This simple strategy avoids conglomerations of paths.

We keep the complete patch layout updated at every insertion step, and we mark all the patches that do not satisfy the goals specified in \secref{sec:requirements}. If a candidate path splits one patch that does not meet our goals and does not intersect tangentially with any previously inserted path, then we insert it. As in \cite{Pietroni2016}, given two candidate paths, it is trivial to check whether they intersect tangentially by simply testing whether they pass the same vertex through non-orthogonal directions.

We repeat this insertion strategy until all the patches match our goals. We perform a recursive step on each patch if needed. Notice that this step needs new patches to be parameterized at every insertion step to check whether the produced mapping fulfills the bijectivity and bounded distortion requirements, hence the need for a fast and reliable method to parameterize and test the produced patches. We illustrate the steps of this sampling procedure in \figref{fig:sampling}, left.

 \paragraph{Path removal}
Once the path insertion is complete, we have a patch layout satisfying our goals. However, during its construction, there is no easy way to predict whether a candidate path is essential in the final layout in its entirety (since some goals can only be satisfied by a combination of multiple paths). We remove redundant path segments starting from the last inserted path in reverse order. 
\figref{fig:sampling} (right) shows the effect of the removal procedure. In this step, we disable the removal of path segments that form T-junctions to avoid creating darts, as these are treated explicitly in the next step. 

\paragraph{Dart creation.}

We removed all the redundant path segments at the end of the previous step. However, there might be adjacent patches that we can \emph{partially} glue together, maintaining the distortion under the predefined threshold. Such partial cuts can be found in fashion design and are usually referred to as \textit{darts}. In traditional pattern making, a tailor introduces darts to better shape the body's curves. By removing a wedge-shaped piece of fabric and sewing both sides together, an experienced tailor effectively introduces angle deficiency on the garment and thus creates a curved shape from a flat pattern.

\begin{figure}[t]
    \begin{tabular}{cc}
        \includegraphics[height=0.2\linewidth]{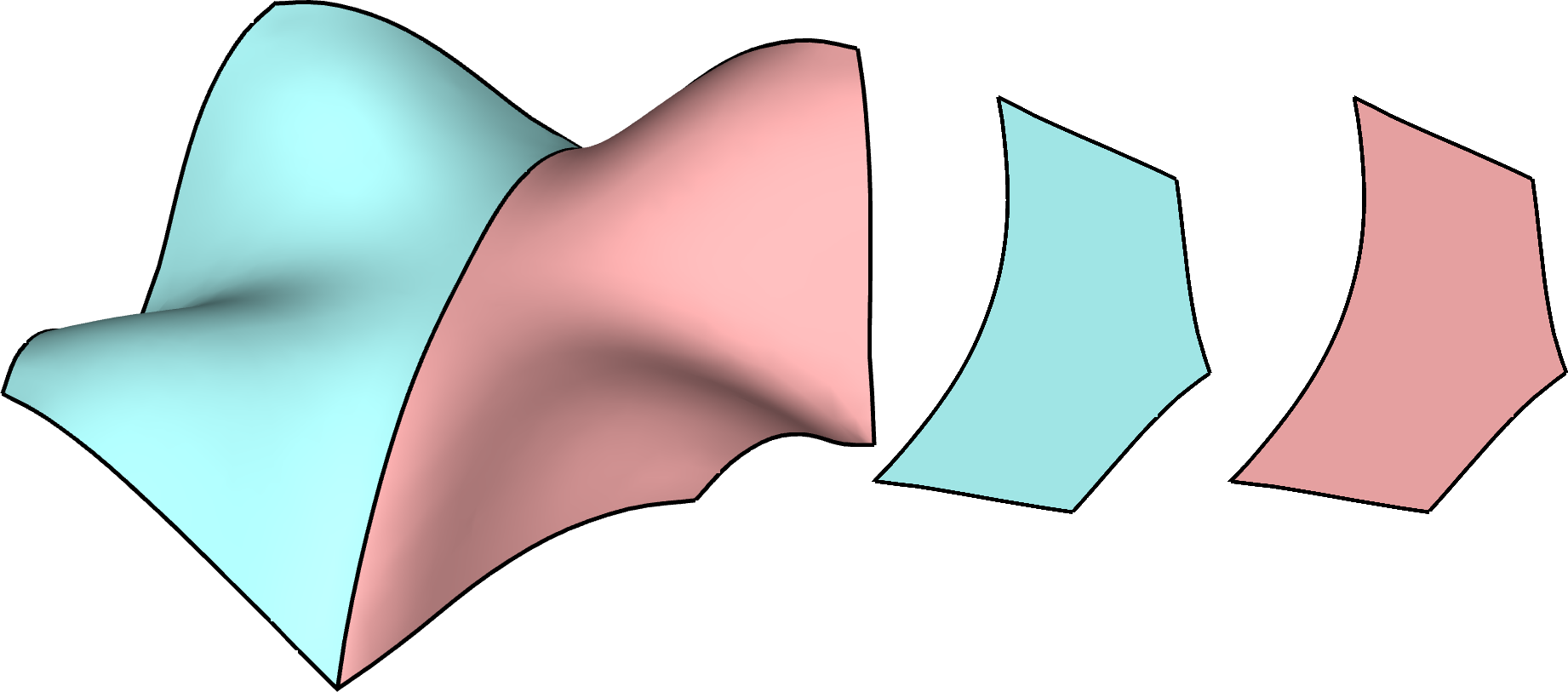}&
        \includegraphics[height=0.2\linewidth]{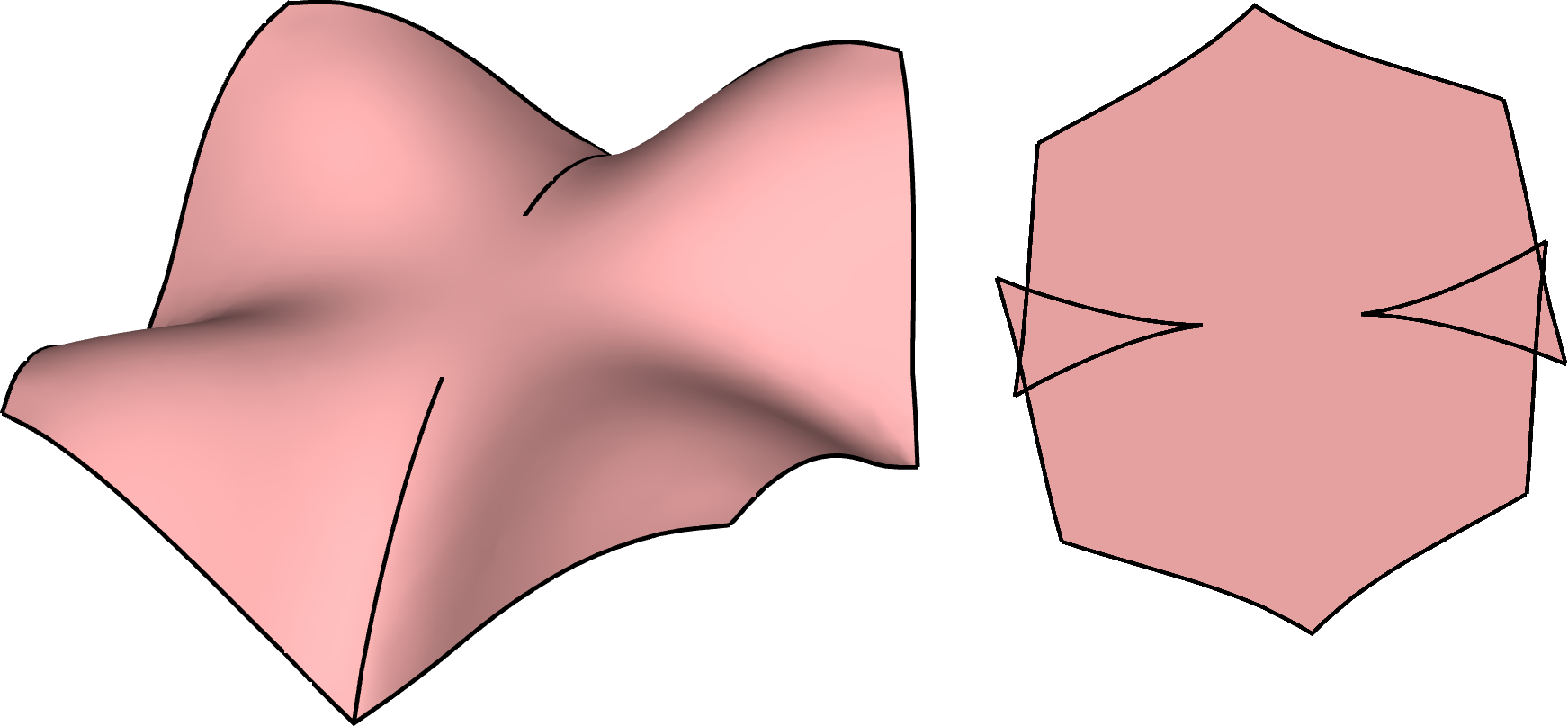}
    \end{tabular}
    \caption{Creating darts starting from areas with negative curvature tends to generate overlaps in the $UV$ mapping. }
    \label{fig:wrong_dart}
\end{figure}

To introduce darts, we first split all paths into their segments, at the intersections with the other paths. Then we sort the different path segments by considering the Gaussian curvature of the mesh region they span (as in field computation, we extract curvature at a low scale using the method proposed by \citet{Pan2010}). Ideally, we want to merge starting from the regions with lower Gaussian curvature, as such seams are more likely to be merged with low distortion. Similarly, areas with negative Gaussian curvature (saddles) should be merged later, as they most likely generate overlaps in the $UV$ domain (see \figref{fig:wrong_dart}).
Then we select the first path segment and split it uniformly into a number of subpaths. Following the same intuition, we start merging from the lower curvature side, and we continue as long as the produced distortion is below the threshold.

 \paragraph{Symmetry.}
 \begin{figure}[t]
    \begin{tabular}{cc}
        \includegraphics[width=0.3\linewidth]{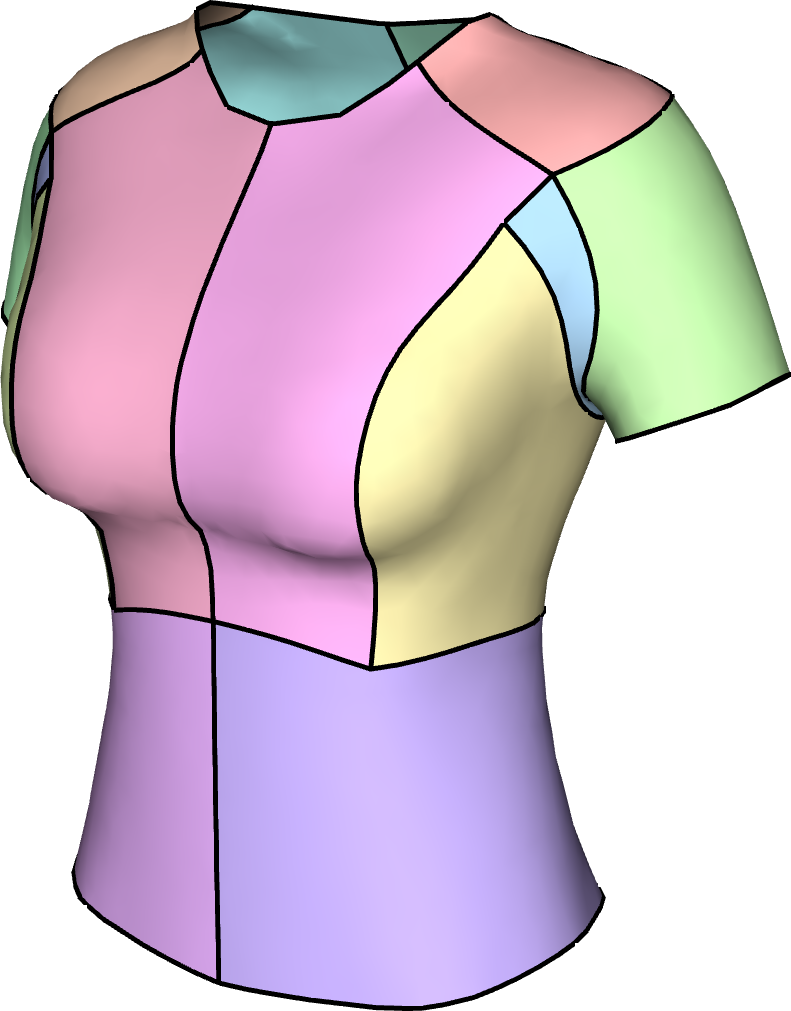}&
        \includegraphics[width=0.3\linewidth]{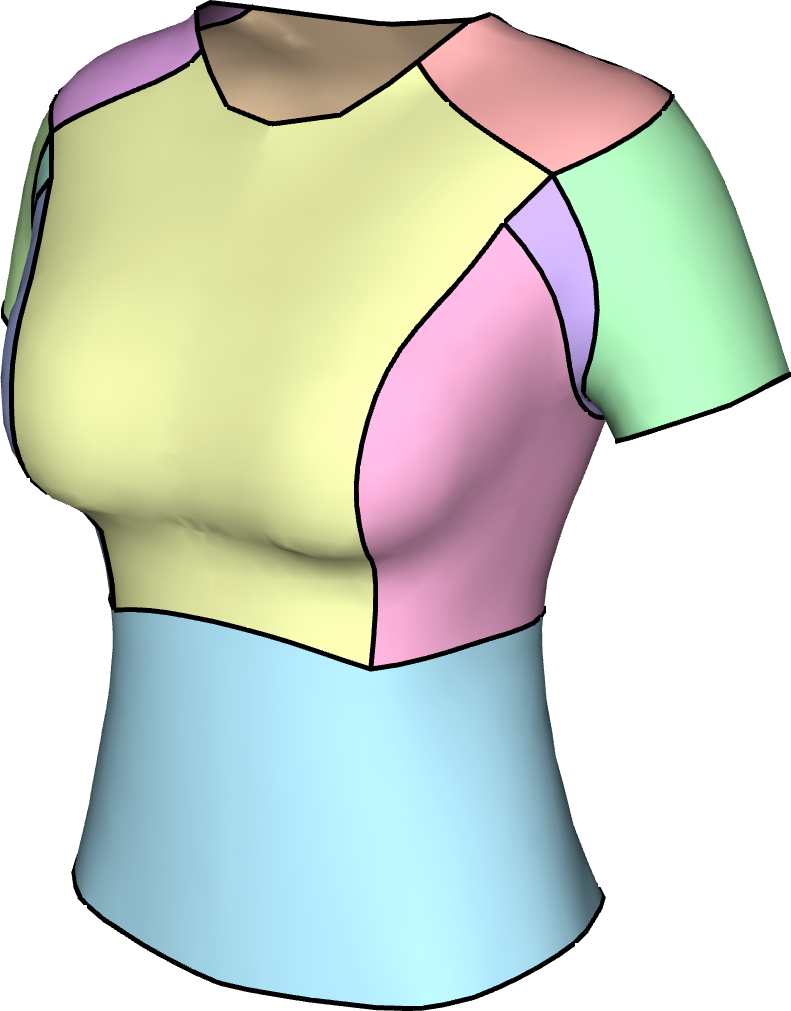}
    \end{tabular}
    \caption{The seam on the global symmetry plane can be safely removed at the end of the  layout generation process.
    }
    \label{fig:symm_removal}
\end{figure}

The generation of the patch layout is conducted on one side of the mesh using a symmetry plane (shown in red in \figref{fig:pipeline}a). We first copy the patch layout on the other symmetric side, then we start the path removal and dart insertion process  for the paths laying on the symmetry plane. This way, we keep the symmetric distribution and, at the same time, remove unwanted seams along the symmetry plane (see \figref{fig:symm_removal}). 

\subsection{Anisotropic textile parameterization}
\label{sec:param}

\begin{figure}[t]
\centering
  \includegraphics[width=\linewidth]{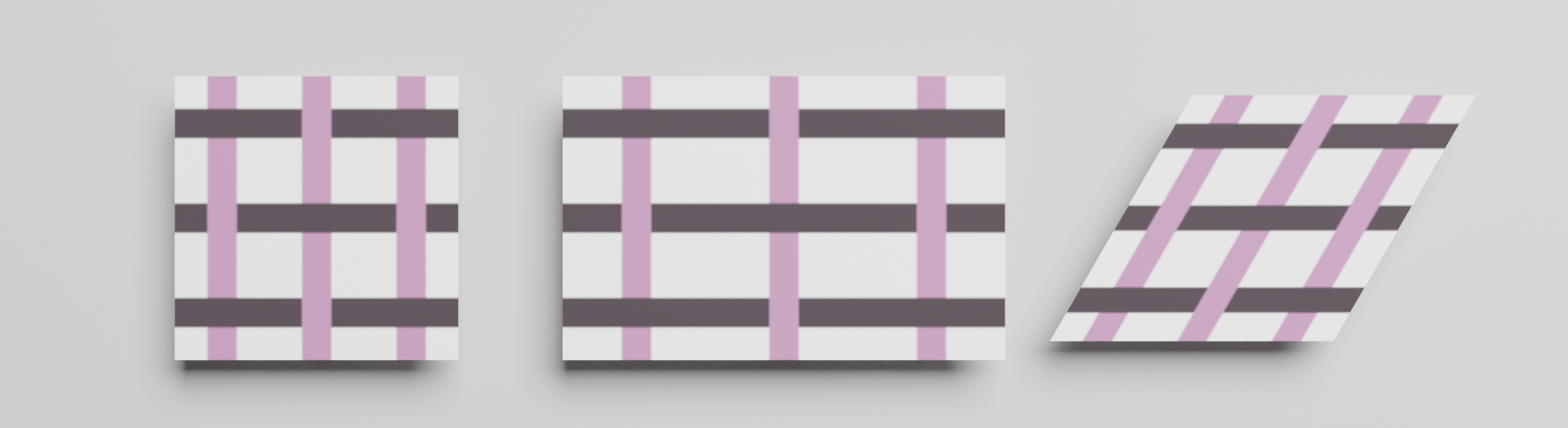}
    \caption{Woven net (left) undergoing stretch (middle) and shear (right). 
    }
  \label{fig:woven}
\end{figure}

Woven fabric can be modeled as a regular grid of threads, where the two axes are called warp and weft (or collectively `grain'). We assume the common case where the grid is orthogonal, and thus represent it by the $UV$ axes in the flat domain, but it is possible to model arbitrary angles between the warp and the weft directions \cite{McCartney2000,MCCARTNEY2005}. Ideally, the cloth undergoes only bending when the flat pattern is draped on the 3D body, but this implies that the 3D garment shape is piecewise developable, which is not always possible. We hence need to measure deviation from developability in a way that is consistent with the woven structure properties. General parameterization distortion measures are isotropic, or invariant to rotations, but this ignores the fact that the threads are nearly inextensible, while a certain, limited angular distortion is permitted, which means that the cloth can stretch diagonally to the warp and weft ($UV$) directions, but much less so along those directions. Hence the intrinsic parameterization distortion measure needs to be anisotropic. We penalize stretch along the $U$ and $V$ directions separately from shear. Note that by shear we mean angular distortion of the fabric grid, while the thread lengths are preserved, so it is not an area-preserving shear transformation (see \figref{fig:woven}).

\begin{figure}[b]
\centering
  \includegraphics[width=0.45\textwidth]{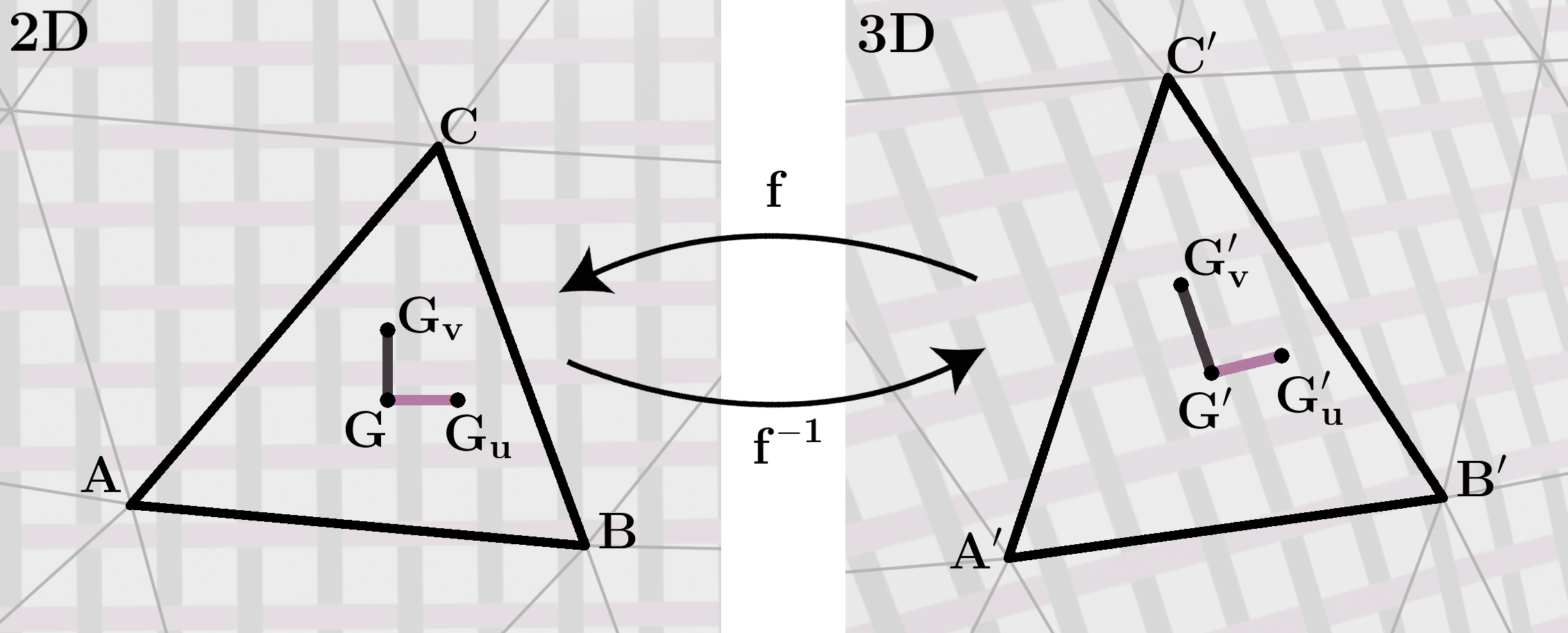}
    \caption{Per-triangle flattening, modeled as an affine map $f$.}
  \label{fig:tri_transform}
\end{figure}

Consider a reference triangle $A'B'C'$ in $XYZ$ coordinates mapped to a distorted triangle $ABC$ in $uv$ coordinates, see \figref{fig:tri_transform}. A unique affine transformation $f$ maps $A'B'C'$ to $ABC$.
To penalize stretch of the fabric grain, we can consider the scaling induced by the inverse transformation $f^{-1}: \R^2 \rightarrow \R^3$, which can be interpreted as the stretch of a canonical unit frame $\mathcal{G}$ in the $UV$ plane, 
undergoing the transformation $f^{-1}$, as illustrated in \figref{fig:tri_transform}. The stretch values $s_u$ and $s_v$ can be expressed using the columns of the Jacobian of $f^{-1}$ (i.e., the tangents of the parameterization), denoted $J = [J_1\ J_2] \in \R^{3\times 2}$:
\begin{align}
    s_u = \|J_{1}\| \text{ , } s_v = \|J_{2}\|.
\end{align}

\paragraph{Stretch.}
We could define the stretch penalty as the deviations of $s_u, s_v$ from $1$. However, these scaling factors are non-linear functions of the variables in our problem, namely the $UV$ coordinates of the triangle vertices $ABC$, leading to a non-quadratic stretch measure. We therefore invert the problem and consider the stretching of the \emph{image} of frame $\mathcal{G}$, namely $J(\mathcal{G})$, when $f$ is applied.

For simplicity, we express the frame $\mathcal{G}$ explicitly by three points: the 2D triangle centroid $G$, and the points $G_u = G + (1,0), \ G_v = G + (0,1)$ (see \figref{fig:tri_transform}).
The 3D grain directions $J(\mathcal{G})$ can be expressed by transforming these three points by $f$ into $G', G'_u, G'_v$. We denote by $(\alpha_X, \beta_X, \gamma_X)$ the barycentric coordinates of any point $X$ w.r.t.\ triangle $ABC$. Then we have:
\begin{align}
&G' = \alpha_G A' + \beta_G B' + \gamma_G C', \\
&G'_u = \alpha_{G_u} A' + \beta_{G_u} B' + \gamma_{G_u} C', \\
&G'_v = \alpha_{G_v} A' + \beta_{G_v} B' + \gamma_{G_v} C'.
\end{align}
This allows us to measure the lengths $s_u = \|G'_u - G'\|$ and $s_v = \|G'_v - G'\|$ on the 3D triangle. If the mapping $f$ does not stretch the $J_1$ direction, then $\|G_u - G\| = s_u$, and equivalently for $J_2$ and $v$. We denote the $UV$ coordinates of our helper points as $G = (u_G, v_G), G_u = (u_{G_u}, v_{G_u}), G_v = (u_{G_v}, v_{G_v})$, and then the stretch of $J_1$ caused by $f$ is $\|G_u - G\|$, which is equal to
$ |u_{G_u} - u_G| = (\alpha_{G_u} - \alpha_G) u_A +
(\beta_{G_u} - \beta_G) u_B + ( \gamma_{G_u} - \gamma_G) u_C,$
and similarly for the other direction.
We thus define the per-triangle stretch energy terms as follows:
\begin{align}
& E_{\mathrm{stretch},u}(ABC) = \\
\nonumber
& \phantom{a} \omega_\mathrm{stretch} 
\left[ s_u  - 
\left(( \alpha_{G_u} - \alpha_G ) u_A 
+ ( \beta_{G_u} - \beta_G ) u_B 
+ ( \gamma_{G_u} - \gamma_G ) u_C\right)\right]^2, \\
\nonumber
& E_{\mathrm{stretch},v}(ABC) = \\
\nonumber
& \phantom{a} \omega_\mathrm{stretch} 
\left[ s_v  - 
\left(( \alpha_{G_v} - \alpha_G ) v_A 
+ ( \beta_{G_v} - \beta_G ) v_B 
+ ( \gamma_{G_v} - \gamma_G ) v_C\right)\right]^2. 
\end{align}
The total energy $E_{\mathrm{stretch},u}$ is then defined as the sum of per-triangle energies, and similarly for $E_{\mathrm{stretch},v}$.

\paragraph{Shear and rigidity.}
A direct expression for shear is the angle between the parameterization tangents $J_1$ and $J_2$, with the constraint that the tangents maintain their length, but this is again not a quadratic in the variables $u_A, v_A, \ldots$ Instead, we propose to measure deviation from isometry, implicitly penalizing shear in the same manner as in as-rigid-as-possible (ARAP) parameterization~\cite{Liu2008}:
\begin{align}
\label{eq:E_rigid}
    E_\mathrm{rigid}(ABC) = \omega_\mathrm{rigid} \sum_{e \in ABC} (u_e - u_{e'})^2 + (v_e - v_{e'})^2,
\end{align}
where $e$ denotes a triangle edge in triangle $ABC$, $e'$ is the corresponding edge in triangle $M(A'B'C')$ and $M$ is the best-fit rigid transformation that aligns $A'B'C'$ to $ABC$, computed using Procrustes~\cite{SorkineRabinovich:SVD-rotations:2016}. 
The total rigidity measure $E_\mathrm{rigid}$ sums up \eqnref{eq:E_rigid} over all triangles. 
Although this ARAP measure mixes shear with isotropic stretch, it tends to produce well conditioned results thanks to the Procrustean step and works well for our purpose.

\paragraph{Seam reflection symmetry.}
Pattern pieces are sewn together along seams by placing one part onto the corresponding part and stitching along a curve. This implies the existence of a perfect reflection between the matching borders that get sewn together. We represented matching seams in the $UV$ domain as two sets of duplicated vertices $\mathcal{P} = (p_1, \ldots, p_n)$ and $\mathcal{Q} = (q_1, \ldots, q_n)$, where $p_i$ and $q_i$ correspond to the same 3D location but are mapped to different locations in 2D. 

To measure the deviation of the matching seams from a perfect reflection, we first compute the best-fit reflection transformation $M$ between $\mathcal{P}$ and $\mathcal{Q}$, similarly to Wolff et al.~  \shortcite{Wolff:Symmetry:VMV2019}. We achieve this by simply switching the sign of the determinant during procrustean analysis.

If the $UV$ seams are reflection-symmetric, $M\mathcal{P}$ and $\mathcal{Q}$ coincide, as well as $\mathcal{P}$ and $M^{-1}\mathcal{Q}$. Otherwise, we define target points as:
\begin{equation}
\forall i \in [1\ldots n],\ \  q_{t,i} = \frac{M p_i + q_i}{2} \text{ , } p_{t,i} = \frac{p_i + M^{-1} q_i}{2}.
\end{equation}
The seam's reflection symmetry energy $E_\mathrm{seam}$ is then defined as:
\begin{equation}
\label{eq:seam_symmetry}
E_\mathrm{seam} = \omega_\mathrm{seam} \left( \sum_{p_i \in \mathcal{P}} (p_i - p_{t,i})^2 + \sum_{q_i \in \mathcal{Q}} (q_i - q_{t,i})^2 \right).
\end{equation}

In our general framework, patches are cut and flattened progressively. Consequently, upon flattening a given patch, the energy $E_\mathrm{seam}$ may not be defined for all seams if the mirroring patch was not flattened previously. To address this, we jointly optimize $E_\mathrm{seam}$ for all patches during the final parameterization.

\begin{figure}[t]
    \subfloat[$\omega_\mathrm{stretch} = 5, \omega_\mathrm{rigid} = 1$]{\includegraphics[width=0.45\textwidth]{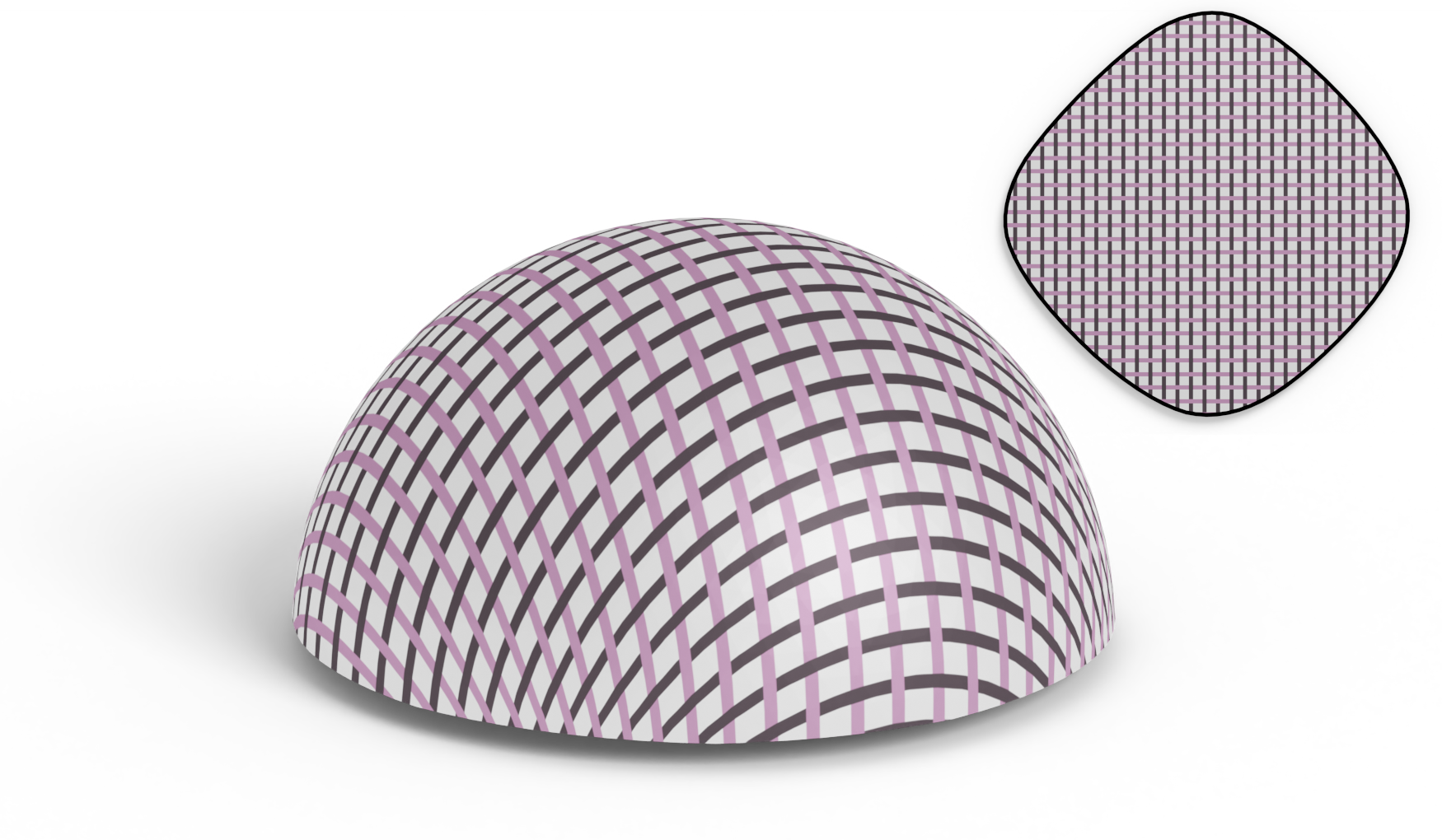}}\\
    \subfloat[$\omega_\mathrm{stretch} = 0, \omega_\mathrm{rigid} = 1$]{\includegraphics[width=0.45\textwidth]{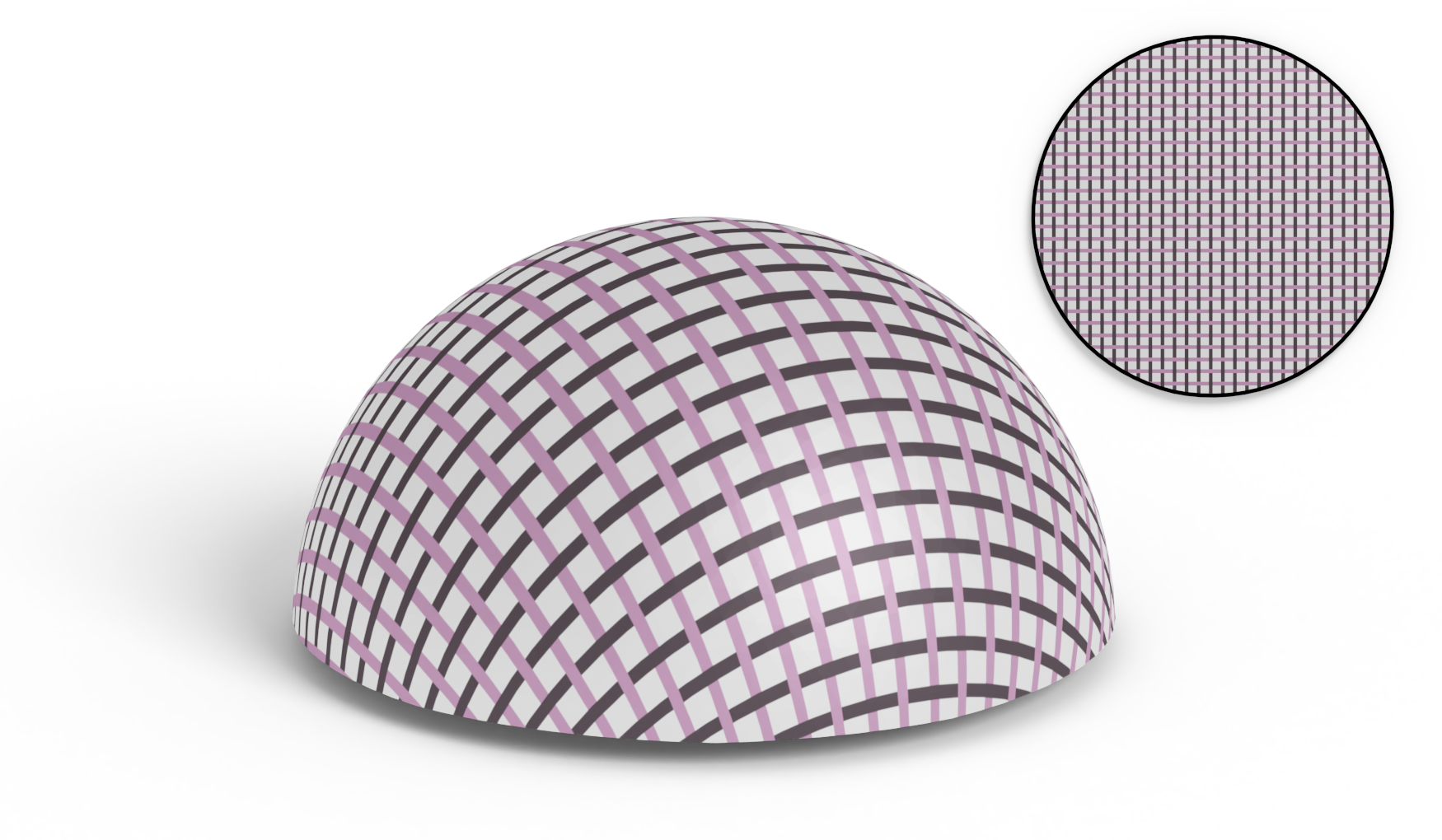}}
    \caption{Influence of the energy terms weights on the parameterization.}
    \label{fig:shear_weight}
\end{figure}

\paragraph{Dart symmetry.}
Darts are similar to seams, but the two matching parts belong to the same pattern piece and meet at the tip, so the tip vertex is not duplicated. Ideally, this vertex lies on the dart's symmetry axis in 2D. We use a similar energy to \eqnref{eq:seam_symmetry} for darts but modify the target point definition.
We define the set of midpoints $\mathcal{R}$, consisting of $r_i = (p_i + q_i) /2$, and find the target symmetry axis by computing the best fitting line to $\mathcal{R}$ while being constrained to pass through the tip vertex.
Then, we compute symmetric point sets $\overline{\mathcal{P}}$ and $\overline{\mathcal{Q}}$ and define an energy term $E_\mathrm{dart}$ equivalently to $E_\mathrm{seam}$, weighted by $\omega_\mathrm{dart}$ and with target positions
\begin{equation}
\forall i \in [1\ldots n], \ \ q_{t,i} = \frac{\overline{p_i} + q_i}{2} \text{ , } p_{t,i} = \frac{p_i + \overline{q_i}}{2}.
\end{equation}

\begin{figure}[t]
\begin{tabular}{@{}cc@{}}
        \includegraphics[width=0.45\linewidth]{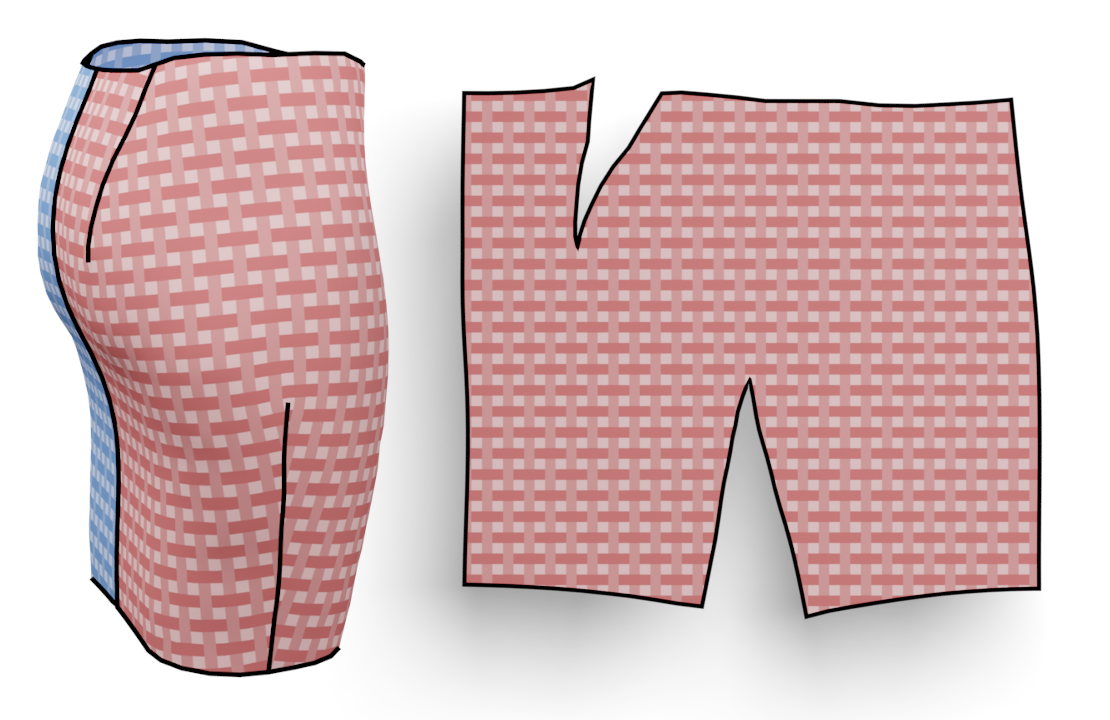}&
        \includegraphics[width=0.45\linewidth]{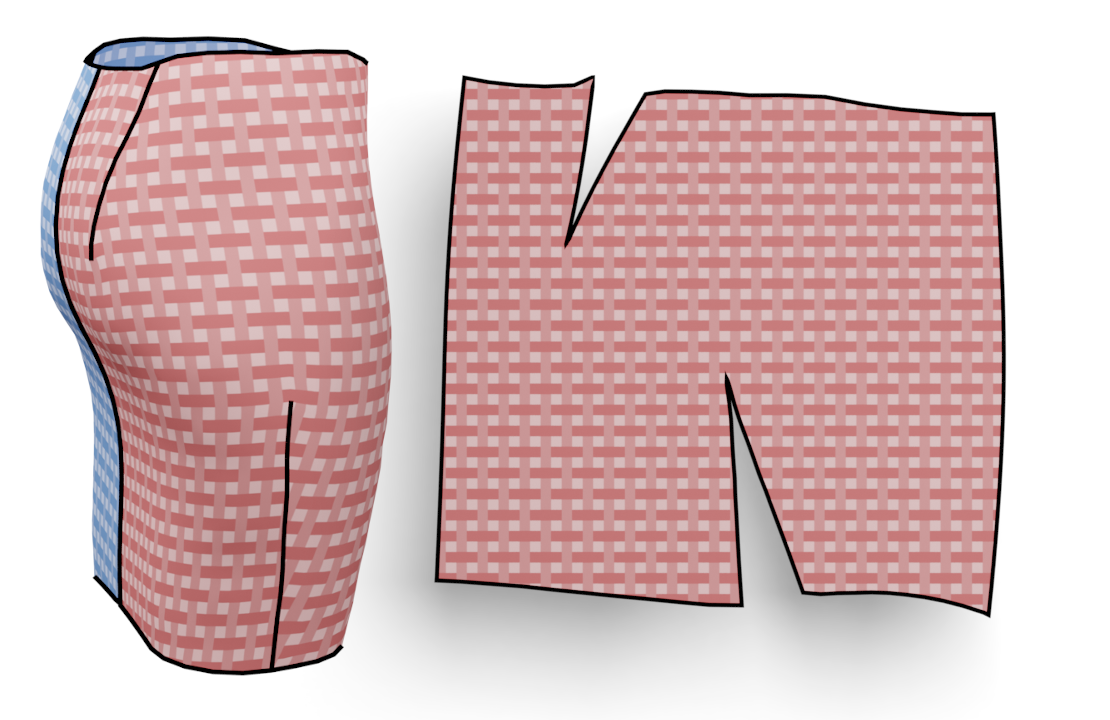}\\
        \small $\omega_\mathrm{dart} = 0 \text{ , } \omega_\mathrm{seam} = 0 $&
        \small $\omega_\mathrm{dart} = 5 \text{ , } \omega_\mathrm{seam} = 5 $
\end{tabular}
    \caption{Dart and seam reflection symmetry (\secref{sec:param}) is essential for producing workable patterns with reflective cuts.}
    \label{fig:global_constr}
\end{figure}

\begin{figure}[t]
    \includegraphics[width=0.95\linewidth]{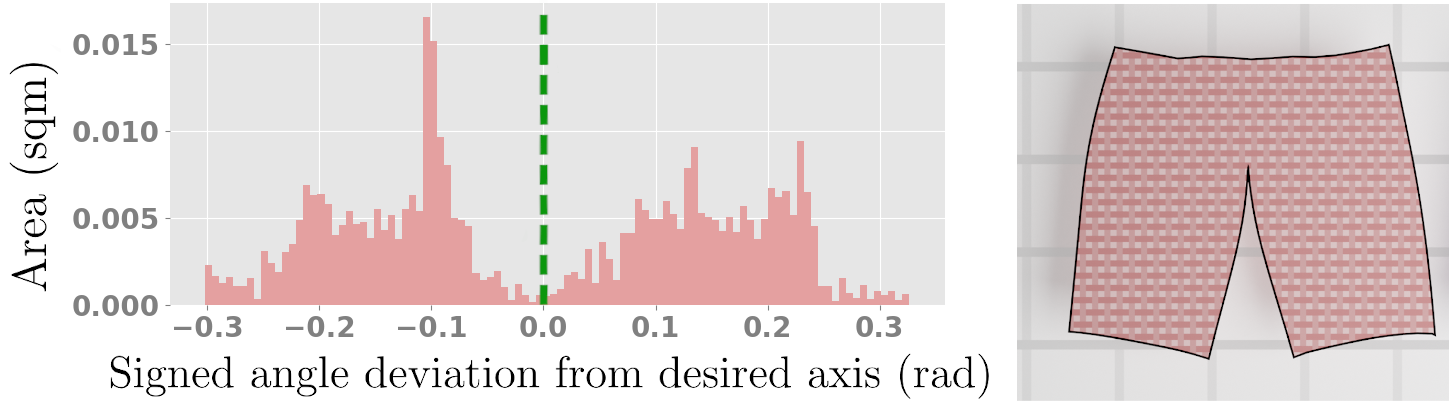}
    \caption{
    A skirt pattern piece (\figref{fig:requirements}, top) is split by a dart, resulting in conflicting desired grain alignment on each side. The global rotation chosen by our parameterization accommodates for both sides, despite being optimal for only a small area. On the left we show a histogram of the signed angle between the warp direction and the prescribed desired alignment.}
    \label{fig:align_graph}
\end{figure}

\paragraph{Grain alignment.} Traditional garment patterns align the grain (warp) direction with the vertical axis in the final garment, or the ``centerline'' direction of a body part, such as along the sleeves. This ensures predictable and symmetric draping. In some cases, the tailor may perform a \textit{bias cut} instead, favoring another alignment axis in specific regions in order to allow shearing on the main stress direction. In our framework, we allow the definition of a desired 3D alignment axis ${a'}$ and rotate the 2D patches such that their warp aligns with ${a'}$. This is first performed on a per-triangle basis by computing the projection 
of ${a'}$ on the triangle plane in 3D and transporting it to ${a}$ on the 2D triangle. This defines a per-triangle desired axis, and we compute the best-fit global axis $a_\mathrm{opt}$ that the $V$ direction should align to. Instead of including this in the energy formulation, we simply perform this alignment step in each iteration. \figref{fig:align_graph} shows an example of grain alignment.

\paragraph{Optimization.} 
We employ a local-global iteration strategy similar to~\cite{ARAP07,Liu2008} to compute a $UV$ mapping of a given pattern piece by minimizing the distortion measure 
\begin{equation}
E_\mathrm{textile} = E_\mathrm{stretch,u} + E_\mathrm{stretch,v} + E_\mathrm{rigid} + E_\mathrm{seam} + E_\mathrm{dart}
\end{equation}
and accounting for grain alignment.
As the initial guess we use the least squares conformal map of \cite{Levy02}, orienting the 2D patch such that the $V$ axis matches the prescribed grain direction. We then compute the expressions in all the energy terms as described above (the local step), and solve for the updated parameterization of the pattern piece by minimizing  $E_\mathrm{textile}$ w.r.t.\ the $UV$ coordinates of the mesh vertices.  This global step amounts to efficiently solving a sparse linear system, because $E_\mathrm{textile}$ is quadratic in the $UV$'s. We iterate the local and global steps, reinstating the grain alignment in each local step. For all tested meshes, 5 iterations were sufficient to compute a satisfactory estimate and 20 iterations to converge. 

The fabric properties are expressed by the weights $\omega$;  the default values in our examples are $(\omega_\mathrm{stretch}, \omega_\mathrm{rigid}, \omega_\mathrm{dart}, \omega_\mathrm{seam}) = (5, 1, 5, 5)$.
\figref{fig:shear_weight} shows the effects of varying the distortion weights, while \figref{fig:global_constr} illustrates the seam and dart reflection symmetry constraints. 

\begin{figure}[b]
    \includegraphics[width=0.95\linewidth]{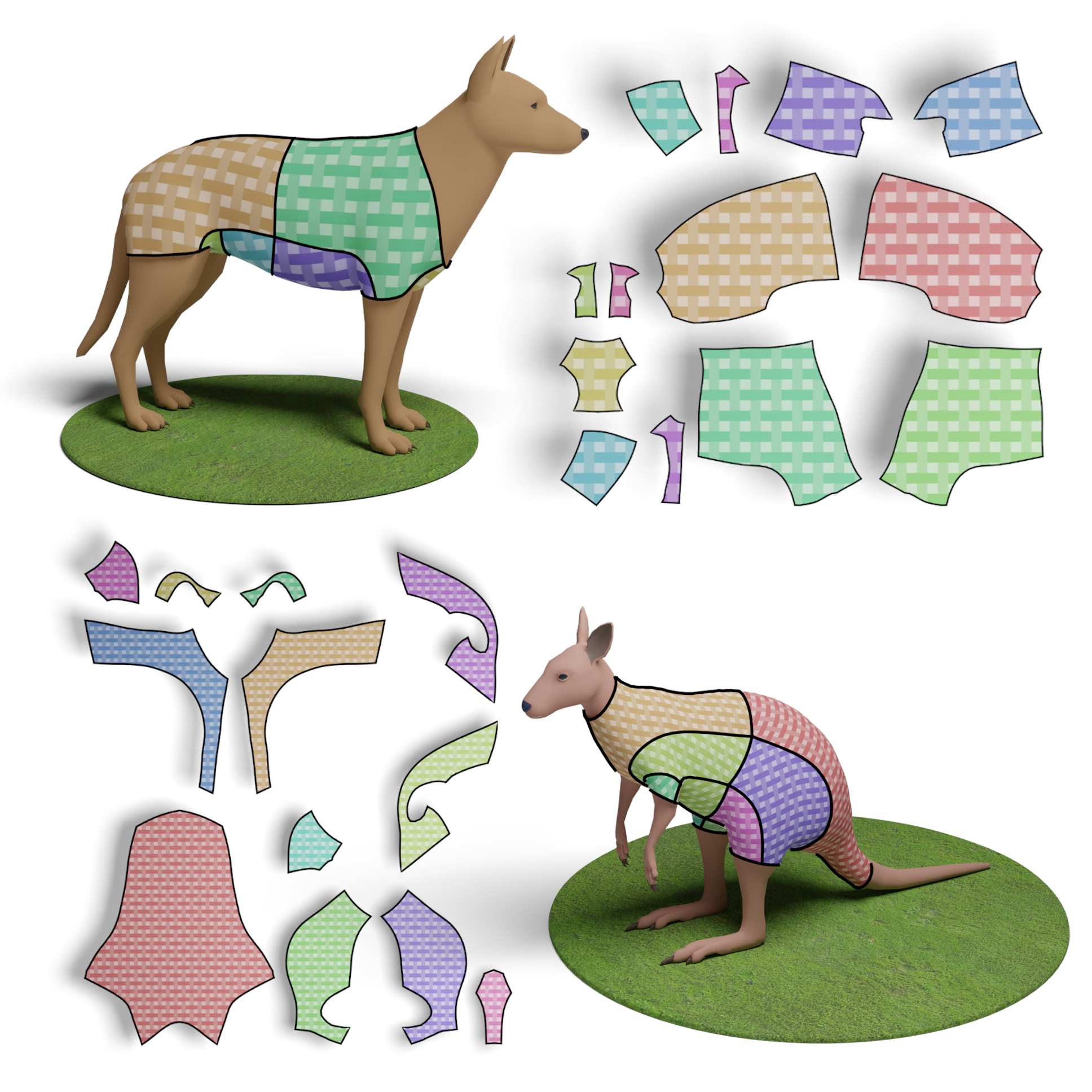}
    \caption{Our method works also for non-humanoid shapes, such as a dog (top) or a kangaroo (bottom).}
    \label{fig:animals}
\end{figure}

\begin{figure*}[t]
    \includegraphics[width=0.90\linewidth]{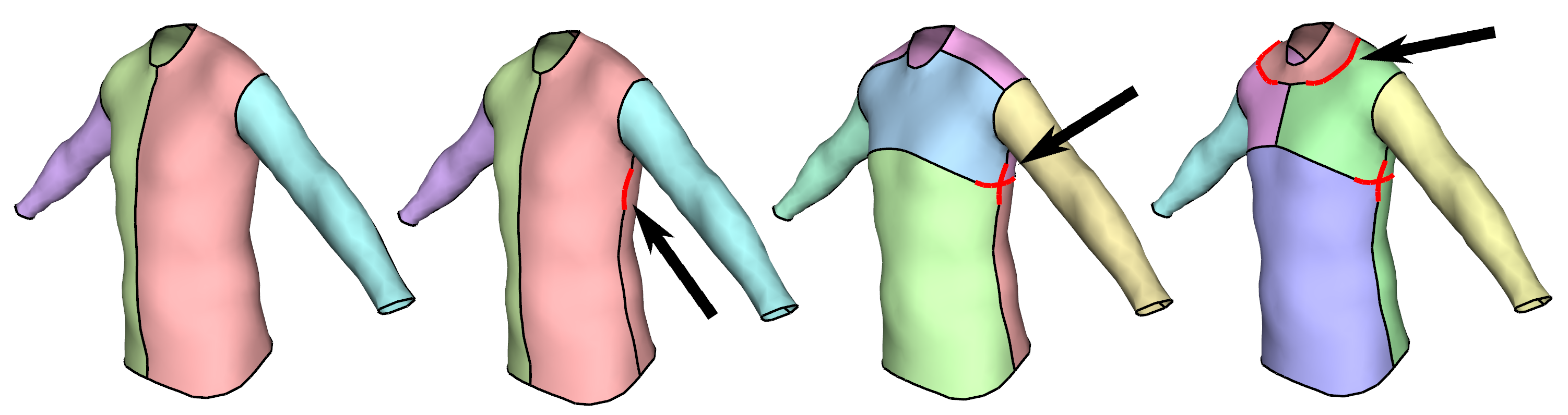}
    \caption{A sequence of interactive adjustments of to the patch layout. The designer adds sketches (in red), and our method computes seams that interpolate the sketched curves while generating a valid sewing pattern.}
    \label{fig:interactive}
\end{figure*}

Our parameterization naturally generalizes to the case where multiple poses are considered, as only the right side of the system in the global step depends on the 3D mesh. At each iteration, we thus compute the right-hand side for each pose considered, and use the average in the solve.

Note that our parameterization method does not guarantee bijectivity. Self-intersections can thus occur during patch layout computation when cuts are introduced in non-satisfactory locations. Our patch decomposition method takes advantage of this by discarding cuts that lead to non-injective patches. This significantly speeds up our exploration of possible cuts, as only promising solutions are considered. The final pattern produced is guaranteed to contain no overlaps, as all the cuts passed our self-intersection test. 

\begin{figure*}[t]
    \begin{tabular}{@{}c@{}c@{}c@{}}
    \begin{tabular}{@{}c@{}}
        \begin{tabular}{@{}c@{}c@{}}
        \includegraphics[width=0.15\linewidth]{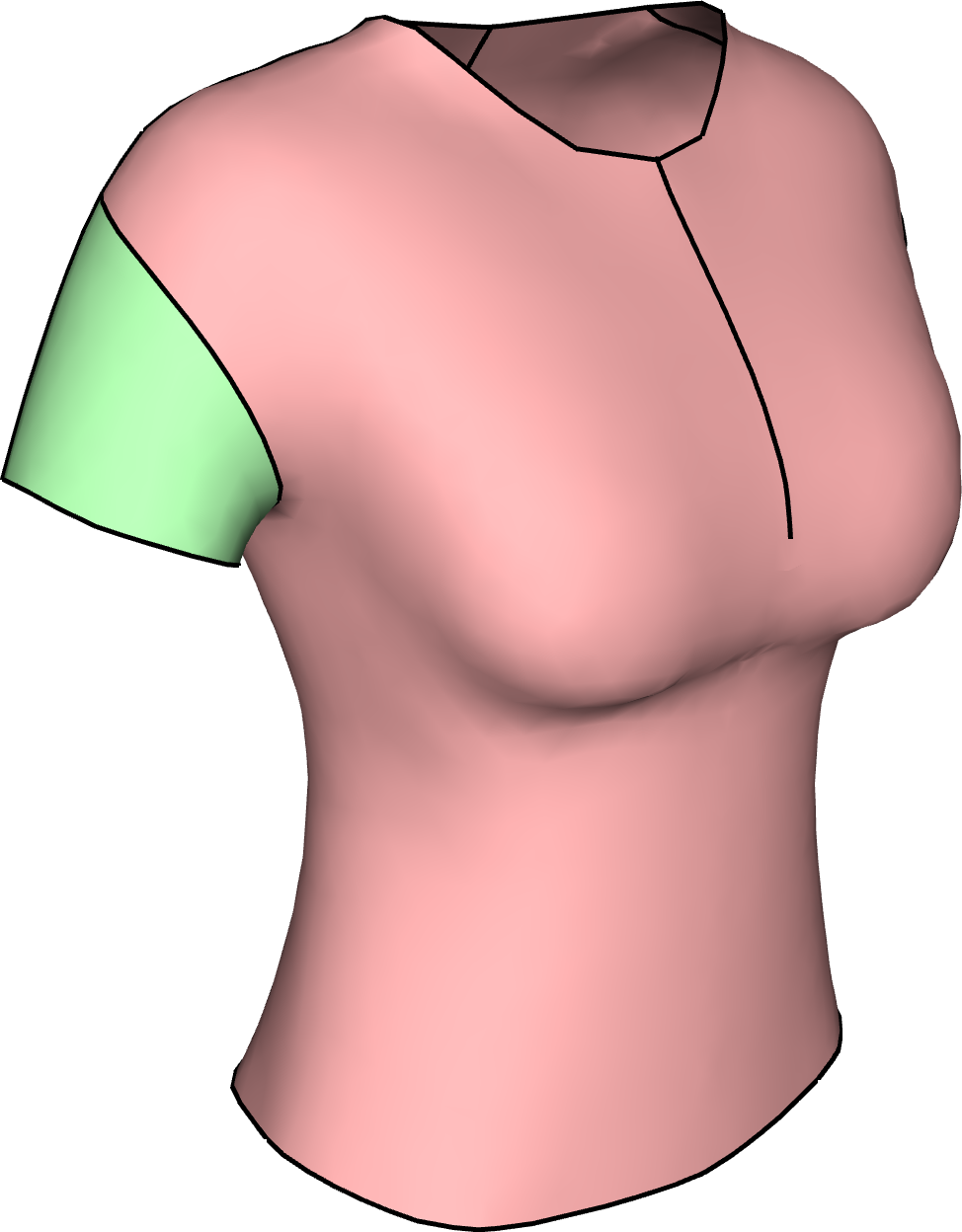}&
        \includegraphics[width=0.15\linewidth]{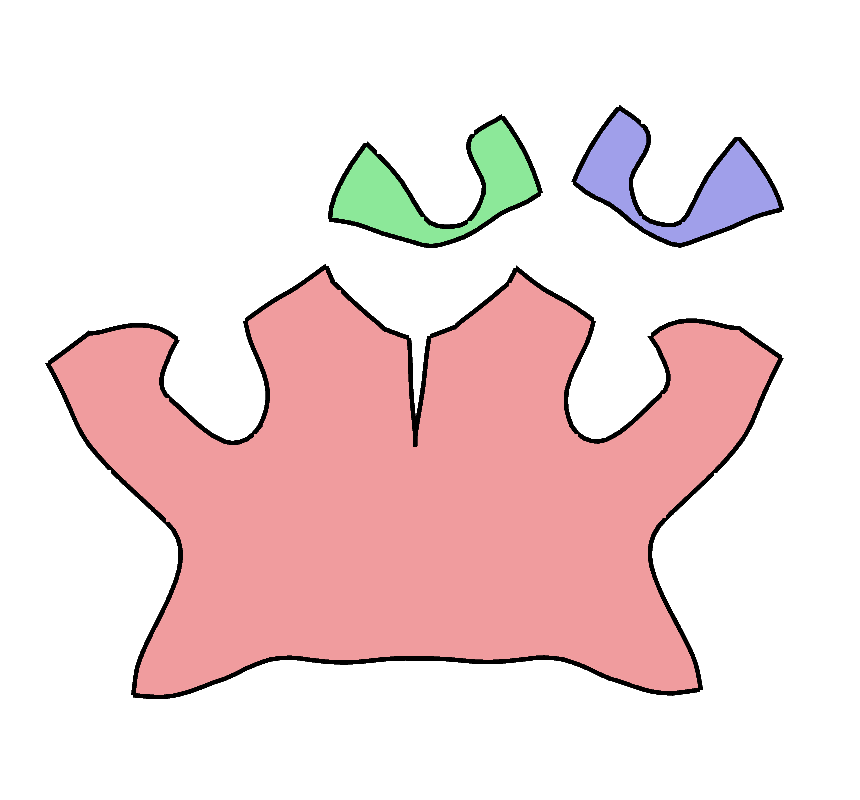}
        \end{tabular}\\
        \small $C=8$ , $s_\mathrm{max} = 0.05$
    \end{tabular}&
    \begin{tabular}{@{}c@{}}
        \begin{tabular}{@{}c@{}c@{}}
        \includegraphics[width=0.15\linewidth]{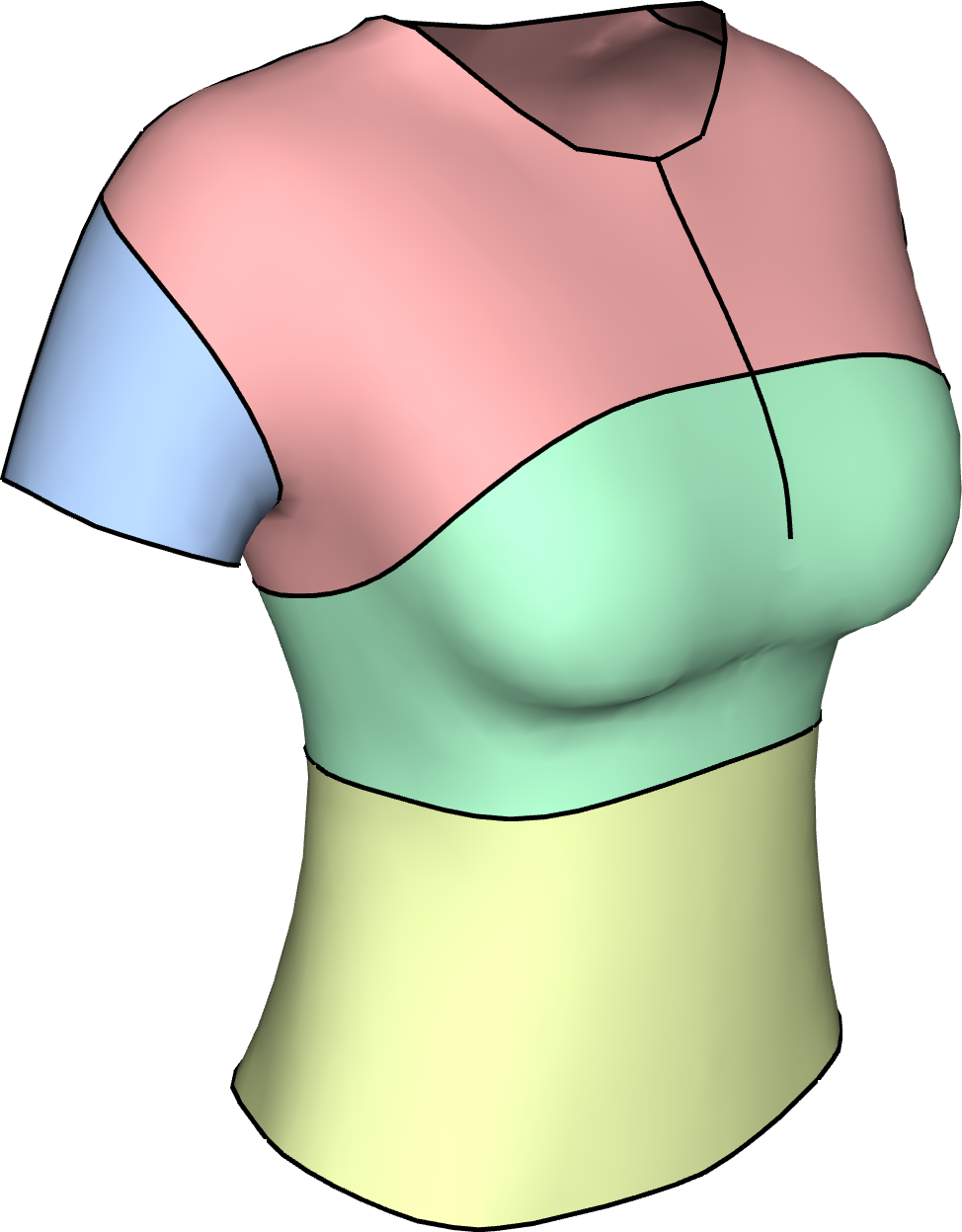}&
        \includegraphics[width=0.15\linewidth]{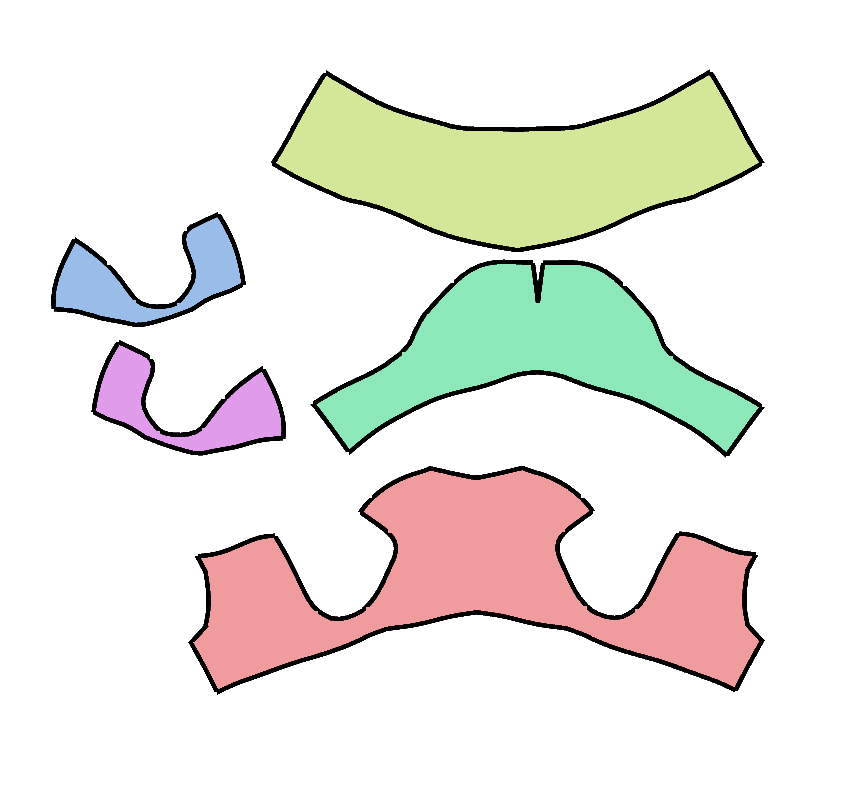}
        \end{tabular}\\
        \small $C=8$ , $s_\mathrm{max} = 0.04$
    \end{tabular}
    \begin{tabular}{@{}c@{}}
        \begin{tabular}{@{}c@{}c@{}}
        \includegraphics[width=0.15\linewidth]{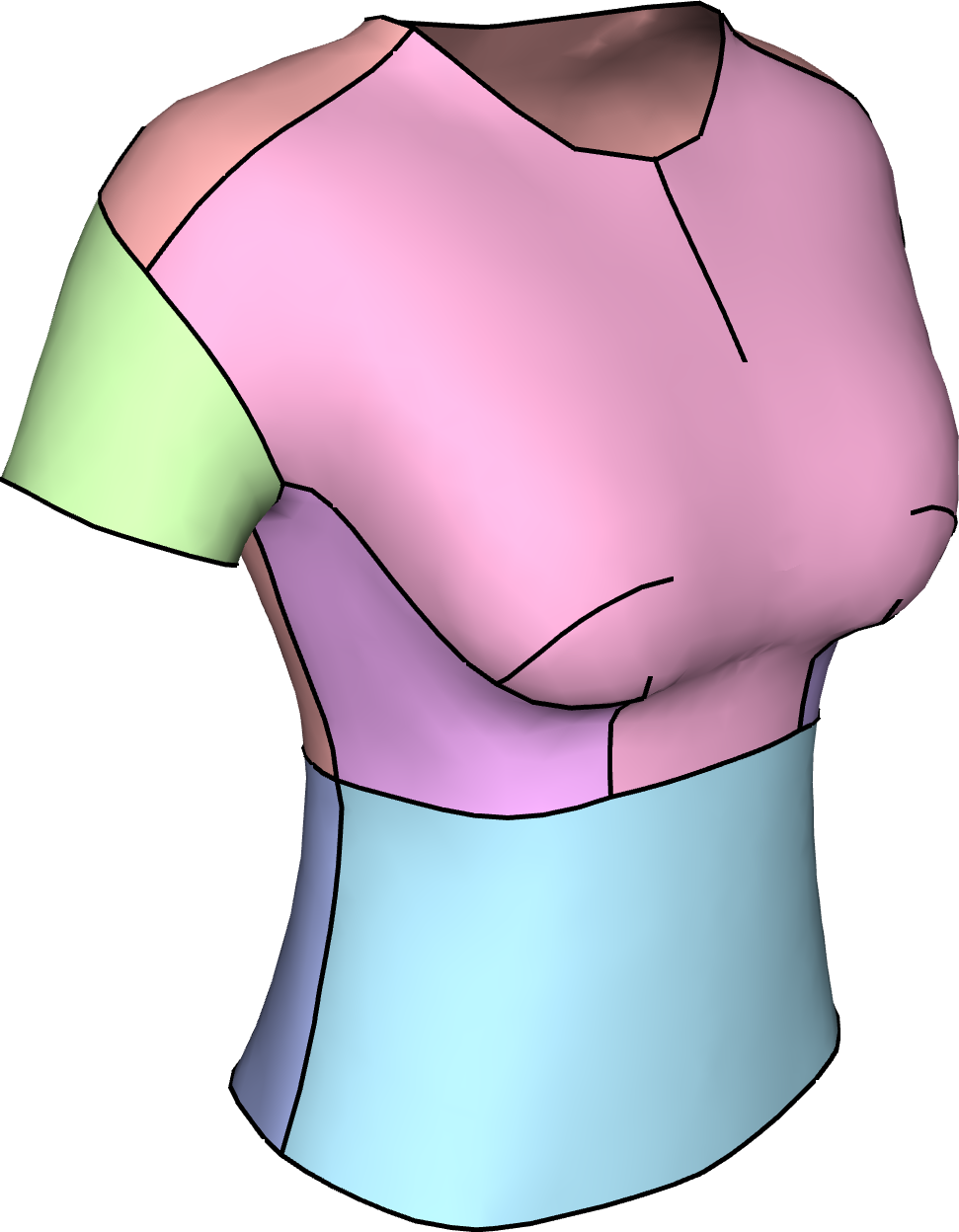}&
        \includegraphics[width=0.15\linewidth]{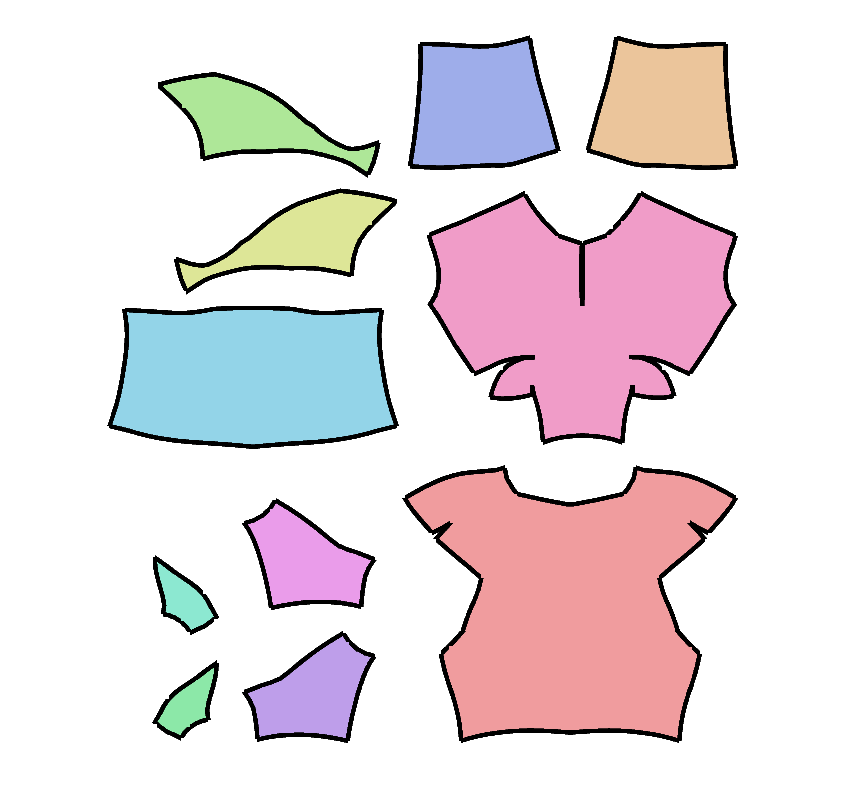}
        \end{tabular}\\
        \small $C=8$ , $s_\mathrm{max} = 0.02$
    \end{tabular}\\
    \begin{tabular}{@{}c@{}}
        \begin{tabular}{@{}c@{}c@{}}
        \includegraphics[width=0.15\linewidth]{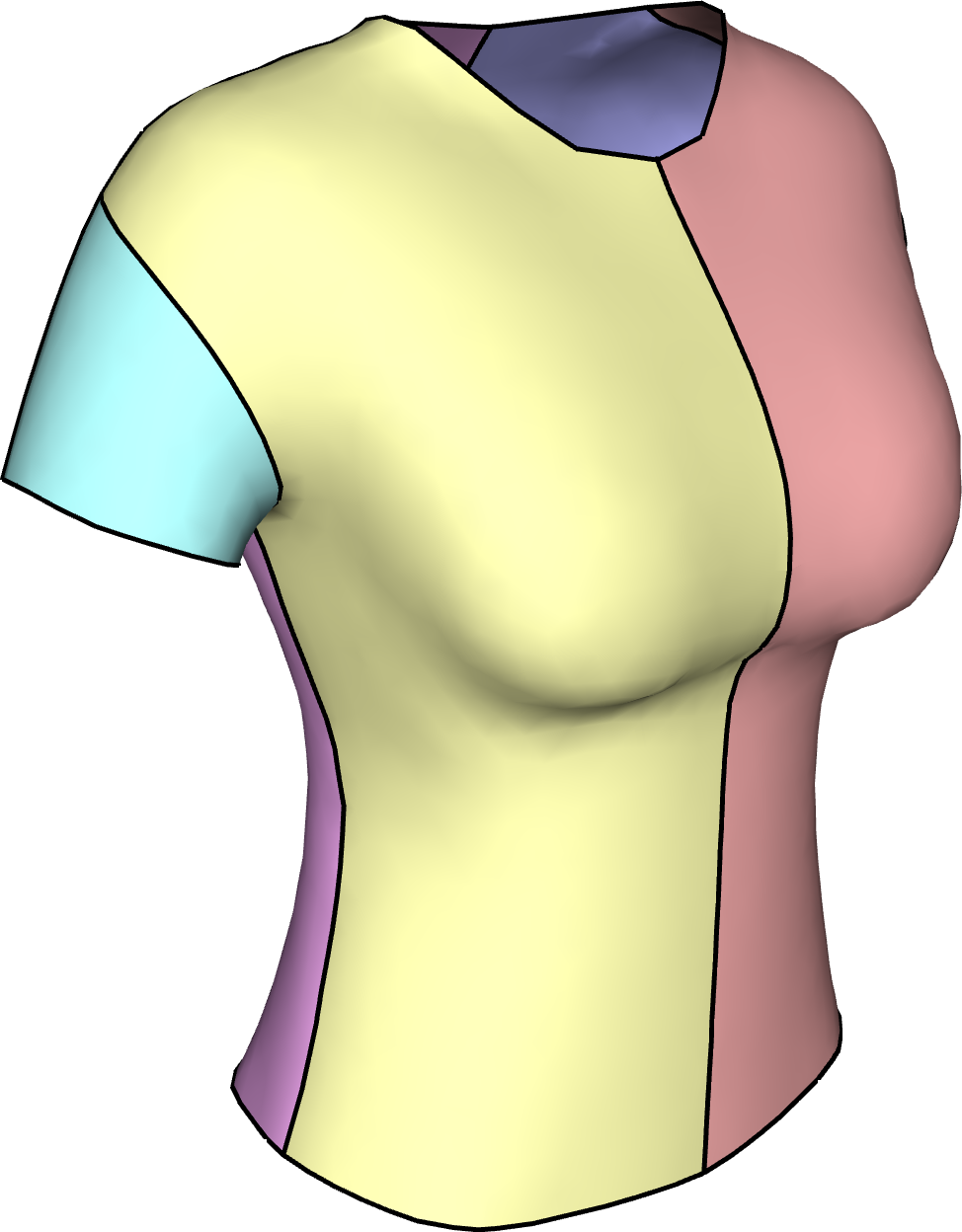}&
        \includegraphics[width=0.15\linewidth]{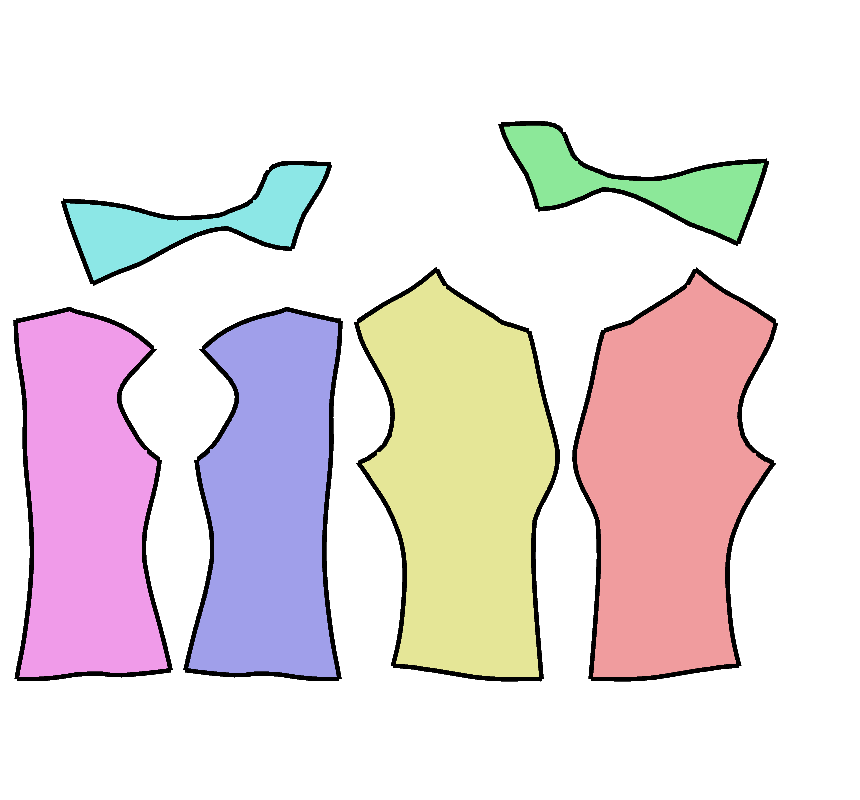}
        \end{tabular}\\
        \small $C=6$ , $s_\mathrm{max} = 0.05$
    \end{tabular}&
    \begin{tabular}{@{}c@{}}
        \begin{tabular}{@{}c@{}c@{}}
        \includegraphics[width=0.15\linewidth]{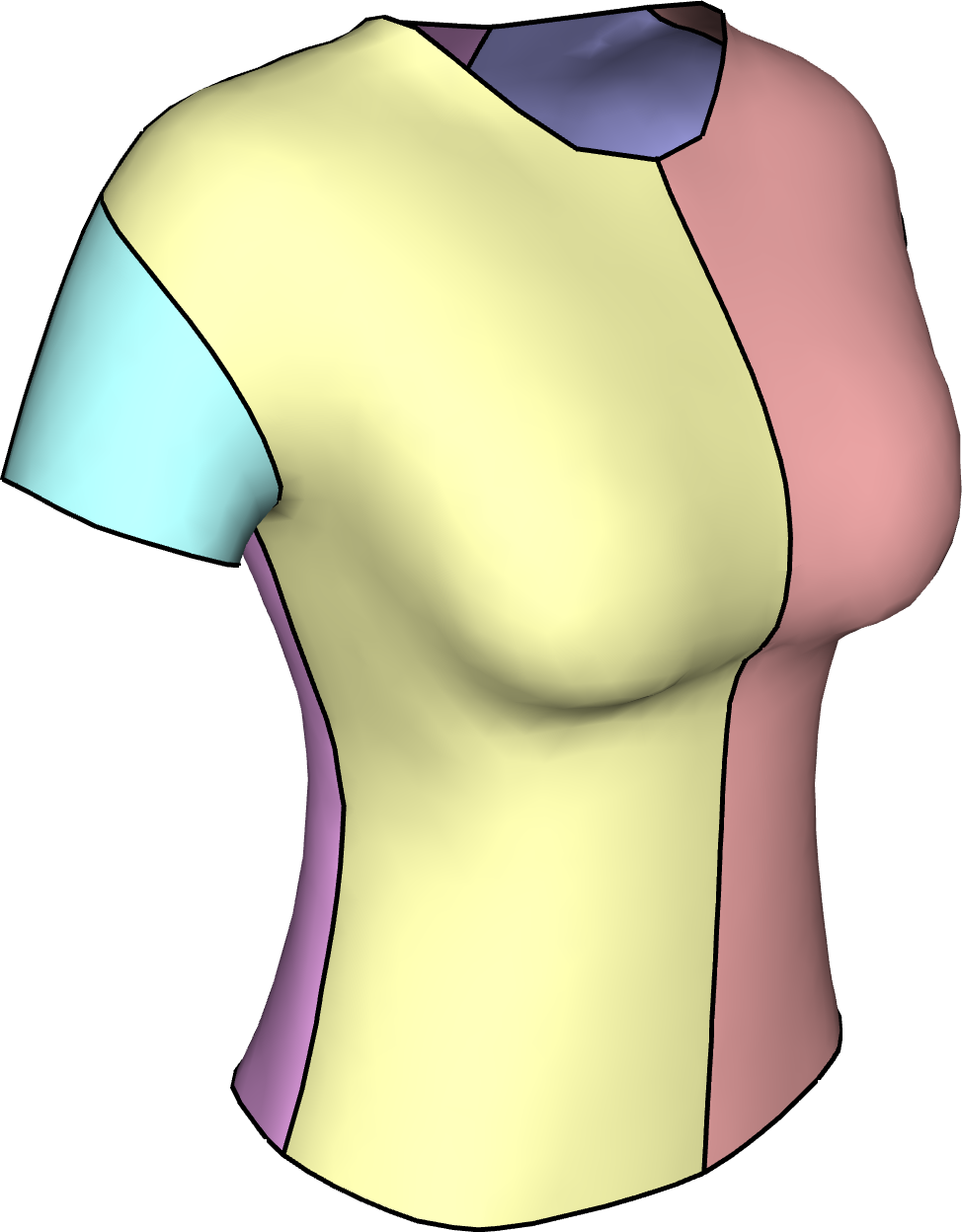}&
        \includegraphics[width=0.15\linewidth]{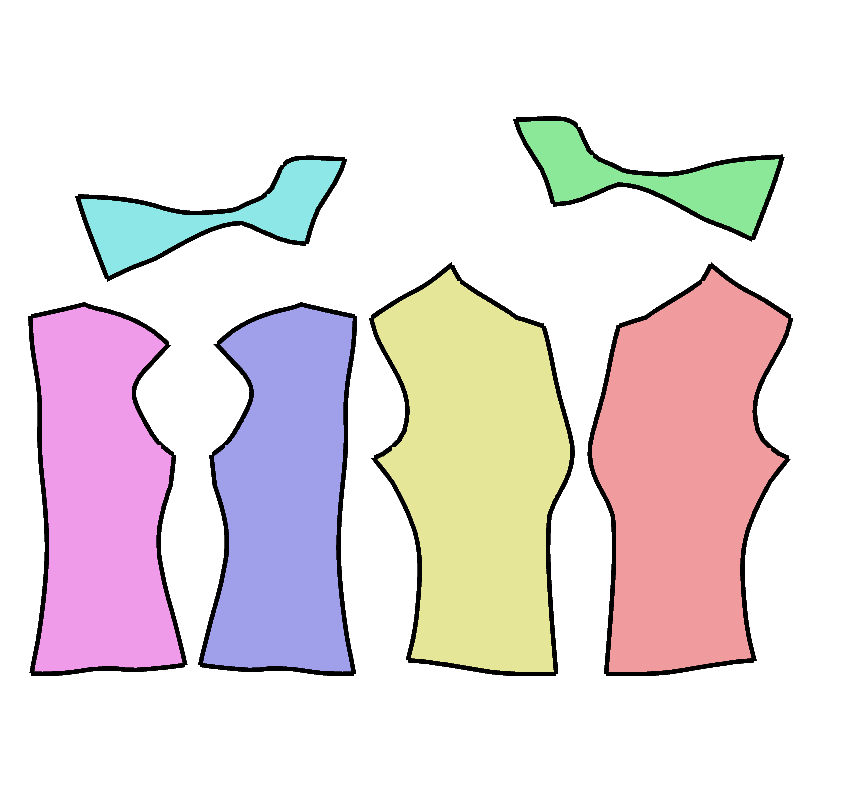}
        \end{tabular}\\
        \small $C=6$ , $s_\mathrm{max} = 0.04$
    \end{tabular}
    \begin{tabular}{@{}c@{}}
        \begin{tabular}{@{}c@{}c@{}}
        \includegraphics[width=0.15\linewidth]{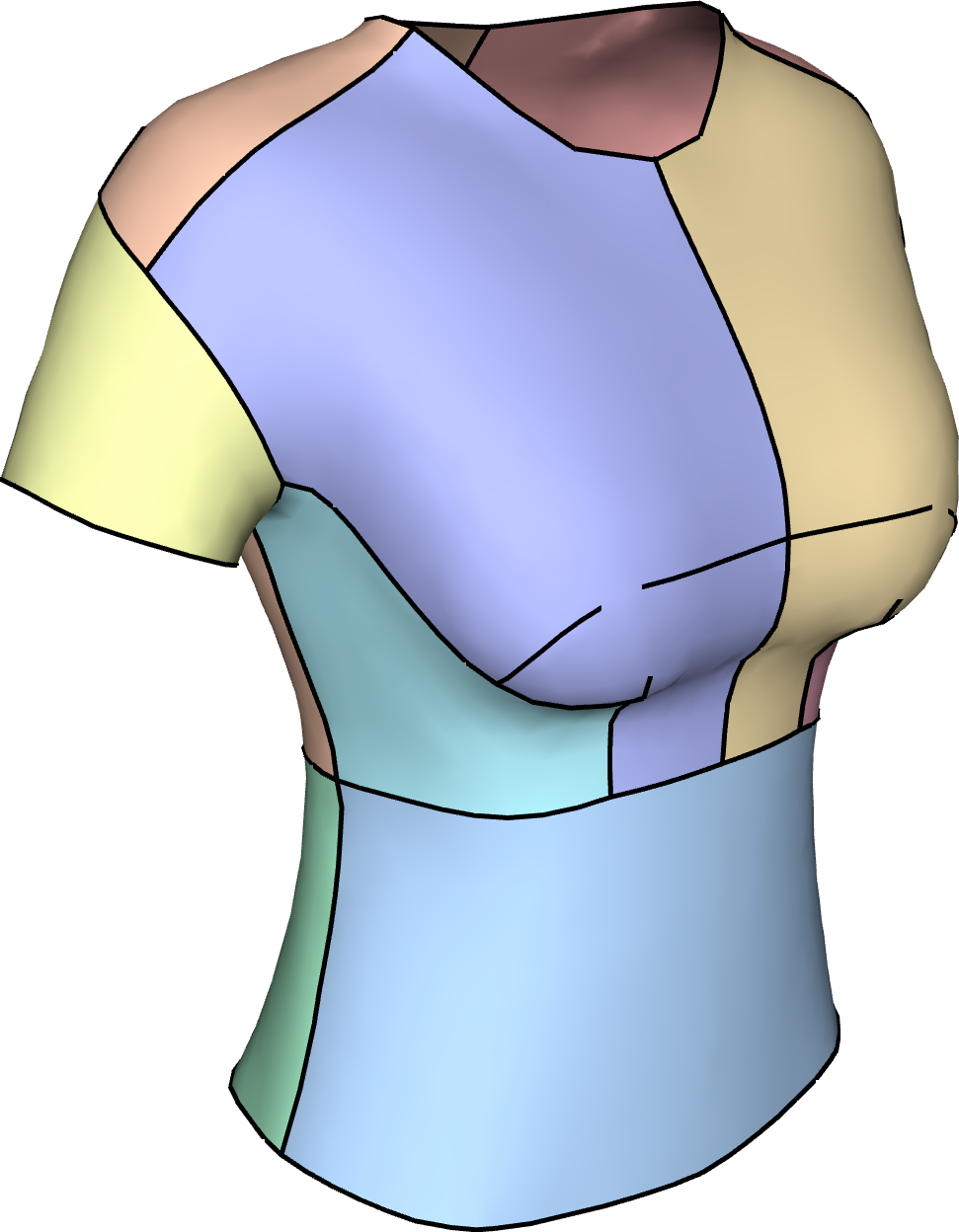}&
        \includegraphics[width=0.15\linewidth]{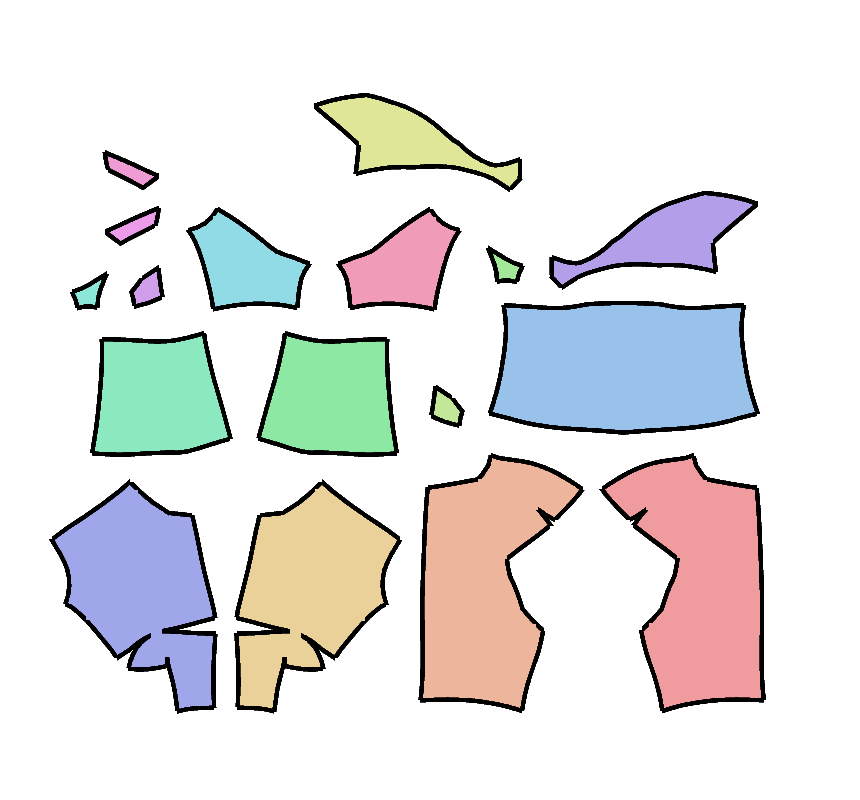}
        \end{tabular}\\
        \small $C=6$ , $s_\mathrm{max} = 0.02$
    \end{tabular}
\end{tabular}
    \caption{Automatic pattern layout generation for a shirt model, varying the two parameters that control the decomposition: the maximum number of corners per patch $C$ and the maximum stretch $s_\mathrm{max}$.}
    \label{fig:parameters}
\end{figure*}

\begin{figure}[th]
\begin{tabular}{@{}cc@{}}
        \includegraphics[width=0.25\linewidth]{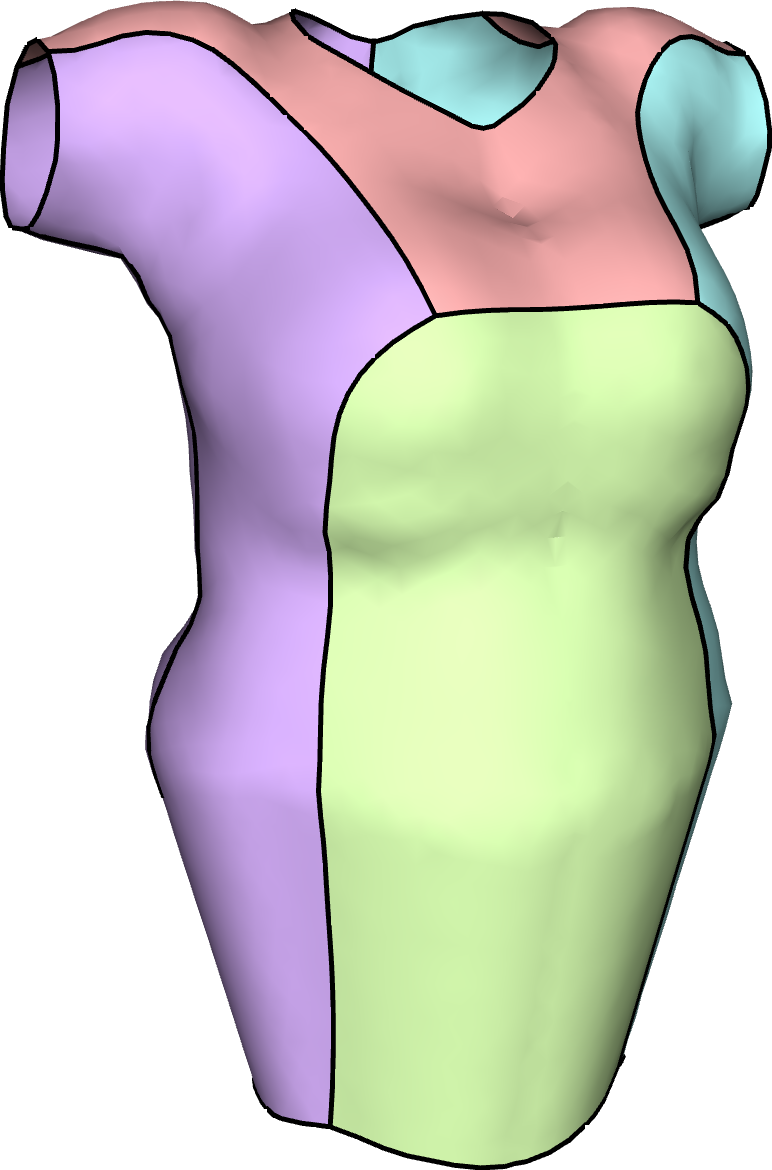}&
        \includegraphics[width=0.45\linewidth]{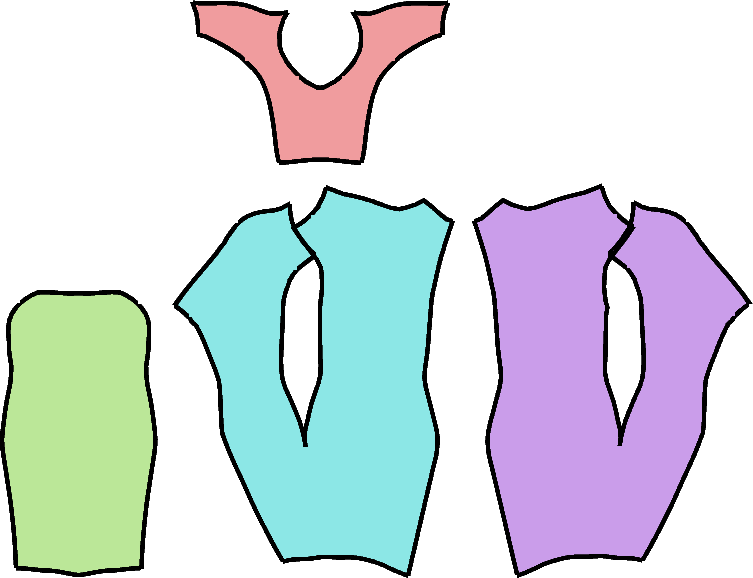}\\
\end{tabular}
    \caption{A tight-fitting dress pattern for a scanned body shape.}
    \label{fig:diff_body}
\end{figure}

\section{Results}
\label{sec:results}

\begin{figure}[t]
\begin{tabular}{@{}ccc@{}}
       \includegraphics[width=0.3\linewidth]{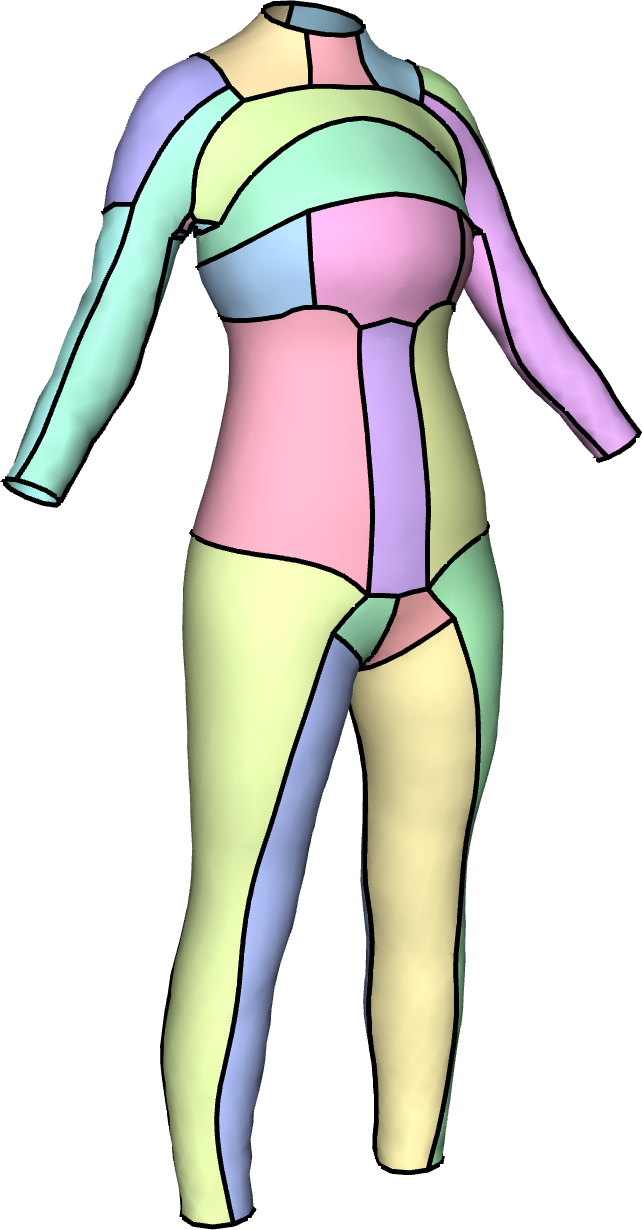}&
       \includegraphics[width=0.3\linewidth]{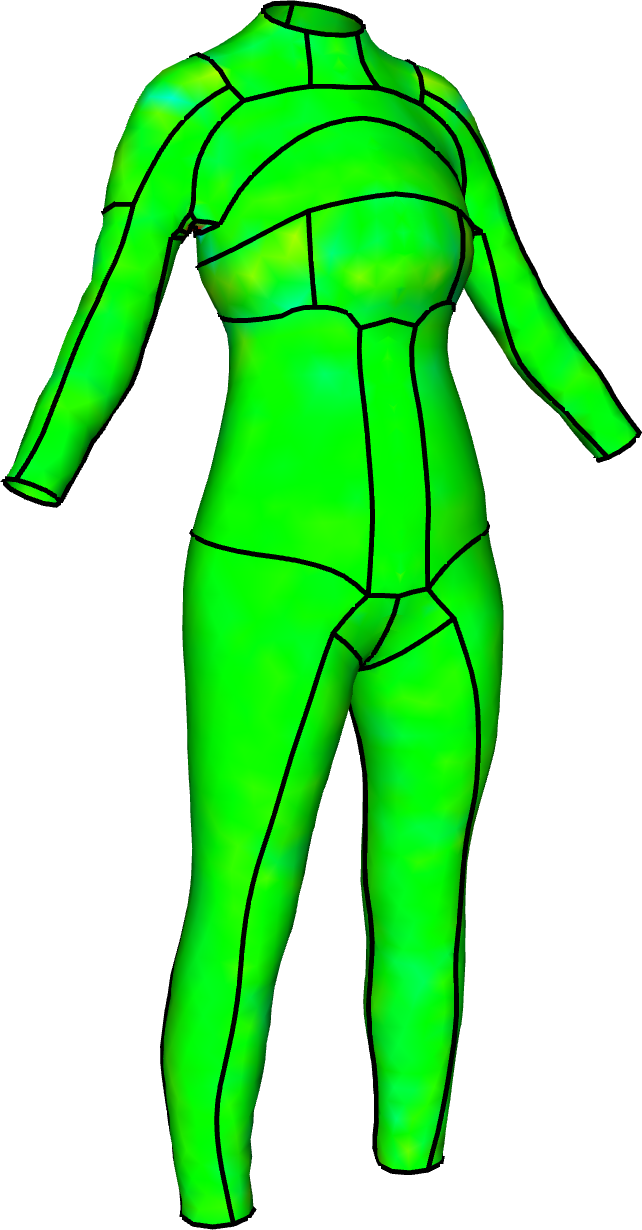}&
       \includegraphics[width=0.3\linewidth]{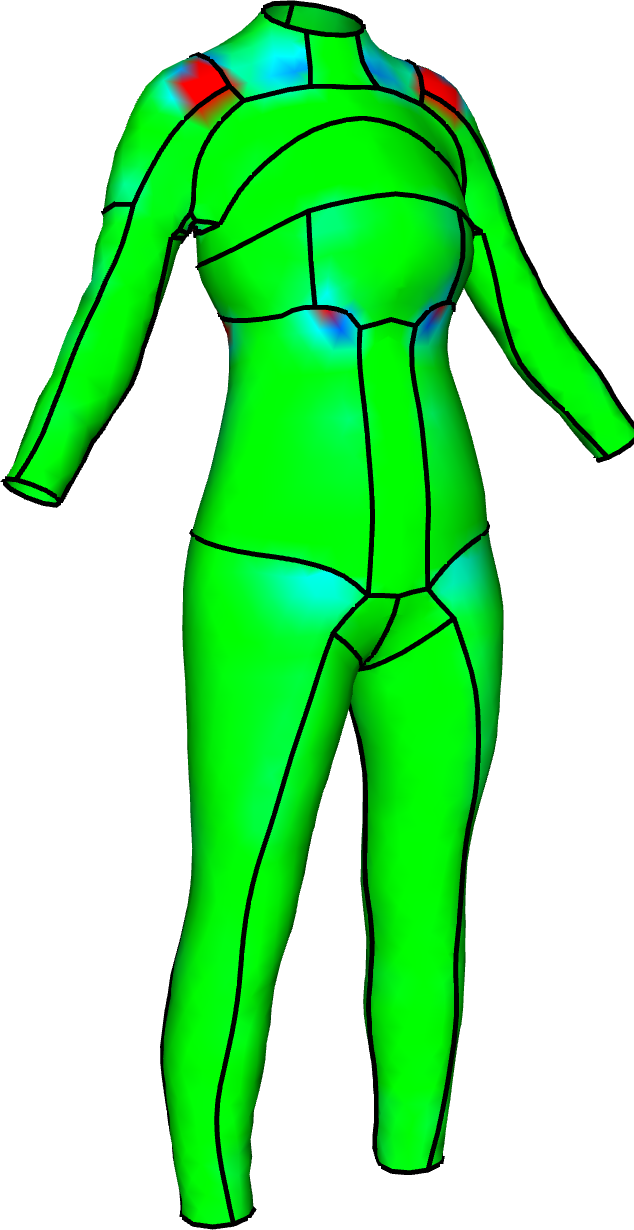}\\
       \includegraphics[width=0.3\linewidth]{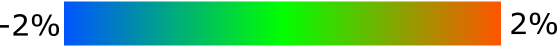}  & 
       \small fabric stress & \small ARAP energy
\end{tabular}
    \caption{We measure the fabric stress and the ARAP energy on one of our patterns, showing that we achieve good results in terms of both measures.}
    \label{fig:stress}
\end{figure}

We integrate our framework into an interactive editor that allows the designer to control the result and explore the different sewing patterns generated by our system. 
The main two main parameters exposed to the user are the maximum number of corners per pattern piece, $C$, and the maximum allowed stretch $s_\mathrm{max}$. \figref{fig:parameters} shows the effect of these parameters on a simple example. The higher the number of permitted corners, the fewer patches appear in the patch decomposition. However, the patches are then more complex than the ones generated with fewer corners and potentially require more handling when sewing. Similarly, the maximum allowed distortion is also related to the number of patches inserted. Intuitively, more patches are needed to comply with a stricter distortion threshold, as seen in \figref{fig:parameters}. In \figref{fig:stress}, we show how our system is able to match the maximal prescribed stretch, and we also measure the parameterization distortion for each pattern piece using the ARAP energy $E_\mathrm{rigid}$ \cite{Liu2008}, showing that we  achieve good results also in terms of isometry.

The designer can also adjust the patch layout using sketches on the 3D garment surface, see \figref{fig:interactive} and the accompanying video. The initial cross-field is updated after each sketch to allow the patch decomposition to conform to the newly inserted constraints. We handle sketches in the same manner as \citet{Pietroni2021} handle sharp features. Our method allows real-time editing for small models (up to 3000 triangles) and takes a few seconds to process larger models.

\begin{figure*}[t]
    \begin{tabular}{c c|c c c c }
    \includegraphics[height=0.22\linewidth]{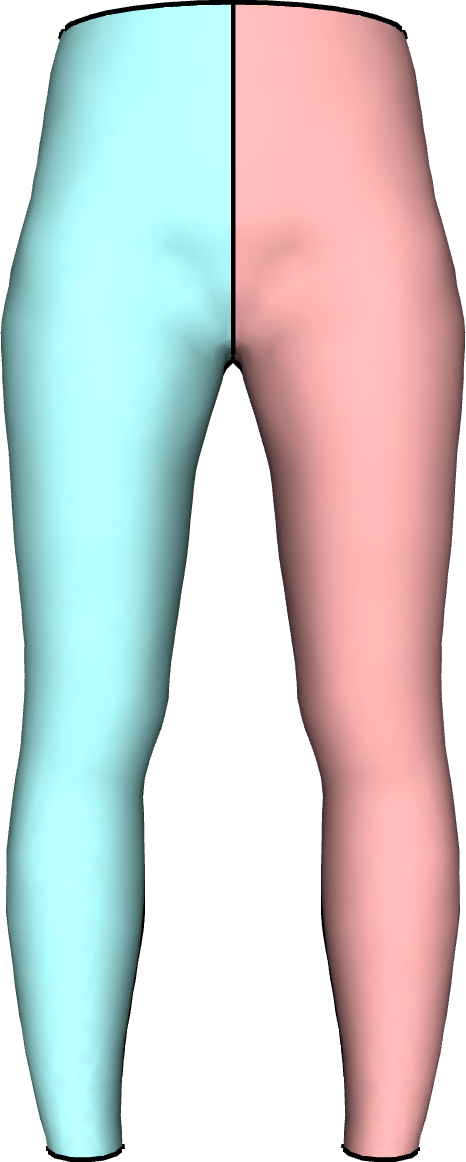}&
    \includegraphics[height=0.22\linewidth]{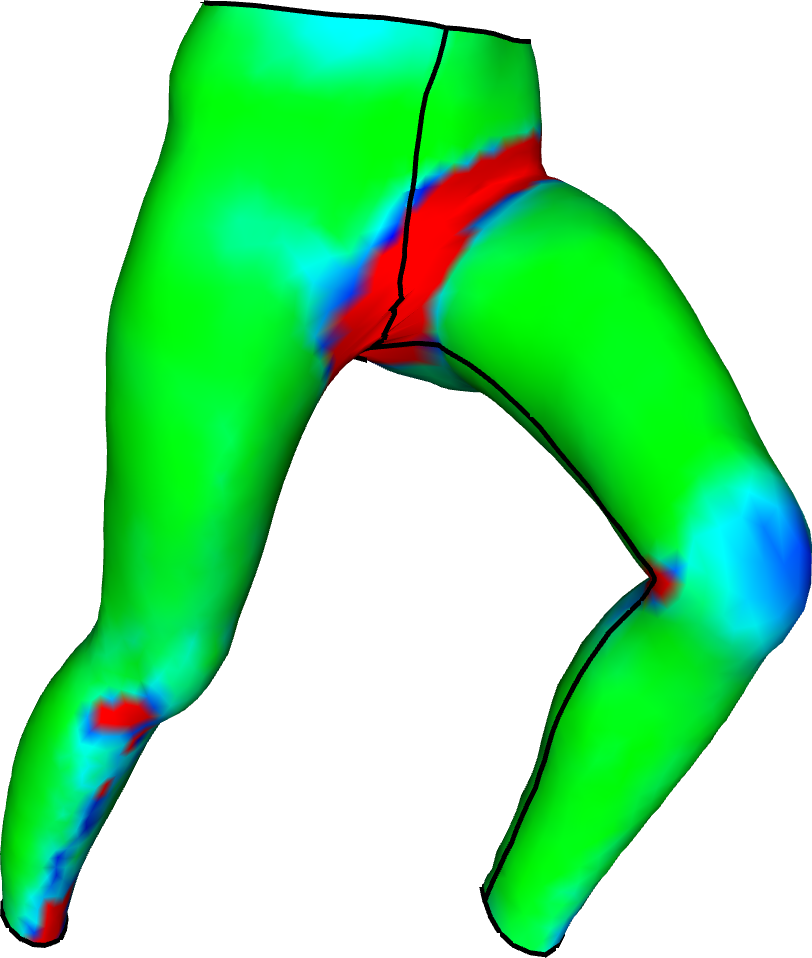}&
    \includegraphics[height=0.22\linewidth]{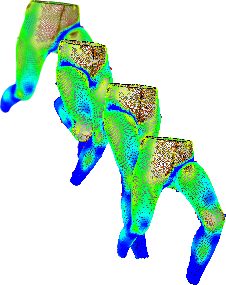}&
    \includegraphics[height=0.22\linewidth]{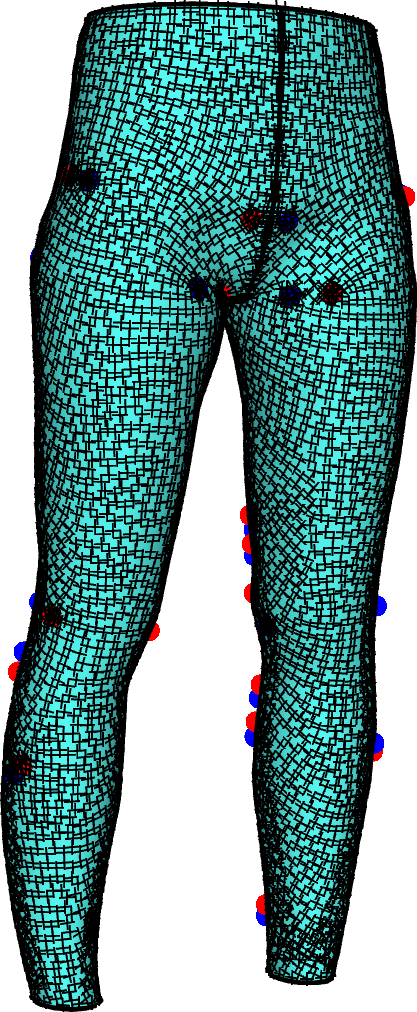}&
    \includegraphics[height=0.22\linewidth]{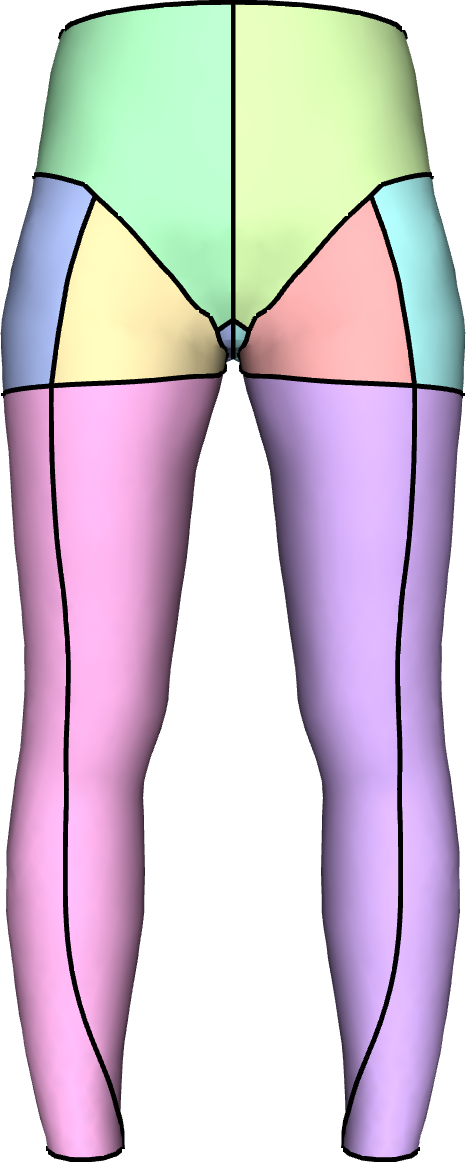}&
    \includegraphics[height=0.22\linewidth]{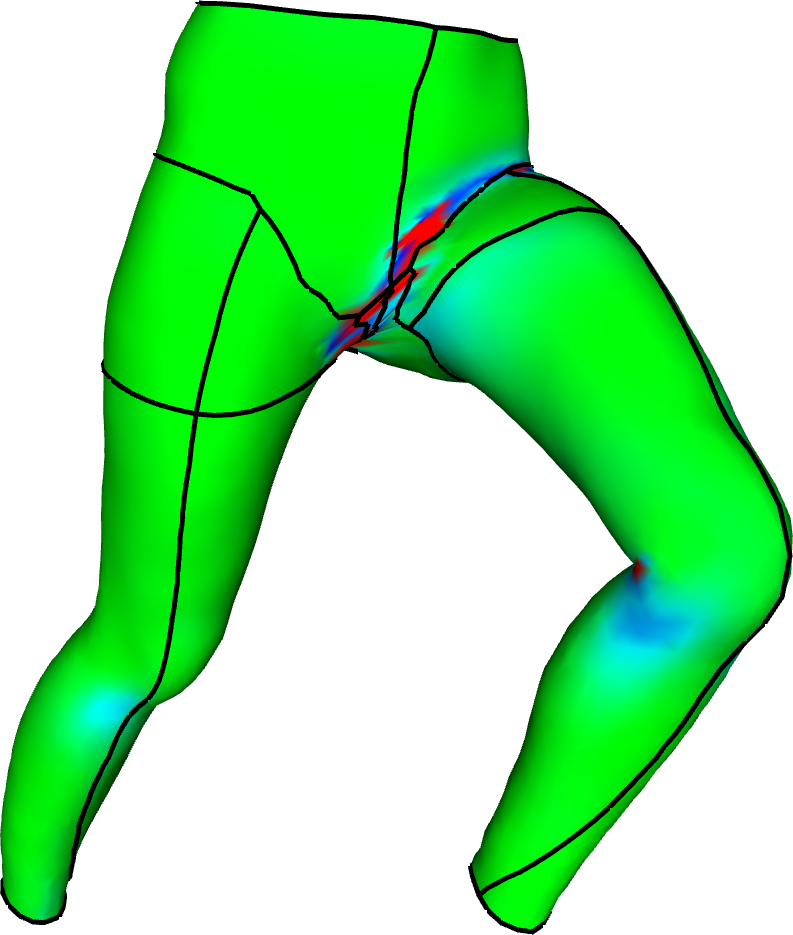}
    \end{tabular}
    \caption{For animated models, we use the multi-scale principal curvature directions integrated though the frames to derive the cross-field. Similarly, we add per-frame energy terms in the patch parameterization. Having multiple target poses of the garment helps the method to place seams that adapt to highly deforming areas, such as the area where the leg connects to the torso (see the result on the right), which would otherwise not be noticed (see the result computed based on the rest shape alone on the left).}
    \label{fig:animated}
\end{figure*}

\begin{figure*}[t]
\begin{tabular}{@{}cc@{}}
    \begin{tabular}{c}
        \includegraphics[width=0.45\linewidth]{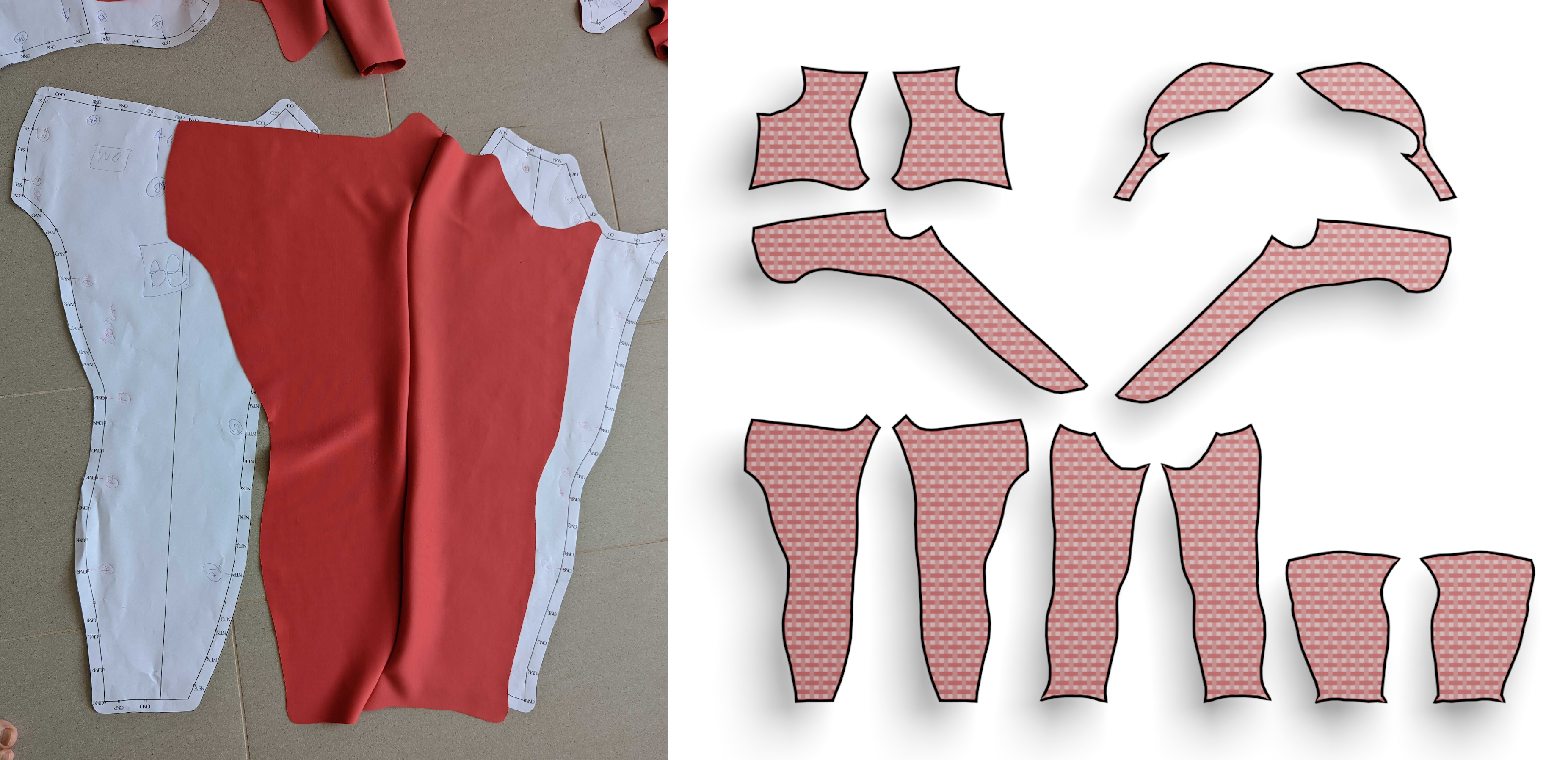}
    \end{tabular}&
    \begin{tabular}{cc}
        \includegraphics[width=0.25\linewidth]{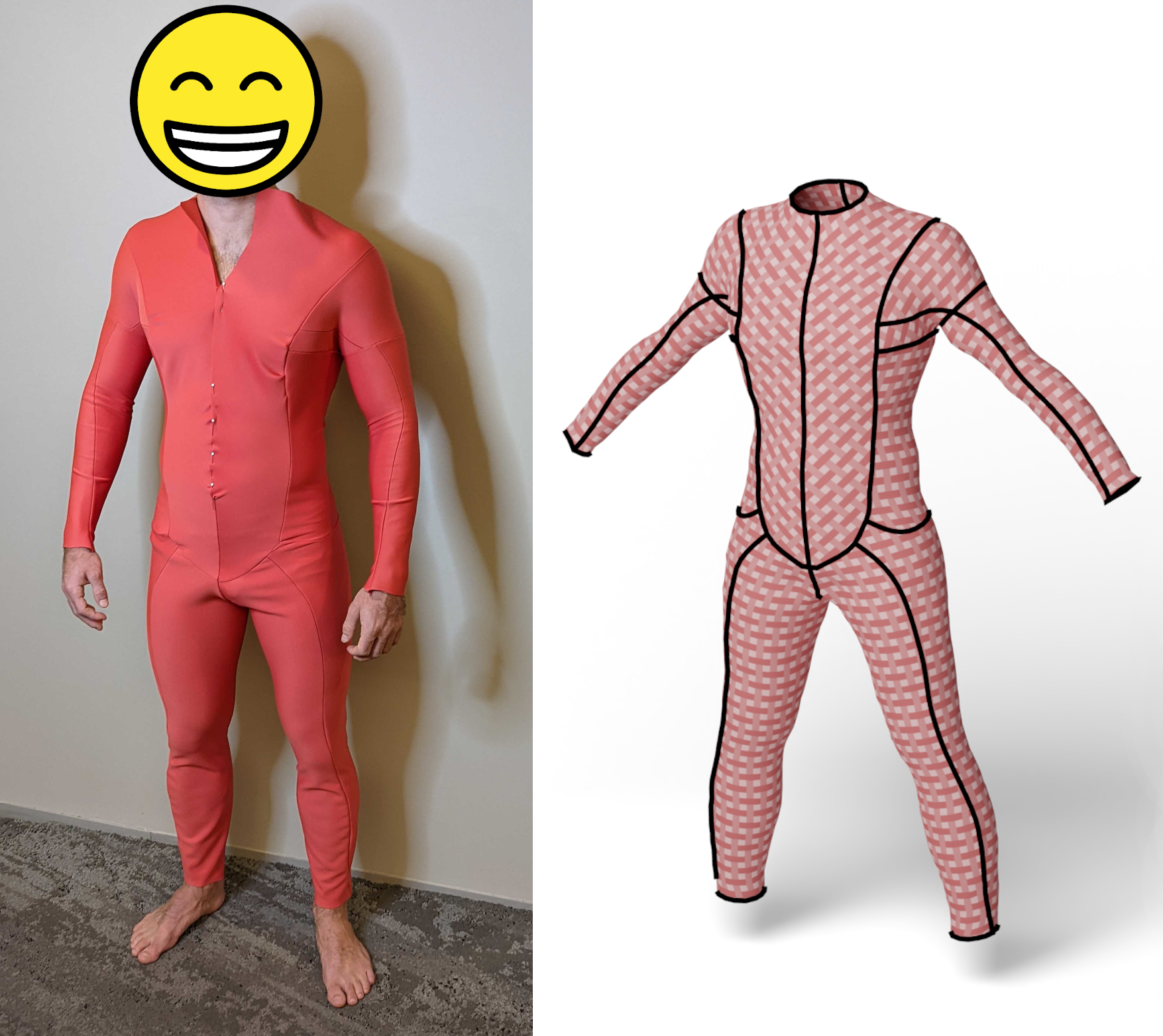}&
        \includegraphics[width=0.25\linewidth]{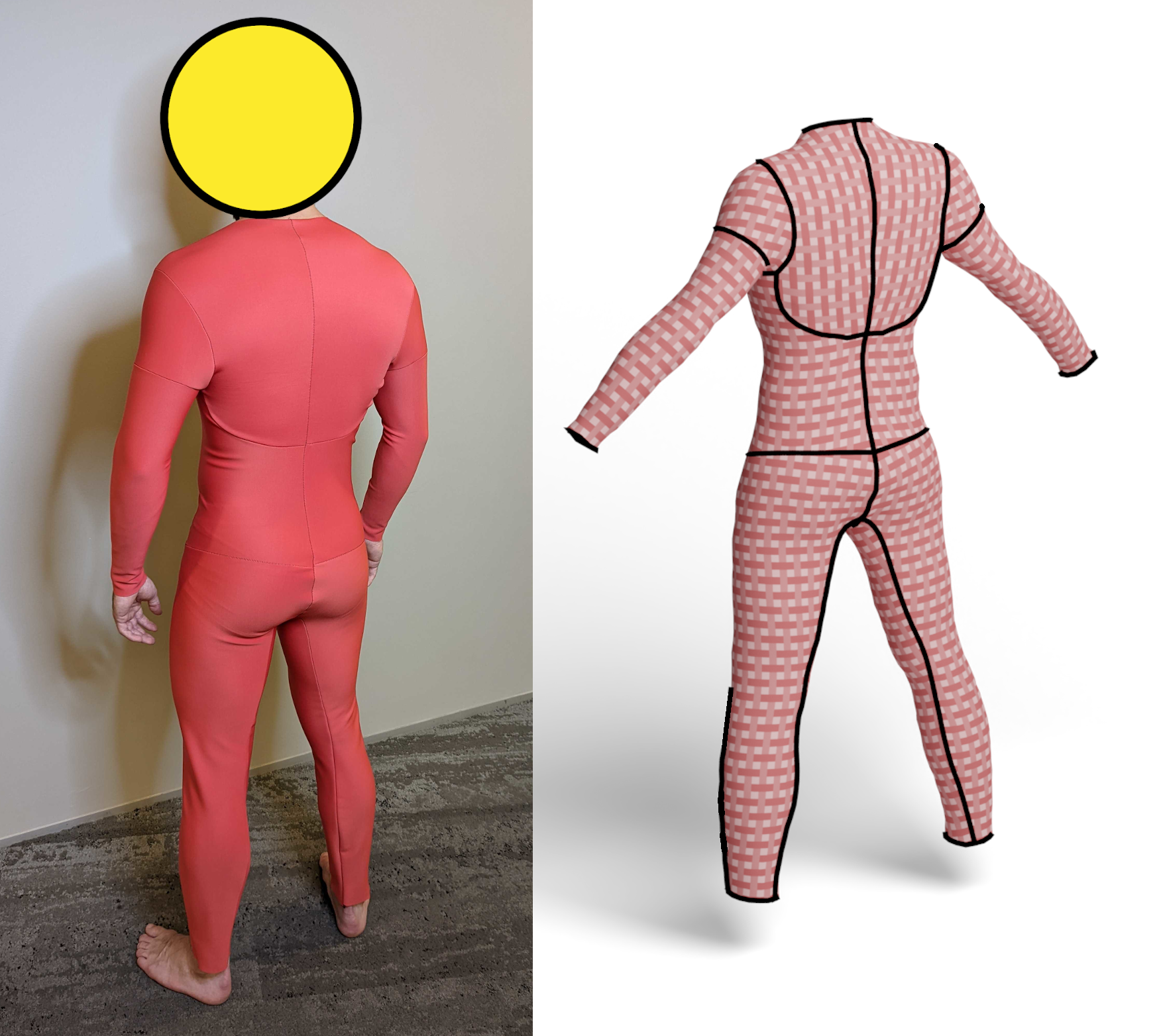}
    \end{tabular}\\
     (a) fabrication from our generated pattern &  (b) final garment\\
\end{tabular}
    \caption{A fabricated wetsuit. The input 3D garment model is based on a 3D scan of the subject. The open collar will be replaced with a zipper in the final garment. It took around 5 hours for a single person to prepare the pieces, cut, and sew.}
    \label{fig:fabrication_wetsuit}
\end{figure*}

\begin{figure}[t]
    \includegraphics[width=0.90\linewidth]{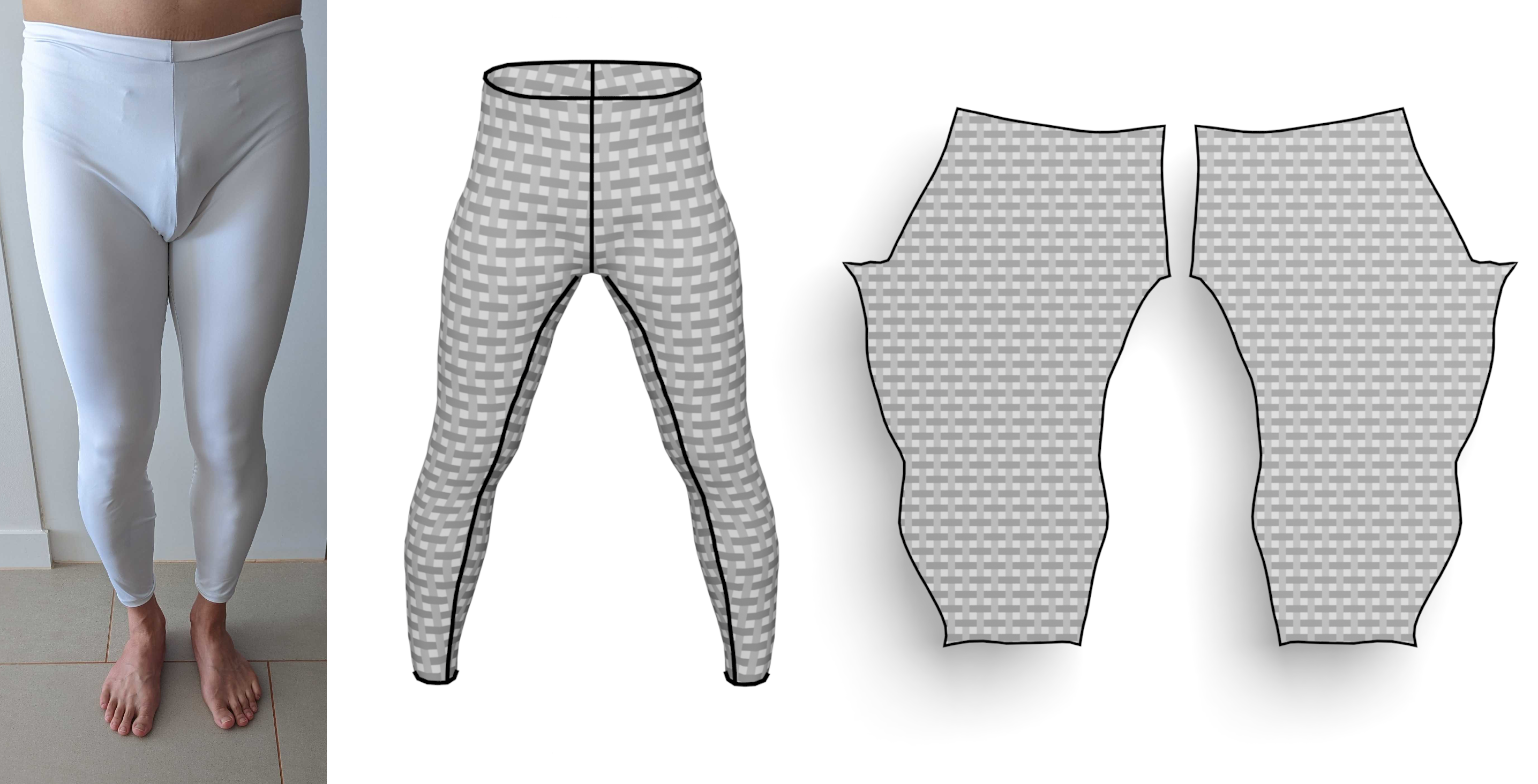}
    \caption{Tight-fitting trousers assembled and fabricated in around one hour.}
    \label{fig:fabrication_trousers}
\end{figure}

Our method works on arbitrary input garments, both loose (\figref{fig:teaser}) and tight-fitting (\figref{fig:diff_body}). Our framework can be used to fabricate personalized garments based on 3D scans (see Figures \ref{fig:fabrication_wetsuit}, \ref{fig:fabrication_trousers}) or even garments for animals (see \figref{fig:animals}). When multiple target poses of the 3D garment are available, our method adapts the sewing pattern to accommodate the motion, see \figref{fig:animated}. 

We fabricated two examples: a wetsuit and a pair of leggings (see Figures \ref{fig:fabrication_wetsuit} and \ref{fig:fabrication_trousers}). Both manufactured garments exhibit an excellent fit, demonstrating the feasibility of our approach. The wetsuit is made of neoprene and assembled with an overlocker machine, whereas the leggings are made from 90\% polyester  and 10\% spandex. The wetsuit pattern has an intricate structure and took around 5 hours for a single person to cut and sew, whereas the leggings only took about an hour. 

\section{Conclusions}
\label{sec:discussion}
We present a pipeline that generates computational sewing patterns directly from 3D digital garments. The key novelty is our algorithm for shape segmentation and patch flattening that explicitly focuses on garment fabrication. By the usage of darts and taking into account the anisotropic properties of woven fabric, grain alignment and seam symmetry, our approach gives the designer the tools to produce (automatically and interactively) the 2D patterns to make the desired garment. We show that our algorithm produces patterns that are indeed suitable for fabrication of a custom, digitally designed garment. Our results also show that the pipeline can be applied to generate dewing patterns for tight-fitting and loose garments, for all types of body shapes, including animals.

\paragraph{Limitations and future work.}
Although our approach is quite robust, it has some limitations. If the input mesh is the result of a simulation or a scan of a physical garment, the wrinkles can introduce noise in the guiding field.
Moreover, since the field tracing approach is greedy, the distortion in the final parameterization of the pattern pieces may differ from the one computed during the patch decomposition stage. This is because the seam reflection constraints cannot be incorporated before all patches are computed. In practice, we never encountered this problem. Also owing to the greedy nature of our method, a slight difference in the input garment or user constraints can result in a significantly different pattern. Finally, in future work we would like to include seam allowance constraints in our framework, to further secure easy and robust fabrication of the resulting sewing patterns.

\bibliographystyle{ACM-Reference-Format}
\bibliography{parafashion}


\begin{thebibliography}{74}


\ifx \showCODEN    \undefined \def \showCODEN     #1{\unskip}     \fi
\ifx \showDOI      \undefined \def \showDOI       #1{#1}\fi
\ifx \showISBNx    \undefined \def \showISBNx     #1{\unskip}     \fi
\ifx \showISBNxiii \undefined \def \showISBNxiii  #1{\unskip}     \fi
\ifx \showISSN     \undefined \def \showISSN      #1{\unskip}     \fi
\ifx \showLCCN     \undefined \def \showLCCN      #1{\unskip}     \fi
\ifx \shownote     \undefined \def \shownote      #1{#1}          \fi
\ifx \showarticletitle \undefined \def \showarticletitle #1{#1}   \fi
\ifx \showURL      \undefined \def \showURL       {\relax}        \fi
\providecommand\bibfield[2]{#2}
\providecommand\bibinfo[2]{#2}
\providecommand\natexlab[1]{#1}
\providecommand\showeprint[2][]{arXiv:#2}

\bibitem[\protect\citeauthoryear{Alldieck, Xu, and Sminchisescu}{Alldieck
  et~al\mbox{.}}{2021}]%
        {imGHUM:2021}
\bibfield{author}{\bibinfo{person}{Thiemo Alldieck}, \bibinfo{person}{Hongyi
  Xu}, {and} \bibinfo{person}{Cristian Sminchisescu}.}
  \bibinfo{year}{2021}\natexlab{}.
\newblock \showarticletitle{imGHUM: Implicit Generative Models of 3D Human
  Shape and Articulated Pose}. In \bibinfo{booktitle}{\emph{Proc.\ ICCV}}.
  \bibinfo{pages}{5461--5470}.
\newblock
\urldef\tempurl%
\url{https://arxiv.org/abs/2108.10842}
\showURL{%
\tempurl}


\bibitem[\protect\citeauthoryear{Bang, Korosteleva, and Lee}{Bang
  et~al\mbox{.}}{2021}]%
        {Bang2021}
\bibfield{author}{\bibinfo{person}{Seungbae Bang}, \bibinfo{person}{Maria
  Korosteleva}, {and} \bibinfo{person}{Sung-Hee Lee}.}
  \bibinfo{year}{2021}\natexlab{}.
\newblock \showarticletitle{Estimating Garment Patterns from Static Scan Data}.
\newblock \bibinfo{journal}{\emph{Computer Graphics Forum}}
  \bibinfo{volume}{40} (\bibinfo{date}{05} \bibinfo{year}{2021}).
\newblock
\urldef\tempurl%
\url{https://doi.org/10.1111/cgf.14272}
\showDOI{\tempurl}


\bibitem[\protect\citeauthoryear{Bartle, Sheffer, Kim, Kaufman, Vining, and
  Berthouzoz}{Bartle et~al\mbox{.}}{2016}]%
        {Bartle2016}
\bibfield{author}{\bibinfo{person}{Aric Bartle}, \bibinfo{person}{Alla
  Sheffer}, \bibinfo{person}{Vladimir~G. Kim}, \bibinfo{person}{Danny~M.
  Kaufman}, \bibinfo{person}{Nicholas Vining}, {and} \bibinfo{person}{Floraine
  Berthouzoz}.} \bibinfo{year}{2016}\natexlab{}.
\newblock \showarticletitle{Physics-Driven Pattern Adjustment for Direct 3D
  Garment Editing}.
\newblock \bibinfo{journal}{\emph{ACM Trans. Graph.}} \bibinfo{volume}{35},
  \bibinfo{number}{4}, Article \bibinfo{articleno}{50} (\bibinfo{date}{jul}
  \bibinfo{year}{2016}), \bibinfo{numpages}{11}~pages.
\newblock
\showISSN{0730-0301}
\urldef\tempurl%
\url{https://doi.org/10.1145/2897824.2925896}
\showDOI{\tempurl}


\bibitem[\protect\citeauthoryear{Binninger*, Verhoeven*, Herholz, and
  Sorkine-Hornung}{Binninger* et~al\mbox{.}}{2021}]%
        {BinningerVerhoeven:GaussThinning:2021}
\bibfield{author}{\bibinfo{person}{Alexandre Binninger*},
  \bibinfo{person}{Floor Verhoeven*}, \bibinfo{person}{Philipp Herholz}, {and}
  \bibinfo{person}{Olga Sorkine-Hornung}.} \bibinfo{year}{2021}\natexlab{}.
\newblock \showarticletitle{Developable Approximation via Gauss Image
  Thinning}.
\newblock \bibinfo{journal}{\emph{Computer Graphics Forum (proceedings of SGP
  2021)}} \bibinfo{volume}{40}, \bibinfo{number}{5} (\bibinfo{year}{2021}),
  \bibinfo{pages}{289--300}.
\newblock
\showISSN{1467-8659}
\urldef\tempurl%
\url{https://doi.org/10.1111/cgf.14374}
\showDOI{\tempurl}


\bibitem[\protect\citeauthoryear{Bogo, Romero, Loper, and Black}{Bogo
  et~al\mbox{.}}{2014}]%
        {bogo2014faust}
\bibfield{author}{\bibinfo{person}{Federica Bogo}, \bibinfo{person}{Javier
  Romero}, \bibinfo{person}{Matthew Loper}, {and} \bibinfo{person}{Michael~J
  Black}.} \bibinfo{year}{2014}\natexlab{}.
\newblock \showarticletitle{{FAUST}: Dataset and evaluation for 3D mesh
  registration}. In \bibinfo{booktitle}{\emph{Proc.\ CVPR}}.
  \bibinfo{pages}{3794--3801}.
\newblock


\bibitem[\protect\citeauthoryear{Bommes, L{\'{e}}vy, Pietroni, Puppo, Silva,
  Tarini, and Zorin}{Bommes et~al\mbox{.}}{2013}]%
        {Bommes2013}
\bibfield{author}{\bibinfo{person}{David Bommes}, \bibinfo{person}{Bruno
  L{\'{e}}vy}, \bibinfo{person}{Nico Pietroni}, \bibinfo{person}{Enrico Puppo},
  \bibinfo{person}{Cl{\'{a}}udio~T. Silva}, \bibinfo{person}{Marco Tarini},
  {and} \bibinfo{person}{Denis Zorin}.} \bibinfo{year}{2013}\natexlab{}.
\newblock \showarticletitle{Quad-Mesh Generation and Processing: {A} Survey}.
\newblock \bibinfo{journal}{\emph{Comput. Graph. Forum}} \bibinfo{volume}{32},
  \bibinfo{number}{6} (\bibinfo{year}{2013}), \bibinfo{pages}{51--76}.
\newblock


\bibitem[\protect\citeauthoryear{Bommes, Zimmer, and Kobbelt}{Bommes
  et~al\mbox{.}}{2009}]%
        {Bommes2009}
\bibfield{author}{\bibinfo{person}{David Bommes}, \bibinfo{person}{Henrik
  Zimmer}, {and} \bibinfo{person}{Leif Kobbelt}.}
  \bibinfo{year}{2009}\natexlab{}.
\newblock \showarticletitle{Mixed-integer quadrangulation}.
\newblock \bibinfo{journal}{\emph{{ACM} Trans. Graph.}} \bibinfo{volume}{28},
  \bibinfo{number}{3} (\bibinfo{year}{2009}), \bibinfo{pages}{77}.
\newblock


\bibitem[\protect\citeauthoryear{Brouet, Sheffer, Boissieux, and Cani}{Brouet
  et~al\mbox{.}}{2012}]%
        {Brouet2012}
\bibfield{author}{\bibinfo{person}{Remi Brouet}, \bibinfo{person}{Alla
  Sheffer}, \bibinfo{person}{Laurence Boissieux}, {and}
  \bibinfo{person}{Marie-Paule Cani}.} \bibinfo{year}{2012}\natexlab{}.
\newblock \showarticletitle{Design Preserving Garment Transfer}.
\newblock \bibinfo{journal}{\emph{ACM Trans. Graph.}} \bibinfo{volume}{31},
  \bibinfo{number}{4}, Article \bibinfo{articleno}{36} (\bibinfo{date}{jul}
  \bibinfo{year}{2012}), \bibinfo{numpages}{11}~pages.
\newblock
\showISSN{0730-0301}
\urldef\tempurl%
\url{https://doi.org/10.1145/2185520.2185532}
\showDOI{\tempurl}


\bibitem[\protect\citeauthoryear{Campen}{Campen}{2017}]%
        {Campen2017survey}
\bibfield{author}{\bibinfo{person}{Marcel Campen}.}
  \bibinfo{year}{2017}\natexlab{}.
\newblock \showarticletitle{Partitioning Surfaces Into Quadrilateral Patches:
  {A} Survey}.
\newblock \bibinfo{journal}{\emph{Comput. Graph. Forum}} \bibinfo{volume}{36},
  \bibinfo{number}{8} (\bibinfo{year}{2017}), \bibinfo{pages}{567--588}.
\newblock


\bibitem[\protect\citeauthoryear{Campen, Bommes, and Kobbelt}{Campen
  et~al\mbox{.}}{2012}]%
        {Campen2012}
\bibfield{author}{\bibinfo{person}{Marcel Campen}, \bibinfo{person}{David
  Bommes}, {and} \bibinfo{person}{Leif Kobbelt}.}
  \bibinfo{year}{2012}\natexlab{}.
\newblock \showarticletitle{Dual loops meshing: quality quad layouts on
  manifolds}.
\newblock \bibinfo{journal}{\emph{{ACM} Trans. Graph.}} \bibinfo{volume}{31},
  \bibinfo{number}{4} (\bibinfo{year}{2012}), \bibinfo{pages}{110:1--110:11}.
\newblock


\bibitem[\protect\citeauthoryear{Chen}{Chen}{1998}]%
        {Chen1998}
\bibfield{author}{\bibinfo{person}{Jocelyn Hua-Chu Chen}.}
  \bibinfo{year}{1998}\natexlab{}.
\newblock \showarticletitle{An investigation into 3-Dimensional Garment Pattern
  Design}.
\newblock \bibinfo{journal}{\emph{PhD Thesis, Nottingham Trent University, UK}}
  (\bibinfo{year}{1998}).
\newblock


\bibitem[\protect\citeauthoryear{Chen, Zhou, Lu, Wang, Bi, and Tan}{Chen
  et~al\mbox{.}}{2015}]%
        {Chen2015}
\bibfield{author}{\bibinfo{person}{Xiaowu Chen}, \bibinfo{person}{Bin Zhou},
  \bibinfo{person}{Feixiang Lu}, \bibinfo{person}{Lin Wang},
  \bibinfo{person}{Lang Bi}, {and} \bibinfo{person}{Ping Tan}.}
  \bibinfo{year}{2015}\natexlab{}.
\newblock \showarticletitle{Garment Modeling with a Depth Camera}.
\newblock \bibinfo{journal}{\emph{ACM Trans. Graph.}} \bibinfo{volume}{34},
  \bibinfo{number}{6}, Article \bibinfo{articleno}{203} (\bibinfo{date}{oct}
  \bibinfo{year}{2015}), \bibinfo{numpages}{12}~pages.
\newblock
\showISSN{0730-0301}
\urldef\tempurl%
\url{https://doi.org/10.1145/2816795.2818059}
\showDOI{\tempurl}


\bibitem[\protect\citeauthoryear{CLO}{CLO}{2022}]%
        {CLO}
\bibfield{author}{\bibinfo{person}{CLO}.} \bibinfo{year}{2022}\natexlab{}.
\newblock \bibinfo{title}{clo3d.com}.
\newblock \bibinfo{howpublished}{\url{https://www.clo3d.com}}.
\newblock


\bibitem[\protect\citeauthoryear{Cordier, Seo, and Magnenat-Thalmann}{Cordier
  et~al\mbox{.}}{2003}]%
        {cordier2003made}
\bibfield{author}{\bibinfo{person}{Frederic Cordier}, \bibinfo{person}{Hyewon
  Seo}, {and} \bibinfo{person}{Nadia Magnenat-Thalmann}.}
  \bibinfo{year}{2003}\natexlab{}.
\newblock \showarticletitle{Made-to-measure technologies for an online clothing
  store}.
\newblock \bibinfo{journal}{\emph{IEEE Computer graphics and Applications}}
  \bibinfo{volume}{23}, \bibinfo{number}{1} (\bibinfo{year}{2003}),
  \bibinfo{pages}{38--48}.
\newblock


\bibitem[\protect\citeauthoryear{Decaudin, Julius, Wither, Boissieux, Sheffer,
  and Cani}{Decaudin et~al\mbox{.}}{2006}]%
        {decaudin2006}
\bibfield{author}{\bibinfo{person}{Philippe Decaudin}, \bibinfo{person}{Dan
  Julius}, \bibinfo{person}{Jamie Wither}, \bibinfo{person}{Laurence
  Boissieux}, \bibinfo{person}{Alla Sheffer}, {and}
  \bibinfo{person}{Marie-Paule Cani}.} \bibinfo{year}{2006}\natexlab{}.
\newblock \showarticletitle{Virtual garments: A fully geometric approach for
  clothing design}.
\newblock \bibinfo{journal}{\emph{Comput.\ Graph.\ Forum}}
  \bibinfo{volume}{25}, \bibinfo{number}{3} (\bibinfo{year}{2006}),
  \bibinfo{pages}{625--634}.
\newblock


\bibitem[\protect\citeauthoryear{Diamanti, Vaxman, Panozzo, and
  Sorkine{-}Hornung}{Diamanti et~al\mbox{.}}{2014}]%
        {Diamanti2014}
\bibfield{author}{\bibinfo{person}{Olga Diamanti}, \bibinfo{person}{Amir
  Vaxman}, \bibinfo{person}{Daniele Panozzo}, {and} \bibinfo{person}{Olga
  Sorkine{-}Hornung}.} \bibinfo{year}{2014}\natexlab{}.
\newblock \showarticletitle{Designing \emph{N}-PolyVector Fields with Complex
  Polynomials}.
\newblock \bibinfo{journal}{\emph{Comput. Graph. Forum}} \bibinfo{volume}{33},
  \bibinfo{number}{5} (\bibinfo{year}{2014}), \bibinfo{pages}{1--11}.
\newblock


\bibitem[\protect\citeauthoryear{Hormann, Lévy, and Sheffer}{Hormann
  et~al\mbox{.}}{2007}]%
        {hormann:MPT:2007}
\bibfield{author}{\bibinfo{person}{Kai Hormann}, \bibinfo{person}{Bruno Lévy},
  {and} \bibinfo{person}{Alla Sheffer}.} \bibinfo{year}{2007}\natexlab{}.
\newblock \showarticletitle{Mesh Parameterization: Theory and Practice}. In
  \bibinfo{booktitle}{\emph{ACM SIGGRAPH Course Notes}}.
\newblock


\bibitem[\protect\citeauthoryear{Huang, Mok, Kwok, and Au}{Huang
  et~al\mbox{.}}{2012}]%
        {HUANG2012680}
\bibfield{author}{\bibinfo{person}{H.Q. Huang}, \bibinfo{person}{P.Y. Mok},
  \bibinfo{person}{Y.L. Kwok}, {and} \bibinfo{person}{J.S. Au}.}
  \bibinfo{year}{2012}\natexlab{}.
\newblock \showarticletitle{Block pattern generation: From parameterizing human
  bodies to fit feature-aligned and flattenable 3D garments}.
\newblock \bibinfo{journal}{\emph{Computers in Industry}} \bibinfo{volume}{63},
  \bibinfo{number}{7} (\bibinfo{year}{2012}), \bibinfo{pages}{680--691}.
\newblock
\showISSN{0166-3615}
\urldef\tempurl%
\url{https://doi.org/10.1016/j.compind.2012.04.001}
\showDOI{\tempurl}


\bibitem[\protect\citeauthoryear{Ion, Rabinovich, Herholz, and
  Sorkine{-}Hornung}{Ion et~al\mbox{.}}{2020}]%
        {Ion:ApproximatingDOGs:2020}
\bibfield{author}{\bibinfo{person}{Alexandra Ion}, \bibinfo{person}{Michael
  Rabinovich}, \bibinfo{person}{Philipp Herholz}, {and} \bibinfo{person}{Olga
  Sorkine{-}Hornung}.} \bibinfo{year}{2020}\natexlab{}.
\newblock \showarticletitle{Shape Approximation by Developable Wrapping}.
\newblock \bibinfo{journal}{\emph{ACM Transactions on Graphics (proceedings of
  SIGGRAPH ASIA)}} \bibinfo{volume}{39}, \bibinfo{number}{6}
  (\bibinfo{year}{2020}).
\newblock
\urldef\tempurl%
\url{https://doi.org/10.1145/3414685.3417835}
\showDOI{\tempurl}


\bibitem[\protect\citeauthoryear{Jiang, Schaefer, and Panozzo}{Jiang
  et~al\mbox{.}}{2017}]%
        {Jiang2017}
\bibfield{author}{\bibinfo{person}{Zhongshi Jiang}, \bibinfo{person}{Scott
  Schaefer}, {and} \bibinfo{person}{Daniele Panozzo}.}
  \bibinfo{year}{2017}\natexlab{}.
\newblock \showarticletitle{Simplicial Complex Augmentation Framework for
  Bijective Maps}.
\newblock \bibinfo{journal}{\emph{ACM Trans. Graph.}} \bibinfo{volume}{36},
  \bibinfo{number}{6}, Article \bibinfo{articleno}{186} (\bibinfo{date}{nov}
  \bibinfo{year}{2017}), \bibinfo{numpages}{9}~pages.
\newblock
\showISSN{0730-0301}
\urldef\tempurl%
\url{https://doi.org/10.1145/3130800.3130895}
\showDOI{\tempurl}


\bibitem[\protect\citeauthoryear{Julius, Kraevoy, and Sheffer}{Julius
  et~al\mbox{.}}{2005}]%
        {D-Charts:2005}
\bibfield{author}{\bibinfo{person}{Dan Julius}, \bibinfo{person}{Vladislav
  Kraevoy}, {and} \bibinfo{person}{Alla Sheffer}.}
  \bibinfo{year}{2005}\natexlab{}.
\newblock \showarticletitle{D-charts: Quasi-developable mesh segmentation}.
\newblock \bibinfo{journal}{\emph{Computer Graphics Forum}}
  \bibinfo{volume}{24}, \bibinfo{number}{3} (\bibinfo{year}{2005}),
  \bibinfo{pages}{581--590}.
\newblock


\bibitem[\protect\citeauthoryear{Korosteleva and Lee}{Korosteleva and
  Lee}{2021}]%
        {GarmentDataset:2021}
\bibfield{author}{\bibinfo{person}{Maria Korosteleva} {and}
  \bibinfo{person}{Sung-Hee Lee}.} \bibinfo{year}{2021}\natexlab{}.
\newblock \showarticletitle{Generating Datasets of 3D Garments with Sewing
  Patterns}. In \bibinfo{booktitle}{\emph{NeurIPS 2021 Datasets and Benchmarks
  Track}}.
\newblock


\bibitem[\protect\citeauthoryear{Kwok, Zhang, Wang, Liu, and Tang}{Kwok
  et~al\mbox{.}}{2015}]%
        {Kwok2015}
\bibfield{author}{\bibinfo{person}{Tsz~Ho Kwok}, \bibinfo{person}{Yan-Qiu
  Zhang}, \bibinfo{person}{Charlie Wang}, \bibinfo{person}{Yong-Jin Liu}, {and}
  \bibinfo{person}{Kai Tang}.} \bibinfo{year}{2015}\natexlab{}.
\newblock \showarticletitle{Styling Evolution for Tight-Fitting Garments}.
\newblock \bibinfo{journal}{\emph{IEEE Transactions on Visualization and
  Computer Graphics}}  \bibinfo{volume}{22} (\bibinfo{date}{01}
  \bibinfo{year}{2015}).
\newblock
\urldef\tempurl%
\url{https://doi.org/10.1109/TVCG.2015.2446472}
\showDOI{\tempurl}


\bibitem[\protect\citeauthoryear{L\'{e}vy, Petitjean, Ray, and
  Maillot}{L\'{e}vy et~al\mbox{.}}{2002}]%
        {Levy2002}
\bibfield{author}{\bibinfo{person}{Bruno L\'{e}vy}, \bibinfo{person}{Sylvain
  Petitjean}, \bibinfo{person}{Nicolas Ray}, {and} \bibinfo{person}{J\'{e}rome
  Maillot}.} \bibinfo{year}{2002}\natexlab{}.
\newblock \showarticletitle{Least Squares Conformal Maps for Automatic Texture
  Atlas Generation}.
\newblock \bibinfo{journal}{\emph{ACM Trans. Graph.}} \bibinfo{volume}{21},
  \bibinfo{number}{3} (\bibinfo{date}{jul} \bibinfo{year}{2002}),
  \bibinfo{pages}{362–371}.
\newblock
\showISSN{0730-0301}
\urldef\tempurl%
\url{https://doi.org/10.1145/566654.566590}
\showDOI{\tempurl}


\bibitem[\protect\citeauthoryear{L{\'{e}}vy, Petitjean, Ray, and
  Maillot}{L{\'{e}}vy et~al\mbox{.}}{2002}]%
        {Levy02}
\bibfield{author}{\bibinfo{person}{Bruno L{\'{e}}vy}, \bibinfo{person}{Sylvain
  Petitjean}, \bibinfo{person}{Nicolas Ray}, {and}
  \bibinfo{person}{J{\'{e}}r{\^{o}}me Maillot}.}
  \bibinfo{year}{2002}\natexlab{}.
\newblock \showarticletitle{Least squares conformal maps for automatic texture
  atlas generation}.
\newblock \bibinfo{journal}{\emph{{ACM} Trans. Graph.}} \bibinfo{volume}{21},
  \bibinfo{number}{3} (\bibinfo{year}{2002}), \bibinfo{pages}{362--371}.
\newblock


\bibitem[\protect\citeauthoryear{Li, Kaufman, Kim, Solomon, and Sheffer}{Li
  et~al\mbox{.}}{2018a}]%
        {Li:2018:OptCuts}
\bibfield{author}{\bibinfo{person}{Minchen Li}, \bibinfo{person}{Danny~M.
  Kaufman}, \bibinfo{person}{Vladimir~G. Kim}, \bibinfo{person}{Justin
  Solomon}, {and} \bibinfo{person}{Alla Sheffer}.}
  \bibinfo{year}{2018}\natexlab{a}.
\newblock \showarticletitle{OptCuts: Joint Optimization of Surface Cuts and
  Parameterization}.
\newblock \bibinfo{journal}{\emph{ACM Transactions on Graphics}}
  \bibinfo{volume}{37}, \bibinfo{number}{6} (\bibinfo{year}{2018}).
\newblock
\urldef\tempurl%
\url{https://doi.org/10.1145/3272127.3275042}
\showDOI{\tempurl}


\bibitem[\protect\citeauthoryear{Li, Sheffer, Grinspun, and Vining}{Li
  et~al\mbox{.}}{2018b}]%
        {Li:2018:FoldSketch}
\bibfield{author}{\bibinfo{person}{Minchen Li}, \bibinfo{person}{Alla Sheffer},
  \bibinfo{person}{Eitan Grinspun}, {and} \bibinfo{person}{Nicholas Vining}.}
  \bibinfo{year}{2018}\natexlab{b}.
\newblock \showarticletitle{FoldSketch: Enriching Garments with Physically
  Reproducible Folds}.
\newblock \bibinfo{journal}{\emph{ACM Transaction on Graphics}}
  \bibinfo{volume}{37}, \bibinfo{number}{4} (\bibinfo{year}{2018}).
\newblock
\urldef\tempurl%
\url{https://doi.org/10.1145/3197517.3201310}
\showDOI{\tempurl}


\bibitem[\protect\citeauthoryear{Liu, Zeng, Bruniaux, Tao, Yao, Li, and
  Wang}{Liu et~al\mbox{.}}{2018}]%
        {LIU2018113}
\bibfield{author}{\bibinfo{person}{Kaixuan Liu}, \bibinfo{person}{Xianyi Zeng},
  \bibinfo{person}{Pascal Bruniaux}, \bibinfo{person}{Xuyuan Tao},
  \bibinfo{person}{Xiaofeng Yao}, \bibinfo{person}{Victoria Li}, {and}
  \bibinfo{person}{Jianping Wang}.} \bibinfo{year}{2018}\natexlab{}.
\newblock \showarticletitle{3D interactive garment pattern-making technology}.
\newblock \bibinfo{journal}{\emph{Computer-Aided Design}}
  \bibinfo{volume}{104} (\bibinfo{year}{2018}), \bibinfo{pages}{113--124}.
\newblock
\showISSN{0010-4485}
\urldef\tempurl%
\url{https://doi.org/10.1016/j.cad.2018.07.003}
\showDOI{\tempurl}


\bibitem[\protect\citeauthoryear{Liu, Zhang, Xu, Gotsman, and Gortler}{Liu
  et~al\mbox{.}}{2008}]%
        {Liu2008}
\bibfield{author}{\bibinfo{person}{Ligang Liu}, \bibinfo{person}{Lei Zhang},
  \bibinfo{person}{Yin Xu}, \bibinfo{person}{Craig Gotsman}, {and}
  \bibinfo{person}{Steven~J. Gortler}.} \bibinfo{year}{2008}\natexlab{}.
\newblock \showarticletitle{A Local/Global Approach to Mesh Parameterization}.
  In \bibinfo{booktitle}{\emph{Proceedings of the Symposium on Geometry
  Processing}} (Copenhagen, Denmark) \emph{(\bibinfo{series}{SGP '08})}.
  \bibinfo{publisher}{Eurographics Association}, \bibinfo{address}{Goslar,
  DEU}, \bibinfo{pages}{1495–1504}.
\newblock


\bibitem[\protect\citeauthoryear{Livesu, Pietroni, Puppo, Sheffer, and
  Cignoni}{Livesu et~al\mbox{.}}{2020}]%
        {LivesuPPSC20}
\bibfield{author}{\bibinfo{person}{Marco Livesu}, \bibinfo{person}{Nico
  Pietroni}, \bibinfo{person}{Enrico Puppo}, \bibinfo{person}{Alla Sheffer},
  {and} \bibinfo{person}{Paolo Cignoni}.} \bibinfo{year}{2020}\natexlab{}.
\newblock \showarticletitle{LoopyCuts: practical feature-preserving block
  decomposition for strongly hex-dominant meshing}.
\newblock \bibinfo{journal}{\emph{ACM Trans.\ Graph.}} \bibinfo{volume}{39},
  \bibinfo{number}{4} (\bibinfo{year}{2020}), \bibinfo{pages}{121}.
\newblock


\bibitem[\protect\citeauthoryear{Loper, Mahmood, Romero, Pons-Moll, and
  Black}{Loper et~al\mbox{.}}{2015}]%
        {SMPL:2015}
\bibfield{author}{\bibinfo{person}{Matthew Loper}, \bibinfo{person}{Naureen
  Mahmood}, \bibinfo{person}{Javier Romero}, \bibinfo{person}{Gerard
  Pons-Moll}, {and} \bibinfo{person}{Michael~J. Black}.}
  \bibinfo{year}{2015}\natexlab{}.
\newblock \showarticletitle{{SMPL}: A Skinned Multi-Person Linear Model}.
\newblock \bibinfo{journal}{\emph{ACM Trans.\ Graph.}} \bibinfo{volume}{34},
  \bibinfo{number}{6} (\bibinfo{date}{Oct.} \bibinfo{year}{2015}),
  \bibinfo{pages}{248:1--248:16}.
\newblock


\bibitem[\protect\citeauthoryear{McCann, Albaugh, Narayanan, Grow, Matusik,
  Mankoff, and Hodgins}{McCann et~al\mbox{.}}{2016}]%
        {McCann2016}
\bibfield{author}{\bibinfo{person}{James McCann}, \bibinfo{person}{Lea
  Albaugh}, \bibinfo{person}{Vidya Narayanan}, \bibinfo{person}{April Grow},
  \bibinfo{person}{Wojciech Matusik}, \bibinfo{person}{Jennifer Mankoff}, {and}
  \bibinfo{person}{Jessica Hodgins}.} \bibinfo{year}{2016}\natexlab{}.
\newblock \showarticletitle{A Compiler for 3D Machine Knitting}.
\newblock \bibinfo{journal}{\emph{ACM Trans. Graph.}} \bibinfo{volume}{35},
  \bibinfo{number}{4}, Article \bibinfo{articleno}{49} (\bibinfo{date}{jul}
  \bibinfo{year}{2016}), \bibinfo{numpages}{11}~pages.
\newblock
\showISSN{0730-0301}
\urldef\tempurl%
\url{https://doi.org/10.1145/2897824.2925940}
\showDOI{\tempurl}


\bibitem[\protect\citeauthoryear{McCartney, Hinds, and Chong}{McCartney
  et~al\mbox{.}}{2005}]%
        {MCCARTNEY2005}
\bibfield{author}{\bibinfo{person}{J. McCartney}, \bibinfo{person}{B.K. Hinds},
  {and} \bibinfo{person}{K.W. Chong}.} \bibinfo{year}{2005}\natexlab{}.
\newblock \showarticletitle{Pattern flattening for orthotropic materials}.
\newblock \bibinfo{journal}{\emph{Computer-Aided Design}} \bibinfo{volume}{37},
  \bibinfo{number}{6} (\bibinfo{year}{2005}), \bibinfo{pages}{631--644}.
\newblock
\showISSN{0010-4485}
\urldef\tempurl%
\url{https://doi.org/10.1016/j.cad.2004.09.006}
\showDOI{\tempurl}
\newblock
\shownote{CAD Methods in Garment Design.}


\bibitem[\protect\citeauthoryear{McCartney, Hinds, Seow, and Gong}{McCartney
  et~al\mbox{.}}{2000}]%
        {McCartney2000}
\bibfield{author}{\bibinfo{person}{J. McCartney}, \bibinfo{person}{B.K. Hinds},
  \bibinfo{person}{B.L. Seow}, {and} \bibinfo{person}{D. Gong}.}
  \bibinfo{year}{2000}\natexlab{}.
\newblock \showarticletitle{An energy based model for the flattening of woven
  fabrics}.
\newblock \bibinfo{journal}{\emph{Journal of Materials Processing Tech.}}
  \bibinfo{volume}{107}, \bibinfo{number}{1-3} (\bibinfo{year}{2000}),
  \bibinfo{pages}{312--318}.
\newblock


\bibitem[\protect\citeauthoryear{Meng, Wang, and Jin}{Meng
  et~al\mbox{.}}{2012}]%
        {meng2012flexible}
\bibfield{author}{\bibinfo{person}{Yuwei Meng}, \bibinfo{person}{Charlie~CL
  Wang}, {and} \bibinfo{person}{Xiaogang Jin}.}
  \bibinfo{year}{2012}\natexlab{}.
\newblock \showarticletitle{Flexible shape control for automatic resizing of
  apparel products}.
\newblock \bibinfo{journal}{\emph{Computer-Aided Design}} \bibinfo{volume}{44},
  \bibinfo{number}{1} (\bibinfo{year}{2012}), \bibinfo{pages}{68--76}.
\newblock


\bibitem[\protect\citeauthoryear{Montes, Thomaszewski, Mudur, and Popa}{Montes
  et~al\mbox{.}}{2020}]%
        {Montes2020}
\bibfield{author}{\bibinfo{person}{Juan Montes}, \bibinfo{person}{Bernhard
  Thomaszewski}, \bibinfo{person}{Sudhir Mudur}, {and} \bibinfo{person}{Tiberiu
  Popa}.} \bibinfo{year}{2020}\natexlab{}.
\newblock \showarticletitle{Computational Design of Skintight Clothing}.
\newblock \bibinfo{journal}{\emph{ACM Trans. Graph.}} \bibinfo{volume}{39},
  \bibinfo{number}{4}, Article \bibinfo{articleno}{105} (\bibinfo{date}{jul}
  \bibinfo{year}{2020}), \bibinfo{numpages}{12}~pages.
\newblock
\showISSN{0730-0301}
\urldef\tempurl%
\url{https://doi.org/10.1145/3386569.3392477}
\showDOI{\tempurl}


\bibitem[\protect\citeauthoryear{Narayanan, Albaugh, Hodgins, Coros, and
  McCann}{Narayanan et~al\mbox{.}}{2018}]%
        {Narayanan2018}
\bibfield{author}{\bibinfo{person}{Vidya Narayanan}, \bibinfo{person}{Lea
  Albaugh}, \bibinfo{person}{Jessica Hodgins}, \bibinfo{person}{Stelian Coros},
  {and} \bibinfo{person}{James McCann}.} \bibinfo{year}{2018}\natexlab{}.
\newblock \showarticletitle{Automatic Machine Knitting of 3D Meshes}.
\newblock \bibinfo{journal}{\emph{ACM Trans. Graph.}} \bibinfo{volume}{37},
  \bibinfo{number}{3}, Article \bibinfo{articleno}{35} (\bibinfo{date}{Aug.}
  \bibinfo{year}{2018}), \bibinfo{numpages}{15}~pages.
\newblock
\showISSN{0730-0301}
\urldef\tempurl%
\url{https://doi.org/10.1145/3186265}
\showDOI{\tempurl}


\bibitem[\protect\citeauthoryear{Narayanan*, Wu*, Yuksel, and
  McCann}{Narayanan* et~al\mbox{.}}{2019}]%
        {Narayanan*2019}
\bibfield{author}{\bibinfo{person}{Vidya Narayanan*}, \bibinfo{person}{Kui
  Wu*}, \bibinfo{person}{Cem Yuksel}, {and} \bibinfo{person}{Jim McCann}.}
  \bibinfo{year}{2019}\natexlab{}.
\newblock \showarticletitle{Visual Knitting Machine Programming}.
\newblock \bibinfo{journal}{\emph{ACM Transactions on Graphics (Proceedings of
  SIGGRAPH 2019)}} \bibinfo{volume}{38}, \bibinfo{number}{4}, Article
  \bibinfo{articleno}{63} (\bibinfo{date}{jul} \bibinfo{year}{2019}),
  \bibinfo{numpages}{13}~pages.
\newblock
\showISSN{0730-0301}
\urldef\tempurl%
\url{https://doi.org/10.1145/3306346.3322995}
\showDOI{\tempurl}
\newblock
\shownote{(*Joint First Authors).}


\bibitem[\protect\citeauthoryear{Nayak and Padhye}{Nayak and Padhye}{2017}]%
        {nayak:2017:automation}
\bibfield{author}{\bibinfo{person}{Rajkishore Nayak} {and}
  \bibinfo{person}{Rajiv Padhye}.} \bibinfo{year}{2017}\natexlab{}.
\newblock \bibinfo{booktitle}{\emph{Automation in Garment Manufacturing}}.
\newblock \bibinfo{publisher}{Woodhead Publishing}.
\newblock


\bibitem[\protect\citeauthoryear{Nuvoli, Hernandez, Esperan{\c c}a, Scateni,
  Cignoni, and Pietroni}{Nuvoli et~al\mbox{.}}{2019}]%
        {QuadMixer}
\bibfield{author}{\bibinfo{person}{Stefano Nuvoli}, \bibinfo{person}{Alex
  Hernandez}, \bibinfo{person}{Claudio Esperan{\c c}a},
  \bibinfo{person}{Riccardo Scateni}, \bibinfo{person}{Paolo Cignoni}, {and}
  \bibinfo{person}{Nico Pietroni}.} \bibinfo{year}{2019}\natexlab{}.
\newblock \showarticletitle{QuadMixer: layout preserving blending of
  quadrilateral meshes}.
\newblock \bibinfo{journal}{\emph{ACM Trans. Graph}} \bibinfo{volume}{38},
  \bibinfo{number}{6} (\bibinfo{year}{2019}), \bibinfo{pages}{180:1--180:13}.
\newblock


\bibitem[\protect\citeauthoryear{Optitex}{Optitex}{2022}]%
        {Optitex}
\bibfield{author}{\bibinfo{person}{Optitex}.} \bibinfo{year}{2022}\natexlab{}.
\newblock \bibinfo{title}{optitex.com}.
\newblock \bibinfo{howpublished}{\url{https://optitex.com}}.
\newblock


\bibitem[\protect\citeauthoryear{Osman, Bolkart, and Black}{Osman
  et~al\mbox{.}}{2020}]%
        {osman2020star-body}
\bibfield{author}{\bibinfo{person}{Ahmed A~A Osman}, \bibinfo{person}{Timo
  Bolkart}, {and} \bibinfo{person}{Michael~J. Black}.}
  \bibinfo{year}{2020}\natexlab{}.
\newblock \showarticletitle{{STAR}: A Sparse Trained Articulated Human Body
  Regressor}. In \bibinfo{booktitle}{\emph{European Conference on Computer
  Vision (ECCV)}}. \bibinfo{pages}{598--613}.
\newblock
\urldef\tempurl%
\url{https://star.is.tue.mpg.de}
\showURL{%
\tempurl}


\bibitem[\protect\citeauthoryear{Panozzo, Lipman, Puppo, and Zorin}{Panozzo
  et~al\mbox{.}}{2012}]%
        {PanozzoLPZ12}
\bibfield{author}{\bibinfo{person}{Daniele Panozzo}, \bibinfo{person}{Yaron
  Lipman}, \bibinfo{person}{Enrico Puppo}, {and} \bibinfo{person}{Denis
  Zorin}.} \bibinfo{year}{2012}\natexlab{}.
\newblock \showarticletitle{Fields on symmetric surfaces}.
\newblock \bibinfo{journal}{\emph{ACM Trans. Graph}} \bibinfo{volume}{31},
  \bibinfo{number}{4} (\bibinfo{year}{2012}), \bibinfo{pages}{111:1--111:12}.
\newblock


\bibitem[\protect\citeauthoryear{Panozzo, Puppo, and Rocca}{Panozzo
  et~al\mbox{.}}{2010}]%
        {Pan2010}
\bibfield{author}{\bibinfo{person}{D. Panozzo}, \bibinfo{person}{Enrico Puppo},
  {and} \bibinfo{person}{L. Rocca}.} \bibinfo{year}{2010}\natexlab{}.
\newblock \showarticletitle{Efficient multi-scale curvature and crease
  estimation}.
\newblock \bibinfo{journal}{\emph{2nd International Workshop on Computer
  Graphics, Computer Vision and Mathematics, GraVisMa 2010 - Workshop
  Proceedings}} (\bibinfo{date}{01} \bibinfo{year}{2010}),
  \bibinfo{pages}{9--16}.
\newblock


\bibitem[\protect\citeauthoryear{Pietroni, Nuvoli, Alderighi, Cignoni, and
  Tarini}{Pietroni et~al\mbox{.}}{2021}]%
        {Pietroni2021}
\bibfield{author}{\bibinfo{person}{Nico Pietroni}, \bibinfo{person}{Stefano
  Nuvoli}, \bibinfo{person}{Thomas Alderighi}, \bibinfo{person}{Paolo Cignoni},
  {and} \bibinfo{person}{Marco Tarini}.} \bibinfo{year}{2021}\natexlab{}.
\newblock \showarticletitle{Reliable Feature-Line Driven Quad-Remeshing}.
\newblock \bibinfo{journal}{\emph{ACM Trans. Graph.}} \bibinfo{volume}{40},
  \bibinfo{number}{4}, Article \bibinfo{articleno}{155} (\bibinfo{date}{jul}
  \bibinfo{year}{2021}), \bibinfo{numpages}{17}~pages.
\newblock
\showISSN{0730-0301}
\urldef\tempurl%
\url{https://doi.org/10.1145/3450626.3459941}
\showDOI{\tempurl}


\bibitem[\protect\citeauthoryear{Pietroni, Puppo, Marcias, Scopigno, and
  Cignoni}{Pietroni et~al\mbox{.}}{2016}]%
        {Pietroni2016}
\bibfield{author}{\bibinfo{person}{Nico Pietroni}, \bibinfo{person}{Enrico
  Puppo}, \bibinfo{person}{Giorgio Marcias}, \bibinfo{person}{Roberto
  Scopigno}, {and} \bibinfo{person}{Paolo Cignoni}.}
  \bibinfo{year}{2016}\natexlab{}.
\newblock \showarticletitle{Tracing Field-Coherent Quad Layouts}.
\newblock \bibinfo{journal}{\emph{Comput. Graph. Forum}} \bibinfo{volume}{35},
  \bibinfo{number}{7} (\bibinfo{year}{2016}), \bibinfo{pages}{485--496}.
\newblock


\bibitem[\protect\citeauthoryear{Pons-Moll, Pujades, Hu, and Black}{Pons-Moll
  et~al\mbox{.}}{2017}]%
        {pons2017clothcap}
\bibfield{author}{\bibinfo{person}{Gerard Pons-Moll}, \bibinfo{person}{Sergi
  Pujades}, \bibinfo{person}{Sonny Hu}, {and} \bibinfo{person}{Michael~J
  Black}.} \bibinfo{year}{2017}\natexlab{}.
\newblock \showarticletitle{{ClothCap}: Seamless 4{D} clothing capture and
  retargeting}.
\newblock \bibinfo{journal}{\emph{ACM Trans.\ Graph.}} \bibinfo{volume}{36},
  \bibinfo{number}{4} (\bibinfo{year}{2017}), \bibinfo{pages}{1--15}.
\newblock


\bibitem[\protect\citeauthoryear{Poranne, Tarini, Huber, Panozzo, and
  Sorkine-Hornung}{Poranne et~al\mbox{.}}{2017}]%
        {Poranne:Autocuts:2017}
\bibfield{author}{\bibinfo{person}{Roi Poranne}, \bibinfo{person}{Marco
  Tarini}, \bibinfo{person}{Sandro Huber}, \bibinfo{person}{Daniele Panozzo},
  {and} \bibinfo{person}{Olga Sorkine-Hornung}.}
  \bibinfo{year}{2017}\natexlab{}.
\newblock \showarticletitle{Autocuts: Simultaneous Distortion and Cut
  Optimization for UV Mapping}.
\newblock \bibinfo{journal}{\emph{ACM Transactions on Graphics (proceedings of
  ACM SIGGRAPH ASIA)}} \bibinfo{volume}{36}, \bibinfo{number}{6}
  (\bibinfo{year}{2017}).
\newblock


\bibitem[\protect\citeauthoryear{Rabinovich, Poranne, Panozzo, and
  Sorkine-Hornung}{Rabinovich et~al\mbox{.}}{2017}]%
        {Rabinovich:SLIM:2017}
\bibfield{author}{\bibinfo{person}{Michael Rabinovich}, \bibinfo{person}{Roi
  Poranne}, \bibinfo{person}{Daniele Panozzo}, {and} \bibinfo{person}{Olga
  Sorkine-Hornung}.} \bibinfo{year}{2017}\natexlab{}.
\newblock \showarticletitle{Scalable Locally Injective Mappings}.
\newblock \bibinfo{journal}{\emph{ACM Transactions on Graphics}}
  \bibinfo{volume}{36}, \bibinfo{number}{2} (\bibinfo{date}{April}
  \bibinfo{year}{2017}), \bibinfo{pages}{16:1--16:16}.
\newblock


\bibitem[\protect\citeauthoryear{Razafindrazaka, Reitebuch, and
  Polthier}{Razafindrazaka et~al\mbox{.}}{2015}]%
        {RazaR15}
\bibfield{author}{\bibinfo{person}{Faniry~H. Razafindrazaka},
  \bibinfo{person}{Ulrich Reitebuch}, {and} \bibinfo{person}{Konrad Polthier}.}
  \bibinfo{year}{2015}\natexlab{}.
\newblock \showarticletitle{Perfect Matching Quad Layouts for Manifold Meshes}.
\newblock \bibinfo{journal}{\emph{Comput. Graph. Forum}} \bibinfo{volume}{34},
  \bibinfo{number}{5} (\bibinfo{year}{2015}), \bibinfo{pages}{219--228}.
\newblock


\bibitem[\protect\citeauthoryear{Robson, Maharik, Sheffer, and Carr}{Robson
  et~al\mbox{.}}{2011}]%
        {Robson:ContextAwareGarments:2011}
\bibfield{author}{\bibinfo{person}{C. Robson}, \bibinfo{person}{R. Maharik},
  \bibinfo{person}{A. Sheffer}, {and} \bibinfo{person}{N. Carr}.}
  \bibinfo{year}{2011}\natexlab{}.
\newblock \showarticletitle{Context-Aware Garment Modeling from Sketches}.
\newblock \bibinfo{journal}{\emph{Computers \& Graphics (Proc. SMI 2011)}}
  \bibinfo{volume}{35}, \bibinfo{number}{3} (\bibinfo{year}{2011}),
  \bibinfo{pages}{604--613}.
\newblock


\bibitem[\protect\citeauthoryear{Rose, Sheffer, Wither, Cani, and Thibert}{Rose
  et~al\mbox{.}}{2007}]%
        {Rose:DevelopableSurfaces:2007}
\bibfield{author}{\bibinfo{person}{Kenneth Rose}, \bibinfo{person}{Alla
  Sheffer}, \bibinfo{person}{Jamie Wither}, \bibinfo{person}{Marie-Paule Cani},
  {and} \bibinfo{person}{Boris Thibert}.} \bibinfo{year}{2007}\natexlab{}.
\newblock \showarticletitle{Developable surfaces from arbitrary sketched
  boundaries}. In \bibinfo{booktitle}{\emph{Proc.\ Symposium on Geometry
  Processing}}. Eurographics Association, \bibinfo{pages}{163--172}.
\newblock


\bibitem[\protect\citeauthoryear{Sawhney and Crane}{Sawhney and Crane}{2017}]%
        {Sawhney2017}
\bibfield{author}{\bibinfo{person}{Rohan Sawhney} {and} \bibinfo{person}{Keenan
  Crane}.} \bibinfo{year}{2017}\natexlab{}.
\newblock \showarticletitle{Boundary First Flattening}.
\newblock \bibinfo{journal}{\emph{CoRR}}  \bibinfo{volume}{abs/1704.06873}
  (\bibinfo{year}{2017}).
\newblock
\showeprint[arXiv]{1704.06873}
\urldef\tempurl%
\url{http://arxiv.org/abs/1704.06873}
\showURL{%
\tempurl}


\bibitem[\protect\citeauthoryear{Sharp and Crane}{Sharp and Crane}{2018}]%
        {Sharp:2018:VSC}
\bibfield{author}{\bibinfo{person}{Nicholas Sharp} {and}
  \bibinfo{person}{Keenan Crane}.} \bibinfo{year}{2018}\natexlab{}.
\newblock \showarticletitle{Variational Surface Cutting}.
\newblock \bibinfo{journal}{\emph{j-TOG}} \bibinfo{volume}{37},
  \bibinfo{number}{4} (\bibinfo{year}{2018}).
\newblock


\bibitem[\protect\citeauthoryear{Sheffer, L{\'e}vy, Mogilnitsky, and
  Bogomyakov}{Sheffer et~al\mbox{.}}{2004}]%
        {sheffer:inria-00105689}
\bibfield{author}{\bibinfo{person}{Alla Sheffer}, \bibinfo{person}{Bruno
  L{\'e}vy}, \bibinfo{person}{Maxim Mogilnitsky}, {and}
  \bibinfo{person}{Alexander Bogomyakov}.} \bibinfo{year}{2004}\natexlab{}.
\newblock \showarticletitle{{ABF++ : Fast and Robust Angle Based Flattening}}.
\newblock \bibinfo{journal}{\emph{ACM Trans.\ Graph.}} \bibinfo{volume}{24},
  \bibinfo{number}{2} (\bibinfo{year}{2004}), \bibinfo{pages}{311--330}.
\newblock
\urldef\tempurl%
\url{https://hal.inria.fr/inria-00105689}
\showURL{%
\tempurl}


\bibitem[\protect\citeauthoryear{SizeGermany}{SizeGermany}{2020}]%
        {SizeGermany}
\bibfield{author}{\bibinfo{person}{SizeGermany}.}
  \bibinfo{year}{2020}\natexlab{}.
\newblock \bibinfo{title}{SizeGermany}.
\newblock \bibinfo{howpublished}{\url{https://portal.sizegermany.de}}.
\newblock


\bibitem[\protect\citeauthoryear{Sorkine and Alexa}{Sorkine and Alexa}{2007}]%
        {ARAP07}
\bibfield{author}{\bibinfo{person}{Olga Sorkine} {and} \bibinfo{person}{Marc
  Alexa}.} \bibinfo{year}{2007}\natexlab{}.
\newblock \showarticletitle{{As-Rigid-As-Possible Surface Modeling}}. In
  \bibinfo{booktitle}{\emph{Geometry Processing}},
  \bibfield{editor}{\bibinfo{person}{Alexander Belyaev} {and}
  \bibinfo{person}{Michael Garland}} (Eds.). \bibinfo{publisher}{The
  Eurographics Association}.
\newblock
\showISBNx{978-3-905673-46-3}
\showISSN{1727-8384}
\urldef\tempurl%
\url{https://doi.org/10.2312/SGP/SGP07/109-116}
\showDOI{\tempurl}


\bibitem[\protect\citeauthoryear{Sorkine, Cohen-Or, Goldenthal, and
  Lischinski}{Sorkine et~al\mbox{.}}{2002}]%
        {BoundedDistortParam:2002}
\bibfield{author}{\bibinfo{person}{Olga Sorkine}, \bibinfo{person}{Daniel
  Cohen-Or}, \bibinfo{person}{Rony Goldenthal}, {and} \bibinfo{person}{Dani
  Lischinski}.} \bibinfo{year}{2002}\natexlab{}.
\newblock \showarticletitle{Bounded-distortion piecewise mesh
  parameterization}. In \bibinfo{booktitle}{\emph{Proceedings of IEEE
  Visualization}} (Boston, Massachusetts). \bibinfo{publisher}{IEEE Computer
  Society}, \bibinfo{pages}{355--362}.
\newblock


\bibitem[\protect\citeauthoryear{Sorkine-Hornung and
  Rabinovich}{Sorkine-Hornung and Rabinovich}{2016}]%
        {SorkineRabinovich:SVD-rotations:2016}
\bibfield{author}{\bibinfo{person}{Olga Sorkine-Hornung} {and}
  \bibinfo{person}{Michael Rabinovich}.} \bibinfo{year}{2016}\natexlab{}.
\newblock \bibinfo{title}{Least-Squares Rigid Motion Using SVD}.
\newblock
\newblock
\newblock
\shownote{Technical note.}


\bibitem[\protect\citeauthoryear{Stein, Grinspun, and Crane}{Stein
  et~al\mbox{.}}{2018}]%
        {Stein:2018}
\bibfield{author}{\bibinfo{person}{Oded Stein}, \bibinfo{person}{Eitan
  Grinspun}, {and} \bibinfo{person}{Keenan Crane}.}
  \bibinfo{year}{2018}\natexlab{}.
\newblock \showarticletitle{Developability of triangle meshes}.
\newblock \bibinfo{journal}{\emph{ACM Trans.\ Graph.}} \bibinfo{volume}{37},
  \bibinfo{number}{4} (\bibinfo{year}{2018}).
\newblock


\bibitem[\protect\citeauthoryear{TiltBrush}{TiltBrush}{2022}]%
        {TiltBrush}
\bibfield{author}{\bibinfo{person}{TiltBrush}.}
  \bibinfo{year}{2022}\natexlab{}.
\newblock \bibinfo{title}{www.tiltbrush.com}.
\newblock \bibinfo{howpublished}{\url{https://www.tiltbrush.com/}}.
\newblock


\bibitem[\protect\citeauthoryear{Turquin, Wither, Boissieux, Cani, and
  Hughes}{Turquin et~al\mbox{.}}{2007}]%
        {Turquin:SketchInterface:2007}
\bibfield{author}{\bibinfo{person}{Emmanuel Turquin}, \bibinfo{person}{Jamie
  Wither}, \bibinfo{person}{Laurence Boissieux}, \bibinfo{person}{Marie-Paule
  Cani}, {and} \bibinfo{person}{John~F Hughes}.}
  \bibinfo{year}{2007}\natexlab{}.
\newblock \showarticletitle{A sketch-based interface for clothing virtual
  characters}.
\newblock \bibinfo{journal}{\emph{IEEE Computer Graphics and Applications}}
  \bibinfo{volume}{27}, \bibinfo{number}{1} (\bibinfo{year}{2007}).
\newblock


\bibitem[\protect\citeauthoryear{Umetani, Kaufman, Igarashi, and
  Grinspun}{Umetani et~al\mbox{.}}{2011}]%
        {Umetani:2011}
\bibfield{author}{\bibinfo{person}{Nobuyuki Umetani}, \bibinfo{person}{Danny~M.
  Kaufman}, \bibinfo{person}{Takeo Igarashi}, {and} \bibinfo{person}{Eitan
  Grinspun}.} \bibinfo{year}{2011}\natexlab{}.
\newblock \showarticletitle{Sensitive Couture for Interactive Garment Editing
  and Modeling}.
\newblock \bibinfo{journal}{\emph{ACM Transactions on Graphics (SIGGRAPH
  2011)}} \bibinfo{volume}{30}, \bibinfo{number}{4} (\bibinfo{year}{2011}).
\newblock


\bibitem[\protect\citeauthoryear{Vaxman, Campen, Diamanti, Bommes, Hildebrandt,
  Ben{-}Chen, and Panozzo}{Vaxman et~al\mbox{.}}{2017}]%
        {Vaxman2017}
\bibfield{author}{\bibinfo{person}{Amir Vaxman}, \bibinfo{person}{Marcel
  Campen}, \bibinfo{person}{Olga Diamanti}, \bibinfo{person}{David Bommes},
  \bibinfo{person}{Klaus Hildebrandt}, \bibinfo{person}{Mirela Ben{-}Chen},
  {and} \bibinfo{person}{Daniele Panozzo}.} \bibinfo{year}{2017}\natexlab{}.
\newblock \showarticletitle{Directional field synthesis, design, and
  processing}. In \bibinfo{booktitle}{\emph{SIGGRAPH '17 Courses}}.
  \bibinfo{pages}{12:1--12:30}.
\newblock


\bibitem[\protect\citeauthoryear{Vidaurre, Santesteban, 0001, and
  Casas}{Vidaurre et~al\mbox{.}}{2020}]%
        {Vidarre2020}
\bibfield{author}{\bibinfo{person}{Raquel Vidaurre}, \bibinfo{person}{Igor
  Santesteban}, \bibinfo{person}{Elena~Garces 0001}, {and} \bibinfo{person}{Dan
  Casas}.} \bibinfo{year}{2020}\natexlab{}.
\newblock \showarticletitle{Fully Convolutional Graph Neural Networks for
  Parametric Virtual Try-On}.
\newblock \bibinfo{journal}{\emph{CoRR}}  \bibinfo{volume}{abs/2009.04592}
  (\bibinfo{year}{2020}).
\newblock


\bibitem[\protect\citeauthoryear{Wang, Wang, and Yuen}{Wang
  et~al\mbox{.}}{2005b}]%
        {Wang2005}
\bibfield{author}{\bibinfo{person}{{Charlie C.L.} Wang}, \bibinfo{person}{Yu
  Wang}, {and} \bibinfo{person}{{Matthew M.F.} Yuen}.}
  \bibinfo{year}{2005}\natexlab{b}.
\newblock \showarticletitle{Design automation for customized apparel products}.
\newblock \bibinfo{journal}{\emph{Computer-Aided Design}} \bibinfo{volume}{37},
  \bibinfo{number}{7} (\bibinfo{date}{1 June} \bibinfo{year}{2005}),
  \bibinfo{pages}{675--691}.
\newblock
\showISSN{0010-4485}
\urldef\tempurl%
\url{https://doi.org/10.1016/j.cad.2004.08.007}
\showDOI{\tempurl}


\bibitem[\protect\citeauthoryear{Wang, Tang, and Yeung}{Wang
  et~al\mbox{.}}{2005a}]%
        {Wang:WovenMesh:2005}
\bibfield{author}{\bibinfo{person}{Charlie~CL Wang}, \bibinfo{person}{Kai
  Tang}, {and} \bibinfo{person}{Benjamin~ML Yeung}.}
  \bibinfo{year}{2005}\natexlab{a}.
\newblock \showarticletitle{Freeform surface flattening based on fitting a
  woven mesh model}.
\newblock \bibinfo{journal}{\emph{Computer-Aided Design}} \bibinfo{volume}{37},
  \bibinfo{number}{8} (\bibinfo{year}{2005}), \bibinfo{pages}{799--814}.
\newblock


\bibitem[\protect\citeauthoryear{Wang}{Wang}{2018}]%
        {Wang2018patterns}
\bibfield{author}{\bibinfo{person}{Huamin Wang}.}
  \bibinfo{year}{2018}\natexlab{}.
\newblock \showarticletitle{Rule-Free Sewing Pattern Adjustment with Precision
  and Efficiency}.
\newblock \bibinfo{journal}{\emph{ACM Trans. Graph.}} \bibinfo{volume}{37},
  \bibinfo{number}{4}, Article \bibinfo{articleno}{53} (\bibinfo{date}{jul}
  \bibinfo{year}{2018}), \bibinfo{numpages}{13}~pages.
\newblock
\showISSN{0730-0301}
\urldef\tempurl%
\url{https://doi.org/10.1145/3197517.3201320}
\showDOI{\tempurl}


\bibitem[\protect\citeauthoryear{Wang, Ceylan, Popovic, and Mitra}{Wang
  et~al\mbox{.}}{2018}]%
        {Wang:GarmentShapeSpace:2018}
\bibfield{author}{\bibinfo{person}{Tuanfeng~Y. Wang}, \bibinfo{person}{Duygu
  Ceylan}, \bibinfo{person}{Jovan Popovic}, {and} \bibinfo{person}{Niloy~J.
  Mitra}.} \bibinfo{year}{2018}\natexlab{}.
\newblock \showarticletitle{Learning a Shared Shape Space for Multimodal
  Garment Design}.
\newblock \bibinfo{journal}{\emph{ACM Trans.\ Graph.}} \bibinfo{volume}{37},
  \bibinfo{number}{6} (\bibinfo{year}{2018}), \bibinfo{pages}{1:1--1:14}.
\newblock
\urldef\tempurl%
\url{https://doi.org/10.1145/3272127.3275074}
\showDOI{\tempurl}


\bibitem[\protect\citeauthoryear{Wibowo, Sakamoto, Mitani, and Igarashi}{Wibowo
  et~al\mbox{.}}{2012}]%
        {Wibowo:Dressup:2012}
\bibfield{author}{\bibinfo{person}{Amy Wibowo}, \bibinfo{person}{Daisuke
  Sakamoto}, \bibinfo{person}{Jun Mitani}, {and} \bibinfo{person}{Takeo
  Igarashi}.} \bibinfo{year}{2012}\natexlab{}.
\newblock \showarticletitle{DressUp: a 3D interface for clothing design with a
  physical mannequin}. In \bibinfo{booktitle}{\emph{Proc.\ International
  Conference on Tangible, Embedded and Embodied Interaction}}. ACM,
  \bibinfo{pages}{99--102}.
\newblock


\bibitem[\protect\citeauthoryear{Wolff, Herholz, and Sorkine-Hornung}{Wolff
  et~al\mbox{.}}{2019}]%
        {Wolff:Symmetry:VMV2019}
\bibfield{author}{\bibinfo{person}{Katja Wolff}, \bibinfo{person}{Philipp
  Herholz}, {and} \bibinfo{person}{Olga Sorkine-Hornung}.}
  \bibinfo{year}{2019}\natexlab{}.
\newblock \showarticletitle{Reﬂection Symmetry in Textured Sewing Patterns}.
  In \bibinfo{booktitle}{\emph{Proceedings of the Symposium on Vision, Modeling
  and Visualization (VMV)}}. \bibinfo{publisher}{Eurographics Association}.
\newblock


\bibitem[\protect\citeauthoryear{Wolff, Herholz, Ziegler, Link, Br{\"{u}}gel,
  and Sorkine{-}Hornung}{Wolff et~al\mbox{.}}{2021}]%
        {Katja:2021}
\bibfield{author}{\bibinfo{person}{Katja Wolff}, \bibinfo{person}{Philipp
  Herholz}, \bibinfo{person}{Verena Ziegler}, \bibinfo{person}{Frauke Link},
  \bibinfo{person}{Nico Br{\"{u}}gel}, {and} \bibinfo{person}{Olga
  Sorkine{-}Hornung}.} \bibinfo{year}{2021}\natexlab{}.
\newblock \showarticletitle{3{D} Custom Fit Garment Design with Body Movement}.
\newblock \bibinfo{journal}{\emph{CoRR}}  \bibinfo{volume}{abs/2102.05462}
  (\bibinfo{year}{2021}).
\newblock
\showeprint[arXiv]{2102.05462}
\urldef\tempurl%
\url{https://arxiv.org/abs/2102.05462}
\showURL{%
\tempurl}


\bibitem[\protect\citeauthoryear{Wu, Swan, and Yuksel}{Wu
  et~al\mbox{.}}{2019}]%
        {Wu2019}
\bibfield{author}{\bibinfo{person}{Kui Wu}, \bibinfo{person}{Hannah Swan},
  {and} \bibinfo{person}{Cem Yuksel}.} \bibinfo{year}{2019}\natexlab{}.
\newblock \showarticletitle{Knittable Stitch Meshes}.
\newblock \bibinfo{journal}{\emph{ACM Transactions on Graphics}}
  \bibinfo{volume}{38}, \bibinfo{number}{1}, Article \bibinfo{articleno}{10}
  (\bibinfo{date}{jan} \bibinfo{year}{2019}), \bibinfo{numpages}{13}~pages.
\newblock
\showISSN{0730-0301}
\urldef\tempurl%
\url{https://doi.org/10.1145/3292481}
\showDOI{\tempurl}


\bibitem[\protect\citeauthoryear{Yuksel, Kaldor, James, and Marschner}{Yuksel
  et~al\mbox{.}}{2012}]%
        {Yuksel2012}
\bibfield{author}{\bibinfo{person}{Cem Yuksel}, \bibinfo{person}{Jonathan~M.
  Kaldor}, \bibinfo{person}{Doug~L. James}, {and} \bibinfo{person}{Steve
  Marschner}.} \bibinfo{year}{2012}\natexlab{}.
\newblock \showarticletitle{Stitch Meshes for Modeling Knitted Clothing with
  Yarn-Level Detail}.
\newblock \bibinfo{journal}{\emph{ACM Transactions on Graphics (Proceedings of
  SIGGRAPH 2012)}} \bibinfo{volume}{31}, \bibinfo{number}{3}, Article
  \bibinfo{articleno}{37} (\bibinfo{year}{2012}), \bibinfo{numpages}{12}~pages.
\newblock
\urldef\tempurl%
\url{https://doi.org/10.1145/2185520.2185533}
\showDOI{\tempurl}


\end{thebibliography}

\end{document}